\documentclass[sn-nature,Numbered]{sn-jnl}
\usepackage{graphicx,tabularx}%
\usepackage{multirow}%
\usepackage{amsmath,amssymb,amsfonts}%
\usepackage{amsthm}%
\usepackage{mathrsfs}%
\usepackage[title]{appendix}%
\usepackage[table]{xcolor}
\usepackage{textcomp}%
\usepackage{manyfoot}%
\usepackage{bigints}
\usepackage{booktabs}%
\usepackage{algorithm}%
\usepackage{algorithmicx}%
\usepackage{algpseudocode}%
\usepackage{listings}%
\usepackage{nature_jnlabbrev}
\usepackage[export]{adjustbox}
\usepackage{academicons}
\usepackage{xcolor}
\usepackage{orcidlink}
\usepackage{lineno}
\usepackage{hyperref}
\usepackage[numbers]{natbib}

\usepackage{multibib} 
\newcites{Methods}{Methods References}

\begin{document}
\renewcommand{\figurename}{\hspace{-4pt}}
\renewcommand{\thefigure}{Fig.~\arabic{figure}}
\renewcommand{\theHfigure}{Fig.~\arabic{figure}}
\renewcommand{\tablename}{\hspace{-4pt}}
\renewcommand{\thetable}{Table \arabic{table}}
\renewcommand{\theHtable}{Table \arabic{table}}
\newcommand{\equalcontribsymbol}{\textsuperscript{\dag}}

\title[Article Title]{First JWST thermal phase curves of temperate terrestrial exoplanets reveal no thick atmosphere around TRAPPIST-1 b and c}


\author*[1]{\fnm{Micha\"el} \sur{Gillon}\, \orcidlink{0000-0003-1462-7739}}
\equalcont{These authors contributed equally to this work.}
\email{michael.gillon@uliege.be}

\author*[2,3,1]{Elsa Ducrot\,
\orcidlink{0000-0002-7008-6888}}
\equalcont{These authors contributed equally to this work.}
\email{elsa.ducrot@obspm.fr}

\author[4,5,6]{Taylor J. Bell\, \orcidlink{0000-0003-4177-2149}}

\author[7]{Ziyu Huang\, \orcidlink{0000-0002-8624-1264}}

\author[8, 9]{Andrew Lincowski\, \orcidlink{0000-0003-0429-9487}}

\author[10]{Xintong Lyu\,\orcidlink{0009-0004-9766-036X}}

\author[11,12]{Alice Maurel}

\author[13,14]{Alexandre Revol\,\orcidlink{0009-0004-9766-036X}}

\author[8]{Eric Agol\,
\orcidlink{0000-0002-0802-9145}}

\author[13,14]{Emeline Bolmont\,\orcidlink{0000-0001-5657-4503}}

\author[7]{Chuanfei Dong\, \orcidlink{0000-0002-8990-094X}}

\author[15,16,17]{Thomas J. Fauchez\,\orcidlink{0000-0002-5967-9631}}

\author[12]{Daniel D.B. Koll\,\orcidlink{0000-0002-9076-6901}}

\author[18]{Jérémy Leconte\, \orcidlink{https://orcid.org/0000-0002-3555-480X}}

\author[8]{Victoria S. Meadows\,\orcidlink{0000-0002-1386-1710}}

\author[18]{Franck Selsis\, \orcidlink{0000-0001-9619-5356}}

\author[12,18]{Martin Turbet\, \orcidlink{0000-0003-2260-9856}}

\author[2]{Benjamin Charnay\,}

\author[1]{Laetitia Delrez\,\orcidlink{0000-0001-6108-4808}}

\author[18,20]{Brice-Olivier Demory\,\orcidlink{0000-0002-9355-5165}}

\author[21,22]{Aaron Householder\, \orcidlink{0000-0002-5812-3236}}

\author[23]{Sebastian Zieba\,\orcidlink{0000-0003-0562-6750}}

\author[21]{David Berardo\, \orcidlink{0000-0001-6298-412X}}

\author[3]{Achrène Dyrek\,\orcidlink{0000-0001-7189-6463}}

\author[24]{Billy Edwards\,\orcidlink{0000-0002-5494-3237}}

\author[21]{Julien de Wit\,\orcidlink{0000-0003-2415-2191}}

\author[5]{Thomas P. Greene\,\orcidlink{0000-0002-8963-8056}}

\author[25,26]{Renyu Hu\,\orcidlink{0000-0003-2215-8485}}

\author[27]{Nicolas Iro\,\orcidlink{0000-0003-2329-418X}}

\author[23]{Laura Kreidberg\, \orcidlink{0000-0003-0514-1147}}

\author[3]{Pierre-Olivier Lagage}

\author[28]{Jacob Lustig-Yaeger\,\orcidlink{0000-0002-0746-1980}}

\author[15]{Aishwarya Iyer\,\orcidlink{https://orcid.org/0000-0003-0971-1709}}

\affil[1]{\orgdiv{Astrobiology Research Unit}, \orgname{Universit\'e de Li\`ege}, \orgaddress{\street{All\'ee du 6 ao\^ut 19}, \city{Li\`ege}, \postcode{4000}, \country{Belgium}}}

\affil[2]{LIRA, Observatoire de Paris, CNRS, Universit\'e Paris Diderot, Universit\'e Pierre et Marie Curie, 5 place Jules Janssen, 92190 Meudon, France}

\affil[3]{AIM, CEA, CNRS, Universit\'e Paris-Saclay, Universit\'e de Paris, F-91191
Gif-sur-Yvette, France}

\affil[4]{Bay Area Environmental Research Institute, NASA Ames Research Center, M.S. 245-6, Moffett Field, 94035 CA}

\affil[5]{Space Science and Astrobiology Division, NASA Ames Research Center, M.S. 245-6, Moffett Field, 94035 CA}

\affil[6]{AURA for the European Space Agency (ESA), Space Telescope Science Institute, 3700 San Martin Drive, Baltimore, MD 21218, USA}

\affil[7]{Department of Astronomy, Boston University, Boston, MA 02215, USA}

\affil[8]{Department of Astronomy and NASA Nexus for Exoplanet System Science, Virtual Planetary Laboratory Team,
University of Washington, Seattle, WA, USA}

\affil[9]{Eastern Wyoming College, Torrington, WY, USA}

\affil[10]{Department of Atmospheric and Oceanic Sciences, Peking University, Beijing, China}

\affil[11]{Sorbonne Universit\'es, UPMC Universit\'e Paris 6 et CNRS, UMR 7095, Institut d’Astrophysique de Paris, 98 bis bd Arago, 75014 Paris, France}

\affil[12]{Laboratoire de M\'et\'eorologie Dynamique/IPSL, CNRS, Sorbonne Universit\'e, Ecole Normale Sup\'erieure, Universit\'e PSL, Ecole Polytechnique, Institut Polytechnique de Paris, 75005 Paris, France}

\affil[13]{D\'epartement d’astronomie de l’Universit\'e de Gen\`eve, Chemin Pegasi, 51, 1290 Sauverny, Switzerland}

\affil[14]{Centre sur la Vie dans l’Univers, Universit\'e de Gen\`eve, Switzerland}

\affil[15]{NASA Goddard Space Flight Center, 8800 Greenbelt Road, Greenbelt, MD 20771, USA}

\affil[16]{Sellers Exoplanet Environment Collaboration (SEEC), NASA Goddard Space Flight Center, USA}

\affil[17]{Integrated Space Science and Technology Institute, Department of Physics, American University, Washington DC, USA}

\affil[18]{Laboratoire d'astrophysique de Bordeaux, Univ. Bordeaux, CNRS, B18N, all\'ee Geoffroy Saint-Hilaire, 33615 Pessac, France}

\affil[19]{Center for Space and Habitability, University of Bern, Gesellschaftsstrasse 6, 3012 Bern, Switzerland.}

\affil[20]{Space Research and Planetary Sciences, Physics Institute, University of Bern, Gesellschaftsstrasse 6, 3012 Bern, Switzerland.}

\affil[21]{Department of Earth, Atmospheric and Planetary Sciences, Massachusetts Institute of Technology, Cambridge, MA 02139, USA}

\affil[22]{Kavli Institute for Astrophysics and Space Research, Massachusetts Institute of Technology, Cambridge, MA 02139, USA}

\affil[23]{Center for Astrophysics, Harvard \& Smithsonian, 60 Garden Street, Cambridge, MA 02138, USA}

\affil[24]{SRON, Netherlands Institute for Space Research, Niels Bohrweg 4, NL-2333 CA, Leiden, The Netherlands}

\affil[25]{Jet Propulsion Laboratory, California Institute of Technology, Pasadena, CA 91109, USA}

\affil[26]{Division of Geological and Planetary Sciences, California Institute of Technology, CA 91125, USA}

\affil[27]{Institut f\"ur Planetenforschung, Deutsches Zentrum f\"ur Luft- und Raumfahrt (DLR), D-12489 Berlin, Germany}

\affil[28]{Johns Hopkins Applied Physics Laboratory, Laurel, Maryland 20723, USA}




\abstract{We report JWST/MIRI 15~$\mu$m phase curves of TRAPPIST-1\,b and c, revealing thermal emission consistent with their irradiation levels, assuming no efficient heat redistribution. We find that TRAPPIST-1\,b shows a high dayside brightness temperature (490~$\pm$~17~K), no significantly detectable nightside emission ($F_{\rm b, Night, max}$ = $39_{-27}^{+55}$ ppm), and no phase offset---features consistent with a low-albedo, airless ultramafic rocky surface. TRAPPIST-1\,c exhibits a lower dayside brightness temperature (369~$\pm$~23~K), and a nightside flux statistically indistinguishable from that of TRAPPIST-1\,b ($F_{\rm c, Night, max}$ = $62_{-43}^{+60}$ ppm). Atmosphere models with surface pressures $\geq$1~bar and efficient greenhouse effects are strongly disfavored for both planets. TRAPPIST-1\,b is unlikely to possess any substantial atmosphere, while TRAPPIST-1\,c may retain a tenuous, greenhouse-poor O$_2$-dominated atmosphere or be similarly airless with a more reflective surface. These results suggest divergent evolutionary pathways or atmospheric loss processes, despite similar compositions. These measurements tightly constrain atmosphere retention in the inner TRAPPIST-1 system.}

\keywords{Exoplanets, Ultracool Dwarfs}



\maketitle

\section{Main}\label{sec1}

Small rocky planets are known to be common in temperate orbits around low-mass M-dwarfs \cite{Ment2023}. An outstanding question concerns the survivability of their planets' atmospheres over billions of years. The TRAPPIST-1 system represents a unique laboratory to empirically address this question with JWST \cite{Gillon2020, Lustig-Yaeger:2019}. Notably, the two inner planets have irradiations large enough ($\sim$4.15 and 2.2 $S_\oplus$) to enable the photometric measurement of their thermal emission with the JWST MIRI instrument. A large thermal emission was measured for the dayside of planet b at 15 $\mu$m  (program GTO 1177) \cite{Greene2023Natur}, suggestive of poor atmospheric heat redistribution, and consistent with the emission of a null-albedo bare-rock scenario or with some low-density atmosphere scenarios \cite{Ih2023}. More recently, a similar measurement was performed with MIRI at 12.8 $\mu$m (program GTO 1279) \cite{ducrot_2024}. The combined analysis of the 12.8 and 15 $\mu$m data showed that they could be well fitted by either an airless planet model with a relatively fresh surface (and albedo $\simeq$ 0.1), or by a thick, pure-CO$_2$ atmosphere with photochemical hazes resulting in CO$_2$ being seen in emission \cite{ducrot_2024}. For planet c's dayside, a relatively high thermal emission was measured with MIRI at 15 $\mu$m (program GO 2304) that disfavored both a dense CO$_2$-rich atmosphere or a null albedo bare rock scenario, and left room for some alternative atmospheric (CO$_2$-poor, low density) and airless (low to moderate albedo surface) scenarios \cite{Zieba2023}, as well as for steam atmospheres \cite{Lincowski2023,Turbet:2023}. As predicted by ref. \cite{Hammond2025}, these initial observations highlight the challenge of definitively detecting—or ruling out—exoplanet atmospheres based solely on their photometric dayside emission. In this context, ref. \cite{Hammond2025} demonstrated that phase curves are essential to detect or rule out the presence of atmospheres through nightside thermal emission.

Thanks to their resonant orbits and fortuitous orientation, planets b and c regularly have their thermal curves `in phase', meaning that their occultations happen at the same time. Their individual phase curves then add up in a constructive way, resulting in a combined phase curve whose amplitude could be boosted up to $>$1000 ppm in the case of poor heat distribution for both planets. To further constrain the presence/absence of atmospheres around the planets, such a combined phase curve of the two planets was observed at 15 $\mu$m with MIRI from Nov. 22 to Nov. 25, 2023 (Program GO 3077), covering a full phase curve of b and a partial (90\%) phase curve of c. These observations were obtained in imaging mode using the BRIGHTSKY (512 $\times$ 512 pixels) detector of MIRI and filter F1500W. In total, 5336 integrations were gathered, with a cadence of 39s, resulting in 59 hours of continuous observations. The goal of these observations was to further constrain the presence/absence of an atmosphere around the planets by assessing the efficiency of heat redistribution to their night sides.

Four independent analyses of all existing MIRI data on TRAPPIST-1 at 15$\mu m$ (secondary eclipses + phase curve) were conducted (see Methods), all yielding consistent results. Among them, the analysis by co-author MG was chosen due to its achieving the smallest standard deviation in the light curve residuals. A brief overview of this `fiducial' analysis is provided in the main text, with comprehensive details of all four analyses available in the Methods section. The images of program 3077, 1177 and 2304 were calibrated using \texttt{Eureka!} \cite{Bell2022} and the JWST pipeline (see Methods for details). The photometric extraction was then performed on the calibrated images with \texttt{IRAF/DAOPHOT} \cite{Stetson1987} and an aperture radius of 4 pixels. The resulting light curves were normalized and outliers were discarded using a 4-$\sigma$ clipping algorithm with a 20-min moving median on the fluxes and external parameters. The top panel of   \ref{fig:lc_raw} shows the resulting raw light curve of program 3077. Its first $\sim$4 hours contains a large structure that we attribute to an instrumental effect due to the stabilization of the detector. The light curve also shows a double flare-like structure. We discarded the data points corresponding to these two structures.

\begin{figure}
   \centering
    \includegraphics[width=0.9\textwidth]{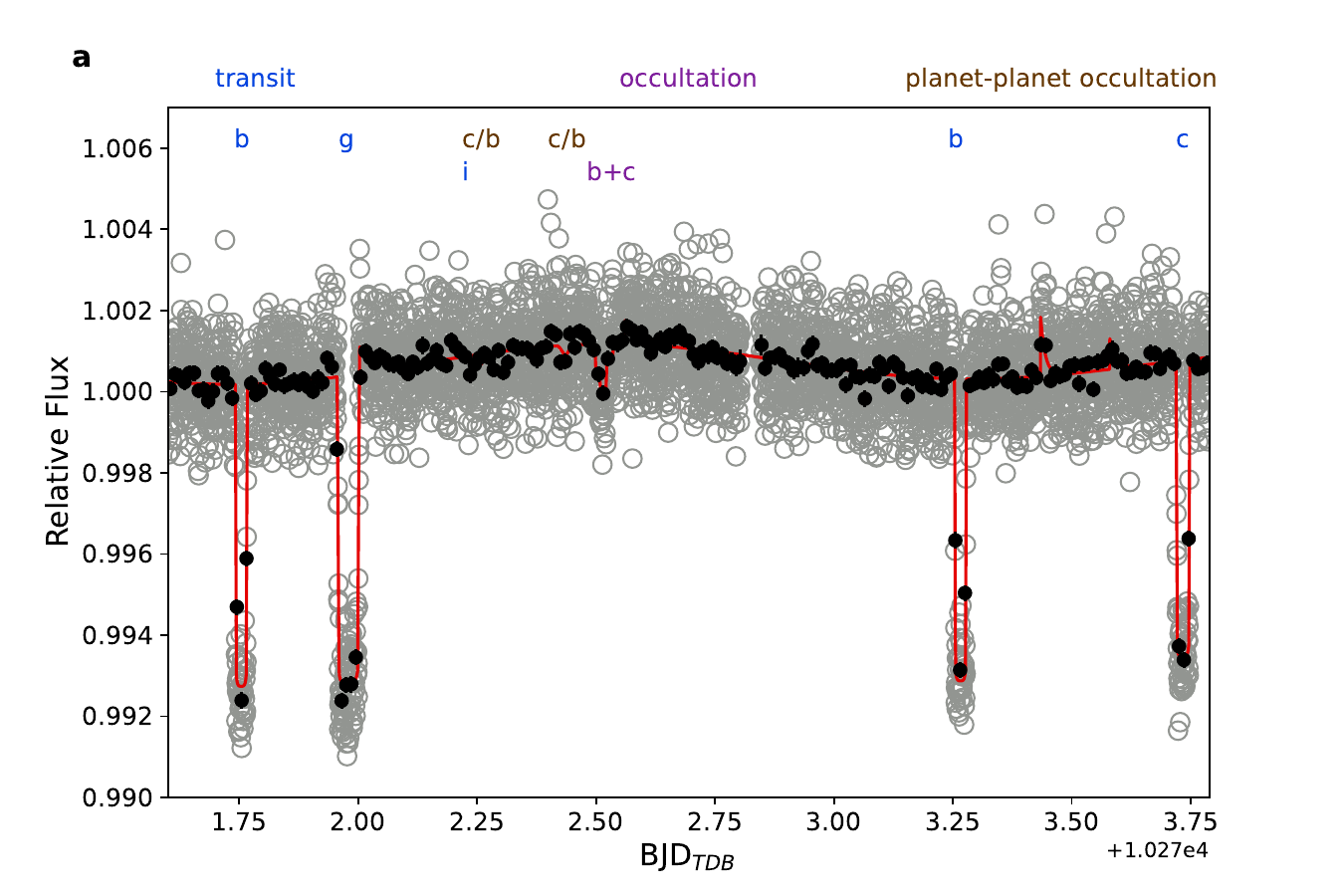}
    \includegraphics[width=0.9\textwidth]{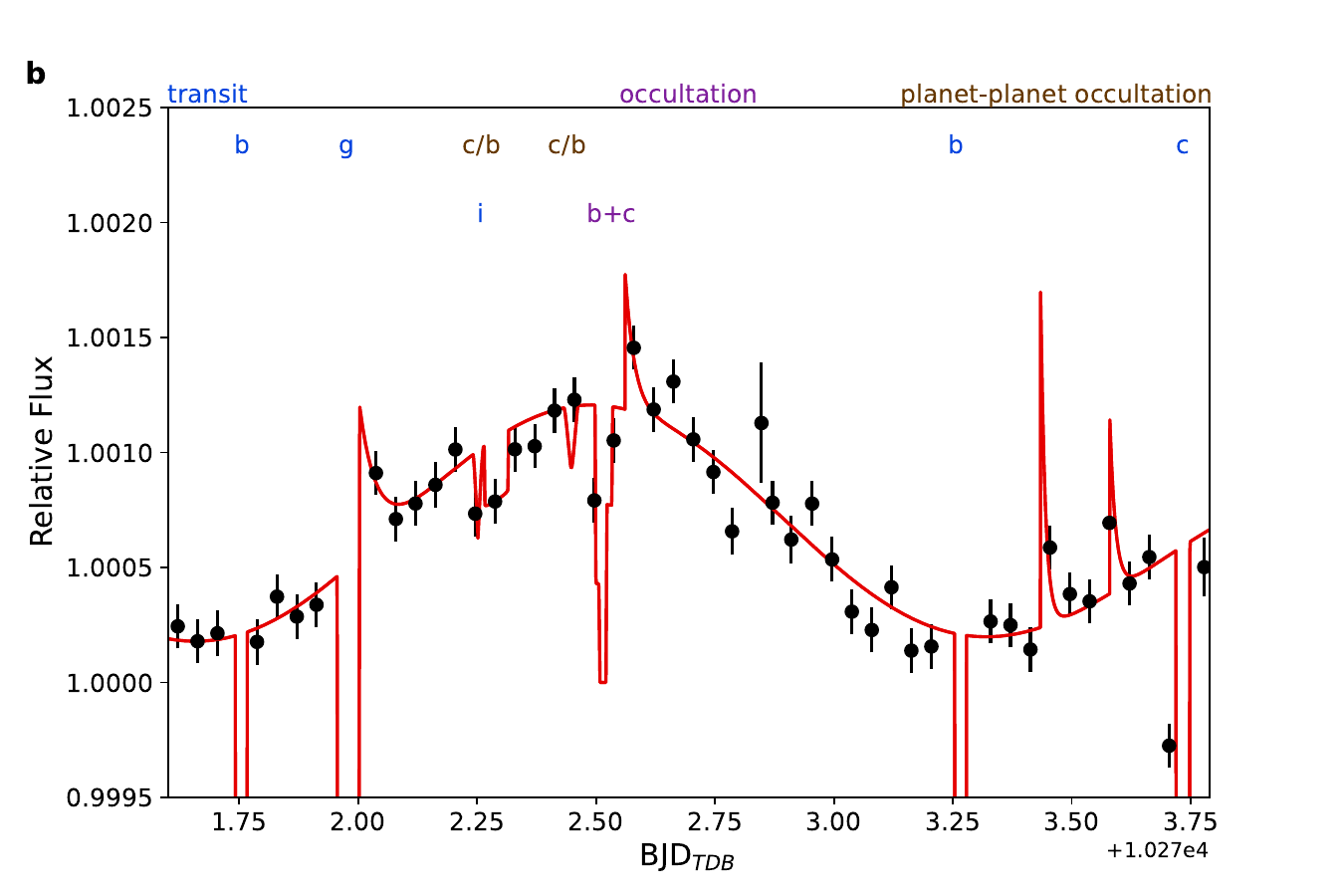}
\caption{\textbf{Detrended Program 3077 light curve.} \textbf{a}. Detrended 3077 program light curve obtained in analysis \#1, with the best-fit planet model deduced from the same analysis superimposed. The initial 4 hours of observations are cut as well as the double flares occurring around 2.8 days (see Methods). Transits, occultations, and planet-planet occultations are shown. Points binned per 0.01d (14.4 min) are shown as black dots, with error bars equal to the standard deviation of the points within the bin. \textbf{b}. Zoom on the phase curve. In addition to transits of b (1.75, 3.25 days), a transit of c (3.75 days), a transit of g (1.90 days), a transit of putative i (2.26 days), occultations of b and c (2.30 and 2.53 days), and planet-planet occultations of b by c (2.25 and 2.46 days), mini-flares are modeled (2, 2.6, 3.45 and 3.6 days). Only points binned per 60 min are shown for clarity.
   }
   \label{fig:lc_det}
\end{figure}

We performed three global Markov Chain Monte Carlo (MCMC) analyses of the 10 light curves of programs 1177, 2304, and 3077 (\ref{baseline}) with the code \texttt{Trafit} \cite{Gillon2010, Gillon2012}. The selected model included the phase curves of planets b and c, two transits of b, one transit of c, one transit of g, the double occultation of b and c, some planet-planet occultations of c by b, four low-amplitude flares (\ref{fig:lc_det} and \ref{fig:miniflare}), and the low-amplitude transit of a putative planet candidate i with an orbital period larger than h (see \ref{fig:lc_det} and \ref{fig:planet_i}). It also included a model for instrumental systematic effects. A quadratic limb-darkening law \cite{MandelAgol2002} was assumed for the star; see Methods for details.

Our nominal analysis \#1 adopted the phase curve model of ref. \cite{Placek_2014} assuming that the planets are tidally locked and may have some significant heat distribution, as expected in the presence of an atmosphere. Our second analysis \# 2 assumed that both planets were airless and adopted the quasi-Lambertian phase-curve model from ref. \cite{Agol2007}. Our third analysis assumed an airless planet phase curve model for b only. The three MCMC analyses assumed informative prior probability distribution functions (PDFs) for some parameters of the system (\ref{tab:priors}). The convergence of the MCMC chains was checked using the Gelman and Rubin statistical test \cite{Gelman1992}.

\begin{figure}
   \centering
    \includegraphics[width=\textwidth]{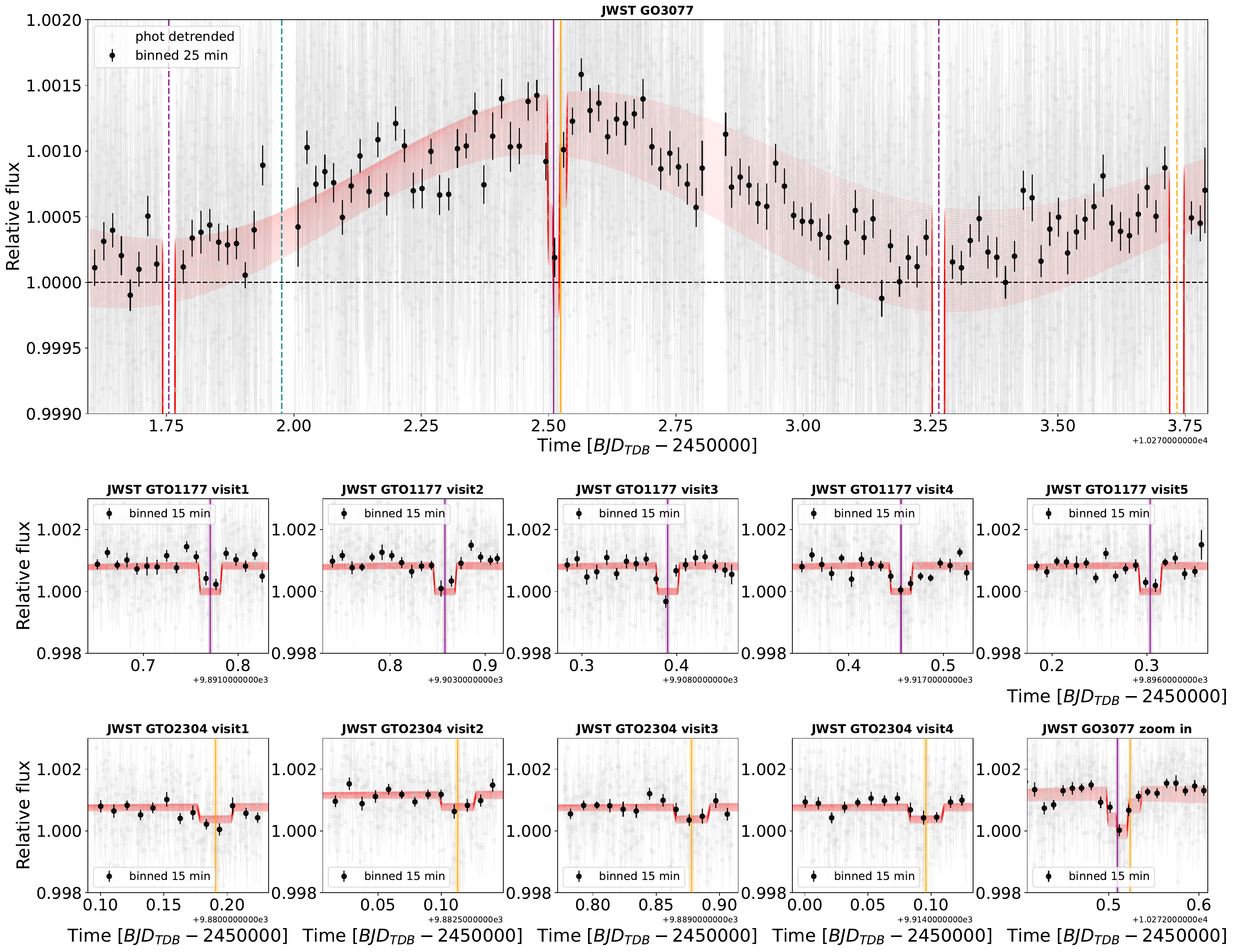}
    \caption{\textbf{Range of best-fit phase curve and eclipse models for TRAPPIST-1\,b and c.} The detrended data are shown with gray dots and binned data with black dots. Red curves show the light curve model for each JWST program that observed TRAPPIST-1 b and c at 15$\mu m$ using MIRI F1500W; the shading spans the range of models drawn from the posterior distribution of the day flux, night flux and phase offset for planets b and c from Analysis \# 1 at 2-$\sigma$. The top panel shows the phase curve from GO 3077; the second row displays the five visits from GO 1177 (PI: Greene); and the third row presents the four visits from GO 2304 (PI: Kreidberg), along with a zoom-in on the double occultation observed in GO 3077. The expected transit and eclipse timings for each planet from \protect\cite{Agol2021} are shown in vertical dashed and plain lines respectively (purple for b, orange for c, and green for g).}
   \label{fig:lc_planet_1}
\end{figure}

\ref{tab:results} shows the median and 1-$\sigma$ errors of the posterior PDFs of the parameters derived in the three MCMC analyses. The posterior PDFs for the day- and night-side fluxes ($F_{b,day}$, $F_{c,day}$, $F_{b,night}$, $F_{c,night}$) as inferred from analysis \#1 are shown in the Methods section for each planet, and for their sum, on \ref{fig:zeta}. \ref{fig:lc_planet_1} shows the range of best fit models for the global analysis \#1 of all existing 15$\mu m$ observations of TRAPPIST-1 (GTO 1177, GO 2304 and GO 3077) when drawing from the posterior distribution of $F_{b,day}$, $F_{c,day}$, $F_{b,night}$, $F_{c,night}$, and the phase curve offsets $\delta_b$, and  $\delta_c$.

\begin{table}[ht]
\begin{tabular}{@{}llll@{}}
\toprule
Parameter & Analysis \#1 & Analysis \#2  & Analysis \#3 \\
\midrule
\textbf{Planet b} &  &  & \\
Mid-transit timing (transit 1) & 271.754380 $\pm$ 0.000088   & 271.754375 $\pm$ 0.000088 & 271.754377 $\pm$ 0.000088 \\
Mid-transit timing (transit 2) & 273.265974 $\pm$ 0.000084 & 273.265972 $\pm$ 0.000084  &  273.265974 $\pm$ 0.000084\\
Semi-major axis $a$ & $0.011513 \pm 0.000078$ au & $0.0115256 \pm 0.000081$ au & $0.0115185 \pm 0.000086$ au \\
Orbital inclination $i$ & $89.70 \pm 0.13$ deg & $89.72 \pm 0.13$ deg & $89.71 \pm 0.14$ deg \\
Impact parameter $b$ & $0.107 \pm 0.047$ $R_\ast$  & $0.103_{-0.050}^{+0.044}$ $R_\ast$ & $0.106_{-0.50}^{+0.47}$ $R_\ast$ \\
Transit depth $dF = (Rp/R_\ast)^2$ & $7316 \pm 117$ ppm  & $7338 \pm 112$ ppm & $7358_{-114}^{+109}$ ppm\\
Eccentricity $e$ & $0.0008_{-0.0006}^{+0.0022}$ & $0.0008_{-0.0006}^{+0.0020}$ & $0.0008_{-0.0006}^{+0.0021}$  \\
Argument of pericenter $\omega$ & $203_{-114}^{+68}$  deg & $181 \pm 90$ deg & $200_{-108}^{+71}$ deg \\
$A = F_{\rm b, D, max}$ & $840 \pm 56$ ppm  & 751 $\pm$ 53 ppm  & $778 \pm 64$ ppm \\
$B = F_{\rm b, N, max}$ & $39_{-27}^{+55}$ ppm  & 0 ppm (fixed) & 0 ppm (fixed) \\
Peak offset $\delta$ & $-6.5 \pm 6.4$ deg & 0 deg (fixed)  & 0 deg (fixed) \\
$\gamma$ & - & $2.44 \pm 0.37$ & $2.61_{-0.33}^{+0.41}$ \\
$T_{\rm brightness, D}$ & $490 \pm 17$ K & $469 \pm 17$ K & $473 \pm 19$ K \\
$T_{\rm brightness, N}$ & $197 \pm 41$ K & 0 K (fixed) & 0 K (fixed) \\
\midrule
\textbf{Planet c} &  &  &  \\
Mid-transit timing & $273.733263 \pm 0.000096$ & $273.733264 \pm 0.000096$  & $273.733269 \pm 0.000096$  \\
($BJD_{TDB}$ - 2460000) & & & \\
Semi-major axis $a$ & $0.01577 \pm 0.00011$ au & $0.01579 \pm 0.00012$ au & $0.01578 \pm 0.00012$ au \\
Orbital inclination $i$ & $89.813 \pm 0.092$ deg & $89.798 \pm 0.087$ deg & $89.793 \pm 0.099$ deg \\
Impact parameter $b$ & $0.092 \pm 0.046$ $R_\ast$ & $0.100 \pm 0.043$ $R_\ast$  & 0.102$\pm$ 0.048 $R_\ast$ \\
Transit depth $dF = Rp/R_\ast$ & $7098_{-149}^{+133}$ ppm  & 7146 $\pm$ 131 ppm & $7134_{-141}^{+124}$ ppm \\
Eccentricity $e$ & $0.0011_{-0.0007}^{+0.0013}$ & $0.0012_{-0.0008}^{+0.0016}$ & $0.0012_{-0.0008}^{+0.0014}$ \\
Argument of pericenter $\omega$ & $161_{-61}^{+85}$ deg & $178 \pm 76$ deg & $160_{-59}^{+86}$ deg \\
$A = F_{\rm c, D, max}$ & $392_{-63}^{+75}$ ppm & $359_{-55}^{+68}$ ppm  &  $413_{-72}^{+66}$ ppm\\
$B = F_{\rm c, N, max}$ & $62_{-43}^{+60}$ ppm & 0 ppm (fixed) &  $102_{-73}^{+97}$ ppm \\
Peak offset $\delta$ & $10_{-22}^{+25}$ deg & 0 deg (fixed)  & $-12_{-15}^{+20}$ deg \\
$\gamma$ & - & $3.34 \pm 0.46$ & - \\
$T_{\rm brightness, D}$ & $369 \pm 23$ K & $357 \pm 21$ K & $374 \pm 25$ K \\
$T_{\rm brightness, N}$ & $220_{-47}^{+38}$ K & 0 K (fixed) & $247_{-60}^{+48}$ K \\
\midrule
\textbf{Planet b + c} &   &  & \\
$A = F_{\rm b+c, D, max}$ & $1231_{-70}^{+93}$ ppm & $1117_{-72}^{+66}$ ppm & $1189 \pm 81$ \\
$B = F_{\rm b+c, N, max}$ & $116_{-57}^{+70}$ ppm & 0 ppm (fixed)  & $102_{-73}^{+97}$ ppm \\
\midrule
\textbf{Planet $g$} &  & &   \\
Mid-transit timing & $271.97948 \pm 0.00013$  &  $271.97955 \pm 0.00011$ &  $271.97954 \pm 0.00012$ \\
Semi-major axis $a$ & $0.04673 \pm 0.00032$ au & $0.04678 \pm 0.00033$ au & $0.04675 \pm 0.00035$ au \\
Orbital inclination $i$ & $89.7459 \pm 0.0087$ deg & $89.7464 \pm 0.0091$ deg & $89.745 \pm 0.010$ deg \\
Impact parameter $b$ & 0.373 $\pm$ 0.012 $R_\ast$ &  0.373 $\pm$ 0.012 $R_\ast$ & 0.374 $\pm$ 0.014 $R_\ast$ \\
Transit depth $dF = Rp/R_\ast$ & $7563_{-124}^{+138}$ ppm & $7726_{-137}^{+185}$ ppm & $7680_{-128}^{+129}$ ppm\\
Eccentricity $e$ & $0.00092_{-0.00065}^{+0.00099}$ & $0.0009_{-0.0006}^{+0.0010}$ & $0.0009_{-0.0007}^{+0.0010}$ \\
Argument of pericenter $\omega$ & $184_{-121}^{+114}$ deg &  $196_{-128}^{+109}$ deg & $181 \pm 120$ deg \\
\midrule
\textbf{Candidate planet $i$} &  & &   \\
Mid-transit timing & $272.2928_{-0.0054}^{+0.0038}$  &  $272.289_{-0.011}^{+0.015}$ &  $272.2921_{-0.0066}^{+0.0050}$ \\
($BJD_{TDB}$ - 2460000) & & & \\
Semi-major axis $a$ & $0.081 \pm 0.010$ au & $0.082 \pm 0.010$au & $0.081 \pm 0.011$ au \\
Orbital inclination $i$ & $89.79_{-0.08}^{+0.12}$ & $89.78_{-0.20}^{+0.13}$ & $89.77 \pm 0.12$ \\
Impact parameter $b$ & $0.57_{-0.31}^{+0.20}$ $R_\ast$ &  $0.57_{-0.35}^{+0.41}$ $R_\ast$ & $0.61 \pm 0.33$ $R_\ast$ \\
Period $P$ & $28.1 \pm 5.0$ d & $28.5 \pm 5.2$ & $28.4_{-6.1}^{+5.1}$ \\
Transit depth $dF = Rp/R_\ast$ & $305_{-80}^{+70}$ ppm & $289_{-117}^{+326}$ ppm & $285_{-109}^{+210}$ ppm\\
Radius $R_p$ & $0.228_{-0.032}^{+0.025} R_\oplus$ & $0.22_{-0.51}^{+0.10} R_\oplus$ & $0.220_{-0.047}^{+0.070} R_\oplus$ \\
\botrule
\end{tabular}
\caption{\textbf{Resulting planetary parameters from the global MCMC analyses of data of programs 1177, 2305, and 3077}. Analysis \# 1: nominal analysis assuming a phase curve model allowing for the possible presence of an atmosphere for both planets. Analysis \#2: airless model for both planets. Analysis \#3: atmosphere model for c, airless model for b. Transit timings unit = BJD$_{TDB}$-2460000.}
 \label{tab:results}
\end{table}

The nominal analysis (\#1) fully rejects a full heat redistribution for both planets. It leads, for planet b, to a night-side flux and a phase curve offset consistent with zero, as expected for a bare rock scenario. The results for c are more ambiguous. They are consistent with an airless scenario, but also leave room for significant heat redistribution (see below). Indeed, the night side flux and the phase curve peak offset are both consistent with zero but also (at 3-$\sigma$) with values as large as 232 ppm ($vs$ $392^{+63}_{-75}$ ppm measured for its occultation depth) and +76 deg, respectively (see Methods). The posterior PDF of the sum of their night-side fluxes is consistent with zero at the 2-$\sigma$ confidence level (see Methods, \ref{fig:zeta}).  The quasi-Lambertian model assumed for both planets in the analysis \#2 fits the data as well as the atmosphere model assumed in the analysis \#1, further supporting an inefficient heat redistribution for both planets. Analysis \#3 does not bring further constraints on the presence/absence of an atmosphere around $c$.

We compute a Bayes factor of 12.8 in favor of the detection of the occultations of planet c by planet b, which is too low to conclude to a decisive detection (Bayes factor $>$100 \citeMethods{bayesfactor}). The transit of the planet candidate i is detected at $\sim$4 sigma in analysis \#1 (transit depth = $305_{-80}^{+70}$ ppm), but at less than 3 sigma in the two other analyses, which is also too low for a robust detection. If real, this planet would have a radius $\sim$0.2 $R_\oplus$, similar to Neptune's moon Triton, which would make it the smallest exoplanet known so far. Given the transit duration $>$2hr, its orbital period should be larger than 20d, but it is very poorly constrained by the data (our inferred posterior PDF closely resembles the assumed prior PDF). Another hypothesis to explain this deviation in the data could be the existence of non-white-noise statistical fluctuation.

To explore the possibility of atmospheres on TRAPPIST-1 b and c, we employed both a day-night climate-photochemical model and a 3-D global climate model (GCM). The day-night climate-photochemical model provides the speed and versatility to model a broad range of potential atmospheric compositions to identify the suite of atmospheres that best fit the phase curves, and the 3-D GCM provides superior modeling of the spatial dependence of temperature, which is particularly useful in reproducing the hot substellar region which dominates the dayside phase curve.

\begin{figure}
    \centering
    \includegraphics[width=0.86\textwidth]{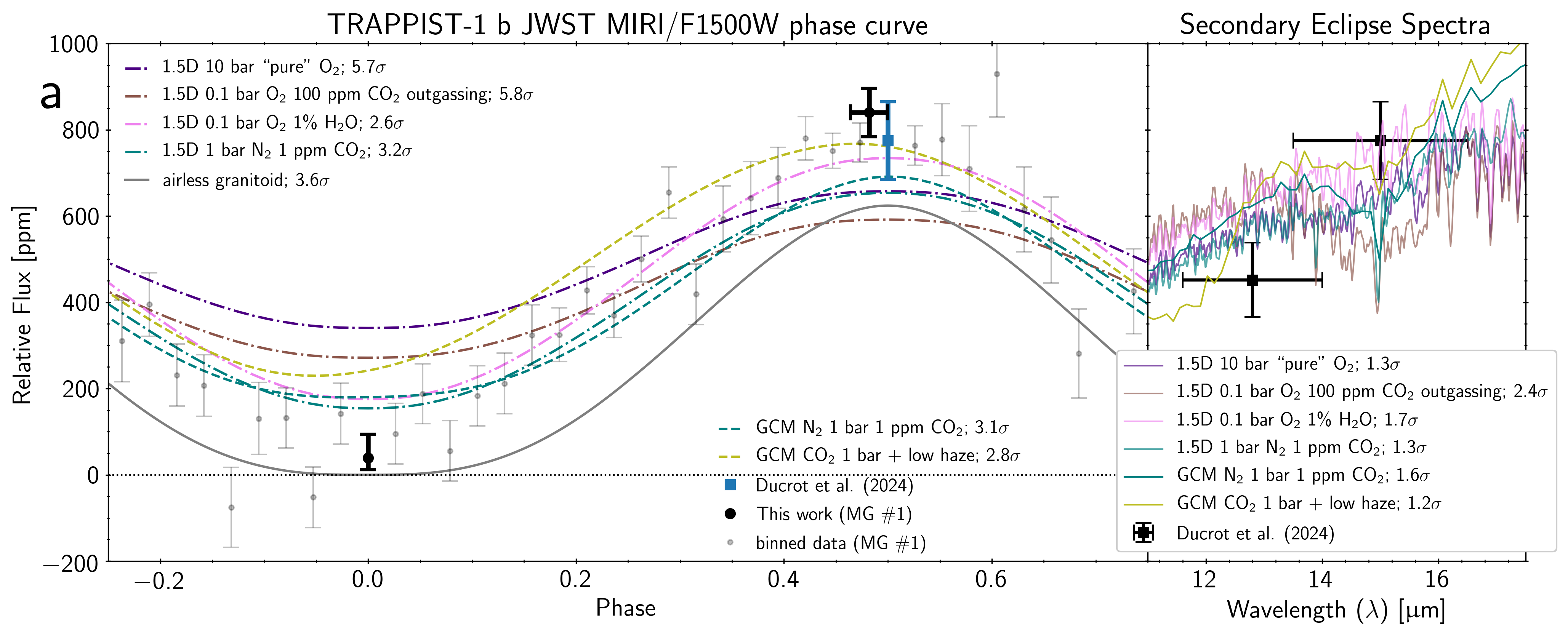}
    \includegraphics[width=0.86\textwidth]{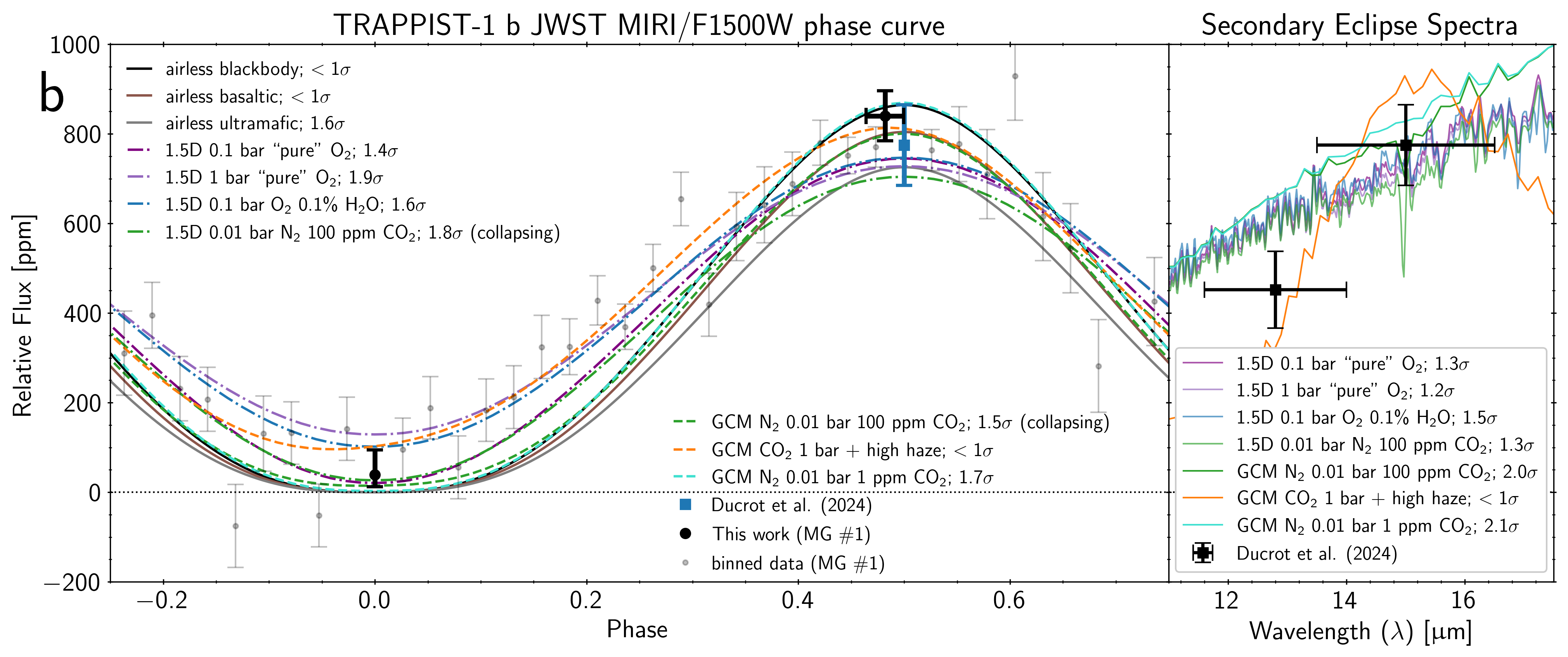}
    \includegraphics[width=0.86\textwidth]{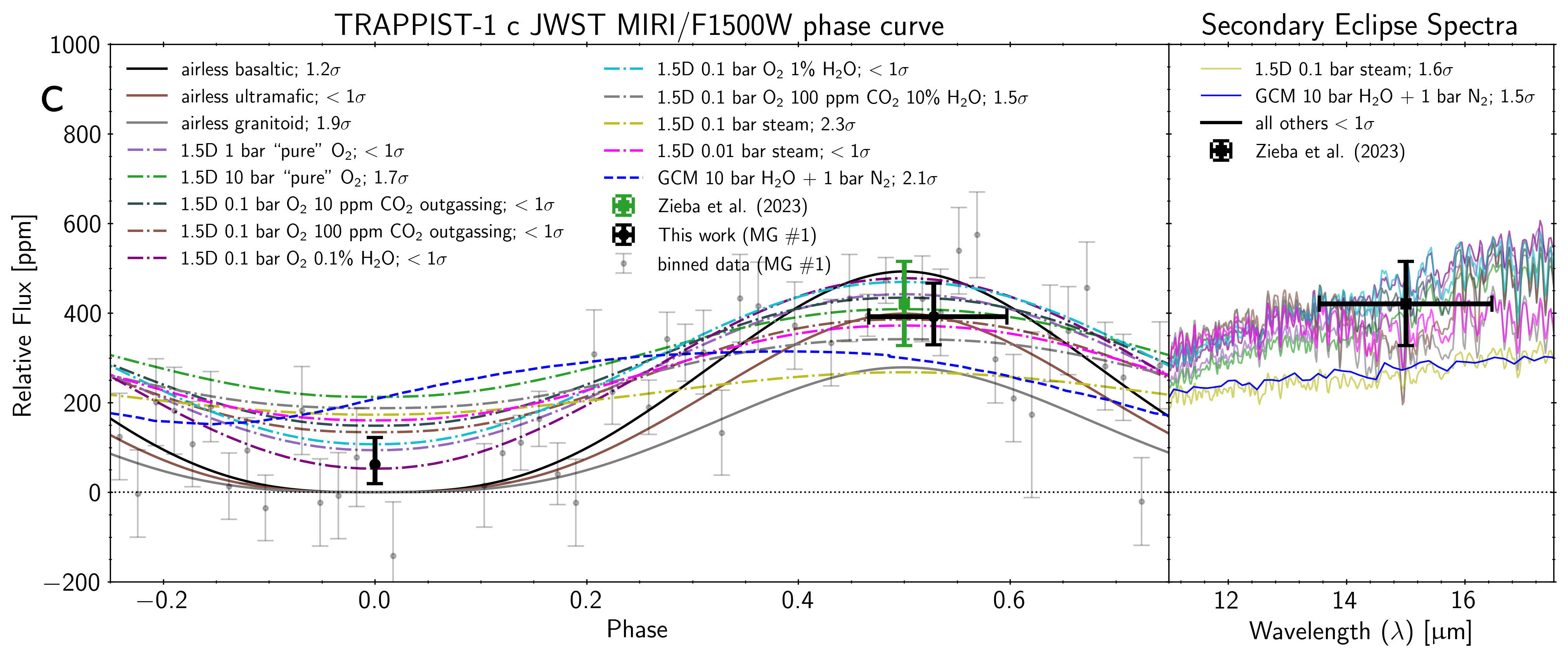}
    \caption{\textbf{Phase-folded phase curves and emission spectra of TRAPPIST-1 b and c.}  \textbf{(a)} \textit{Right panel:} Secondary eclipse depth measurements of TRAPPIST-1\,b from \cite{ducrot_2024}, shown alongside a set of atmospheric models that were previously consistent with the data. \textit{Left panel:} The phase curve of TRAPPIST-1\,b from this work, with the corresponding dayside flux, nightside flux, and phase offset, all shown with their 1$\sigma$ error bars. Goodness of fit reported in $\sigma$ is a joint fit to the day/night phasecurve points, offset, and both data points from \cite{ducrot_2024}. The models on the right that fit the secondary eclipse data are no longer favored once the phase curve data are taken into account.
    \textbf{(b)} Same layout as in (a), but showing models that remain consistent with both the secondary eclipse and the phase curve data. The TRAPPIST-1\,b phase curve measurements are consistent with either a thin oxygen atmosphere with a small amount of water vapor (e.g., 1\% H$_2$O), or a thicker atmosphere largely devoid of greenhouse gases (e.g., a 0.1-bar ``pure'' O$_2$ atmosphere).
    \textbf{(c)} Same format as above, but for TRAPPIST-1\,c. \textit{Right panel:} Secondary eclipse depth measurements from \cite{Zieba2023}, along with atmospheric models that are still consistent with the data. \textit{Left panel:} Phase curve of TRAPPIST-1\,c from this work, with the derived dayside flux, nightside flux, and offset, all with 1$\sigma$ error bars. The signal-to-noise ratio is lower for TRAPPIST-1\,c, which makes it difficult to conclusively rule out any of these cases, although several of them appear to produce better fits including 0.1--1\,bar ``pure'' O$_2$ atmospheres with photochemically generated O$_3$, and a 0.1\,bar O$_2$-dominated atmosphere containing 1\% H$_2$O and 100 ppm CO$_2$ (see Table~\ref{tab:VPL-experiments}).
    For all panels, the data shown correspond to the nominal reduction (Analysis \#1) by MG, binned at $\sim$60-minute intervals.
     }
    \label{fig:VPL-phasecurves}
\end{figure}

\ref{fig:VPL-phasecurves} presents the corrected, phase-folded light curves of TRAPPIST-1 b and c. These curves were derived using models that account for both astrophysical signals and instrumental systematics, employing either MCMC or nested sampling methods, in line with standard practice in the field \cite{Evans2023, Kempton2023, Zhang2024, Bell2024a}. In addition, to isolate each planet’s signal, we subtract the contribution of the other planet from the data.
The figure also shows the emission spectra of TRAPPIST-1 b and TRAPPIST-1 c. Panel a) highlights scenarios that were previously consistent with broadband dayside measurements but are now ruled out by the phase curve for TRAPPIST-1 b, and panel b) shows those that remain viable. Panel c shows the suite of surface and atmospheric scenarios that are still consistent with the data for TRAPPIST-1 c.
We present results from Analysis \#1, which models the phase curves of both planets with sinusoidal shapes. We note that the different analyses (\#1, \#2, and \#3) retrieve slightly different light curve shapes due to differences in model assumptions and parameterizations. For example, Analysis \#1 fits six parameters to model the phase variations—$F_{b,\mathrm{day}}$, $F_{c,\mathrm{day}}$, $F_{b,\mathrm{night}}$, $F_{c,\mathrm{night}}$, $\delta_b$, and $\delta_c$—while Analysis \#2 fits only four: $F_{b,\mathrm{day}}$, $F_{c,\mathrm{day}}$, $\gamma_b$, and $\gamma_c$.
We emphasize that the suitability of atmospheric or surface models to the observed phase curves should not be assessed based on qualitative fits—such as sigma differences in the light curve shape—because such metrics are highly sensitive to the specific modeling framework. The shape of the phase curve alone is not a robust basis for comparison. Instead, meaningful evaluation should rely on the posteriors of the physical parameters sampled in the MCMC analysis. Accordingly, the sigma values shown in \ref{fig:VPL-phasecurves} reflect differences between the inferred dayside flux, nightside flux, and phase offset relative to various atmospheric models.

\ref{fig:VPL-phasecurves} shows the results of self consistent, day--night climate-photochemical-equilibrium atmospheres of a broad range of atmospheric types for TRAPPIST-1 b and c, with layer-by-layer day--night advection using the VPL Climate model in two-column mode (the 1.5D model)  \cite{Lincowski2018,Lincowski2023}.
We computed thermal phase curves (see \ref{fig:VPL-phasecurves}) and occultation spectra for comparison to the data. In panel a, we show that although several 1.5D modeled atmospheres fit the combined 12.8$\mu$m and 15$\mu$m secondary eclipse measurements to within 2.4$\sigma$, they are now ruled out by the day- and night-side measurements of the thermal phase curve, due to excess heat transport to the nightside.  These discarded scenarios include N$_2$ atmospheres of $\geq1$~bar (and with 1~ppm CO$_2$), pure O$_2$ atmospheres of $\geq10$~bar with no extra greenhouse gases, and 0.1~bar O$_2$ atmospheres containing 100ppm or more of CO$_2$.
For the remaining 1.5D cases we examined (panel b), we find that several of our modeled atmospheres are consistent within 1.5$\sigma$ of the combined 12.8$\mu$m and 15$\mu$m secondary eclipse measurements, and also fit the TRAPPIST-1 b phase curves.   Notably, largely transparent atmospheres consisting of 1 bar or less of O$_2$ produce fits to the phase curves at $\leq1.9\sigma$, and a 0.1~bar O$_2$ atmosphere with greenhouse gases, in this case 0.1\% H$_2$O (with photochemical ozone) fits within $\leq1.6\sigma$. Extremely tenuous 0.01 bar N$_2$ atmospheres with 100ppm of CO$_2$ also fit to within 1.8$\sigma$. However, this atmosphere is discarded as it is not stable against collapse on the nightside (see 3D results discussion below).
For TRAPPIST-1~c, the uncertainty on the data preclude conclusively ruling out any of the 1.5D model atmospheres, although some scenarios appear to provide a better fit than others (\ref{fig:VPL-phasecurves} panel c).  Similarly to TRAPPIST-1 b, atmospheres with small but non-zero atmospheric opacity and weak day-night heat transport provide better fits. As is the case for TRAPPIST-1 b, the best fits to the TRAPPIST-1 c thermal phase curve come from atmospheres with minimal greenhouse gases: in this case 1 bar pure O$_2$ (with photochemical ozone), or 0.1 bar O$_2$ with 1\% or less H$_2$O, or less than 100ppm of CO$_2$. A  steam atmosphere up to 0.1 bar is also not currently ruled out by these data.
See Methods for more details about these models and their specific results.
For our $\leq2.5\sigma$ atmosphere fits, the corresponding transit spectral features (\ref{vpl_transit}) from O$_3$ and H$_2$O reach up to 80~ppm for TRAPPIST-1 b. These features are not yet ruled out by the currently available precision of the JWST transit transmission spectra for TRAPPIST-1 b \cite{Lim2023}.

We then performed 3-D Global Climate Model (GCM) simulations of TRAPPIST-1b and c with the \texttt{Generic PCM} \cite{Turbet:2018,Turbet:2023} in order to assess whether phase curves could be used to constrain the atmospheres that are still plausible based on occultations alone \cite{Greene2023Natur,Zieba2023,ducrot_2024} (see details of the calculations in the Methods). For TRAPPIST-1b, this includes thin, residual atmospheres \cite{Ih2023} as well as CO$_2$-rich atmospheres with a hot, hazy stratosphere \cite{ducrot_2024}. We represented here two end-member cases (\textit{Haze high} and \textit{Haze low}) of CO$_2$ atmospheres with maximal and minimal amounts of haze, respectively.
Resolving the 3-D structure of the atmosphere is necessary to accurately represent the energy transport and phase offsets in all atmospheric layers. \ref{fig:VPL-phasecurves} a. and b. (right panels) shows the comparison between the observed phase curve and the synthetic phase curves computed from GCM simulations, carefully selected so that they match the available occultation measurements within 2.4$\sigma$ \cite{Greene2023Natur,Zieba2023,ducrot_2024}. We find that, for TRAPPIST-1 b, among the atmospheric cases still compatible with the occultations (see \ref{fig:eclipse_spectra_GCM}), the values of the dayside flux, nightside flux and peak offset of the phase curves can be used to rule out atmospheric scenarios that produce non-negligible horizontal transport, which includes scenarios with a thick atmosphere (e.g., the 1 bar N$_2$ 1 ppm CO$_2$ case) and CO$_2$-rich atmospheres with hazes that absorbs too deep in the atmosphere (e.g., the \textit{Haze low} scenario, for which we also predict a significant phase curve offset). Some cases can also be discarded (e.g., the N$_2$ 0.01 bar 100 ppm CO$_2$ case) because the GCM predicts CO$_2$ to collapse on the surface \cite{Turbet:2018,Turbet:2020}. Although not all atmospheric scenarios can be theoretically ruled out with all data now available (see  \ref{fig:VPL-phasecurves}a. and \ref{fig:VPL-phasecurves}b.), those that remain (CO$_2$ 1 bar + \textit{Haze high}, N$_2$ 0.01 bar 1ppm CO$_2$) appear to be very fine-tuned, and the most likely hypothesis is that TRAPPIST-1 b has no atmosphere, which is consistent with the prediction from the theoretical study of secondary atmospheric escape \cite{Dong2018} (not considering possible replenishment via out-gassing). For TRAPPIST-1 c, the amplitude of the phase curve can be used to rule out atmospheric scenarios involving efficient heat redistribution as well, including the case of steam atmospheres \cite{Turbet:2023,Lincowski2023} which were compatible with the secondary eclipses of ref. \cite{Zieba2023} (see Methods), but also a possible outcome of atmospheric evolution. Additionally, we computed transit spectra (see Methods, \ref{fig:eclipse_spectra_GCM}) for the atmospheres of TRAPPIST-1 b which are still compatible with all MIRI datasets (GTO 1177, GO 2304, GTO 1279,GO 3077) to assess the impact of using innermost planets (e.g. TRAPPIST-1b) to remove stellar contamination from the transit spectra of other, outer planets \cite{DeWit:2023} as recently demonstrated by ref. \cite{Rathcke2024}. We found that the amplitude of absorption features in TRAPPIST-1 b transit spectra (see \ref{fig:eclipse_spectra_GCM}) is low but non-negligible, in particular for the (unlikely) CO$_2$-rich hazy atmospheric scenario.

Comparing the two sets of modeling results for TRAPPIST-1 b, both the day-night climate-photochemical and 3D GCM models are ruled out by the data for the proposed 1 bar N$_2$ or O$_2$ atmospheres with trace ($\leq$1 ppm) greenhouse gases (CO$_2$ and O$_3$). The data do not reject models with 0.01 bar N$_2$ and 100 ppm CO$_2$; however, the 3D GCM shows that this atmosphere is unstable to collapse, and therefore not a viable scenario.
The self-consistent photochemically-generated hazy atmospheres did not fit the phase curves, and to fit both the phase curve and the 12.8 and 15$\mu$m secondary eclipse data, the 3D GCMs showed that unlikely densities of hydrocarbon haze particles would be required.
For TRAPPIST-1 c, the day-night model showed good fits for 0.1-1 bar pure O$_2$ and 0.1 bars of O$_2$ with small amounts of greenhouse gases (photochemical O$_3$ and 0.1\% H$_2$O). The combined (two-column and 3D GCM) model results disfavor steam-dominated atmospheres from 1 bar to 10 bars.

Next, because the data for TRAPPIST-1\,b are consistent with no atmosphere whatsoever, we model TRAPPIST-1\,b's thermal phase curve using a bare rock model and a database of seven geologically fresh surface materials \cite{Lyu2024}. Note that our models produce noticeably higher albedos than previous surface models published for TRAPPIST-1\,b based on the same database \cite{Ih2023,ducrot_2024}, largely because these previous models were based on \cite{2019ApJ...886..141M} which used an erroneous albedo conversion between spherical albedo and geometric albedo \cite{coy2025}.

Overall, we find that TRAPPIST-1\,b's phase curve indicates the surface should not be very dark, while the planet's secondary eclipse spectrum is able to rule out the blackbody surface. For the eclipse spectrum we combine the 15$\mu$m eclipse depth from analysis \# 2 with the reported secondary eclipse at 12.8$\mu$m from \cite{ducrot_2024}. We find that the spectrum is best fit by ultramafic albedo surfaces(see \ref{X.L spectrum}). Moreover, the planet's phase curve also suggests the planet's surface is likely to be fresh ultramafic (see \ref{X.L bare rock phase}). Although secondary eclipse spectrum doesn't prefer ultramafic over other materials by 3$\sigma$, the fitting is well enough which makes ultramafic most likely to be TRAPPIST-1\,b's composition.
%
One hypothesis to affect the fitting is space weathering, which in the Solar System acts to darken and redden the surface spectra of airless bodies like the Moon and Mercury \cite{pieters2016space}, and which is likely even more efficient for the inner TRAPPIST-1 planets than in the Solar System \cite{Zieba2023}. We find that moderate amounts of space weathering significantly reduce the planet's albedo and warm the dayside surface, thus increasing its phase curve amplitude (see \ref{X.L spectrum}). In this case, adding a few amount of space weathering material on planetary surface, feldspathic and granitoid could also be a good fit on TRAPPIST-1\,b.


This work demonstrates the potential of JWST to assess the presence of dense secondary atmospheres around temperate ($T_{eq} < 400$ K) Earth-sized planets transiting nearby ultracool dwarf stars using thermal phase curves. It shows that it is highly unlikely that TRAPPIST-1\,b has a significant atmosphere. The case of TRAPPIST-1\,c is more ambiguous. On one hand, a full heat redistribution is discarded by the data, and most dense atmosphere ($> 1 bar$) scenarios \cite{Lincowski2023} consistent with the 15 $\mu$m occultation measurement \cite{Zieba2023} are now disfavored, leaving room only for a low opacity atmosphere of 1 bar of pure O$_2$, or thin atmospheres with pressures $< 0.1$ bar and traces of CO$_2$ and water. On the other hand, assuming both planets are airless leads to the conclusion that the surface properties of planet c must differ from those of planet b (which is not unexpected when compared to the variety of planetary surface types observed in the solar system). Indeed, the dayside brightness temperature inferred from analysis \#1 is significantly larger than the equilibrium temperature (assuming a full heat distribution and a null Bond Albedo) for planet b ($490 \pm 17$ K $vs$ 398$\pm$4 K \cite{Ducrot2020}), but not for planet c ($369 \pm 23$ K $vs$ 340$\pm$3 K \cite{Ducrot2020}). This difference could be explained by a larger surface albedo for c, but also by a poor but non-null heat redistribution consistent with a thin atmosphere. This difference justifies the need to further characterize TRAPPIST-1\,c with JWST. Upcoming observations of its thermal emission at 12.8 $\mu$m (program GO 5191) will make it possible to discriminate between more scenarios. In parallel, transmission spectroscopy observations (GO 2589 + GO 2420) will provide additional constraints on the presence of molecules such as CO$_2$ in a putative atmosphere.

\noindent

\vspace{7mm}


\clearpage
\backmatter
\section{Methods}\label{sec11}

\renewcommand{\figurename}{\hspace{-4pt}}
\renewcommand{\thefigure}{Extended Data Fig.~\arabic{figure}}
\renewcommand{\theHfigure}{Extended Data Fig.~\arabic{figure}}
\renewcommand{\tablename}{\hspace{-4pt}}
\renewcommand{\thetable}{Extended Data Table \arabic{table}}
\renewcommand{\theHtable}{Extended Data Table \arabic{table}}
\setcounter{figure}{0}
\setcounter{table}{0}

\subsection*{Observations}\label{sec:observations}

The observations of program 3077 consisted in the continuous observation of TRAPPIST-1 by the MIRI instrument aboard JWST from Nov 22, 2023 19h45 UT to Nov 25 2023 07h02 UT. These observations were obtained in imaging mode within the F1500W filter ($\lambda_\mathrm{effective} = 15.1 \mu m$, width = 3.1 $\mu m$). Unlike for programs GTO 1177 and GO 2304 that used the FULL array of MIRI (1024 $\times$ 1032 pixels), program 3077 used the BRIGHTSKY array (512 pixels $\times$ 512 pixels). The target was put on the center of the array. Because of the smaller group time of the BRIGHTSKY array (0.865s $vs$ 2.775s for the FULL array), 45 groups could be gathered per integration for a similar signal-to-noise (SNR $\sim$840) and cadence ($\sim$39s) than programs 1177 and 2304, resulting in a better sampling of the `ramp' of the detector. JWST detectors are indeed read out using a non-destructive `up-the-ramp' read-out technique, with the read out divided in several groups composing an integration \citeMethods{JWSTpipeline}. For a given integration, a slope is fitted in the counts recorded for all groups to measure the flux of the star in counts per second. The accuracy of this ramp-fitting step depends on the number of groups in the integration, giving a benefit to the selection of the BRIGHTSKY array for the observations. Another benefit is that using a smaller array decreased significantly the data volume, which was convenient for such a long time-series observation. Finally, an extra-benefit of this strategy was a gain of 2.5 hours of overhead time, which allowed more time for the instrument to settle and to maximize the scientific potential of the program.

The start and the end of the observations were based on transit and occultation ephemeris from ref. \citeMethods{Agol2021}, which successfully predicted a double occultation of planets $b$ and $c$ around Nov 24, 2023 00h15 UT. The goal was to observe a full phase curve of planet $b$, including two consecutive transits, in phase with a partial phase curve of planet $c$ (i.e. with concomitant occultations of both planets), including one of its transits. This program was designed based on the analysis of simulated light curves that confirmed its potential to firmly constrain the heat distribution efficiency of both planets and the absence/presence of an atmosphere around them.

Currently, a single JWST exposure (i.e. a block of integrations in JWST jargon) cannot exceed 48 hours. For that reason, we had to cut the observation into two visits of one exposure each. At the start of the second visit, the telescope had to reacquire the target, resulting in a 5 minute and 45 second gap and a small shift of the target on the detector (see \ref{fig:lc_raw}). We selected the number of integrations for both visits to ensure that this gap occurs well away from the double occultation, potential planet-planet occultations, and transits. We thus decided to have one first visit of 1 exposure of 3473 integrations, followed by a shorter visit of 1 exposure of 1863 integrations. Doing so, the gap between visits fell within a relatively signal-free part of the light curve (see \ref{fig:lc_raw}).\\

\subsection*{Data reduction and analysis (MG)}\label{sec:data_reduction}

As in ref. \cite{Zieba2023}, we started the reduction of the data of program 3077 from the raw uncal.fits files available on the MAST website (\url{https://mast.stsci.edu}). We first calibrated the data  using the first stage (going from groups to slopes and applying basic detector-level corrections) of the \texttt{Eureka!} pipeline \cite{Bell2022} and the second stage (flat-fielding, unit conversion to MJy.sr$^{-1}$) of the JWST pipeline, resulting in calibrated calints.fits files. We used a new branch of \texttt{Eureka!} (version 0.11.dev77+g1e43198f.d20240130) developed by co-author TJB and using the version 1.13.4 of the JWST pipeline that included the correction of the 10Hz and 390Hz electromagnetic interference  (EMI) pattern noise \citeMethods{JWSTpipeline}. For stage 1, as in ref. \cite{Zieba2023}, we used the default JWST pipeline settings, except for (1) the ramp-fitting weighting parameter set to \texttt{uniform} and (2) the deactivation of the jump correction.  We also tested starting directly from the calints.fits files processed by the standard JWST pipeline, but the resulting light curves had lower photometric precision. The rest of the reduction was done using a pipeline coded in \texttt{IRAF} and Fortran 2003. It included for each calibrated image (1) a change of unit from MJy.sr$^{-1}$ to recorded electrons, assuming a gain of 3.3 el.s$^{-1}$ (P.-O. Lagage, private comm.), (2) the fit of 1D and 2D Gaussian functions to the point-spread function (PSF) of the star to measure the subpixel position of its centroid and its full width at half maximum (FWHM) in both directions, (3) the measurement of the noise-pixel (as defined in ref. \citeMethods{Deming2015}) within a circle of 4 pixels radius, and (4) the measurement of the stellar and background fluxes using circular and annular apertures, respectively, with \texttt{IRAF/DAOPHOT} \cite{Stetson1987}. Finally, the resulting light curve was normalized and outliers were discarded using a 4-$\sigma$ clipping algorithm with a 20-min moving median  on the fluxes, and a 5--$\sigma$ clipping on the $x$- and $y$-positions, FWHM along the $x$ and $y$ directions, noise-pixels, and background measurements. In total, 2.9\% of the measurements were flagged as outliers and discarded. The light curves obtained with different photometric apertures were compared, and we selected the one obtained with an aperture radius of 4 pixels as it minimized the errors of the measurements. The background was measured in an annulus extending from 30 to 45 pixels from the center of the PSF. For each stellar flux measurement, the corresponding error was computed taking into account the star and background photon noise, the readout noise and the dark noise.

\ref{fig:lc_raw}a shows the resulting raw light curve. Its first $\sim$4 hours contains significant structure. It is concomitant to a ramp-like increase of the background (see \ref{fig:lc_raw}b top-right panel) extending to BJD $\sim$2460271.6. After that time, the background shows a linear increase along the whole run. Based on this behavior of the background, we interpret this initial structure as an instrumental effect due to the stabilization of the detector. We thus discarded this initial part of the light curve in our analysis. The light curve shows also the signature of a double flare-like event between BJD 2460272.805 and 2460272.845 (\ref{fig:miniflare}a). We discarded the corresponding data points. In the end, we discarded 834 points (including 155 outliers) out of the 5336 points of the raw light curve, i.e. 15\%.

The light curve contains transits of planets b (2 transits), c (one transit), and g (one transit), in addition to a double occultation of planets b and c, an occultation of d (which is very shallow and therefore undetected) and some occultations of planet c by planet b (\ref{fig:lc_det}). Its visual inspection also reveals a low-frequency signal that appears to be consistent with the combined phase-curve variations of the two inner planets.


Our data-analysis methodology used the Fortran 2003 code \texttt{Trafit} that includes a revised version of the adaptive Markov chain Monte Carlo code presented in refs. \cite{Gillon2010, Gillon2012} and ref. \citeMethods{Gillon2014}, to perform a global analysis of all available eclipse and phase curve data at 15 $\mu m$ (GTO 1177, GO 2304, GO 3077), adopting the Metropolis-Hasting algorithm \citeMethods{Hasting1979} to sample the posterior probability distributions of the system’s parameters. The assumed model was
composed of the eclipse model from ref. \cite{MandelAgol2002} to represent the transits of planets b, c and g, and the double occultation of b and c, a phase curve model for planet b and c, and a baseline model aiming to represent the other astrophysical and instrumental mechanisms able to produce photometric variations. We tested a large range of baseline models and we adopted the one minimizing the Bayesian Information Criterion (BIC, \citeMethods{Schwarz1978}).  It was composed of the slope, the sum of second-order polynomials of the $x$ and $y$ position of the PSF's center, an initial decreasing ramp modeled as a linear function of the difference in time relative to the first point of the light curve, and a similar ramp model for points taken during the second visit. Our global model also included planet-planet occultations, using the model of ref. \cite{MandelAgol2002} to compute the fraction of each planet occulted by another one at any given time. Our planet occultation models assumed uniform disks for the occulted planets.

For the phase curve model, our nominal analysis (analysis \# 1) adopted the one of ref. \cite{Placek_2014} based on the following formula that assumes that the planets are tidally locked and may have some significant heat distribution, as expected in the presence of an atmosphere:
\begin{gather}
\frac{F_{i, D}(t)}{F_\ast} = \frac{1}{2}(1+cos(\theta(i, t) - \delta(i))) \frac{F_{i, D, max}}{F_\ast} \Omega(i, t) \\
\frac{F_{i, N}(t)}{F_\ast} = \frac{1}{2}(1+cos(\theta(i, t) - \delta(i) - \pi)) \frac{F_{i, N, max}}{F_\ast} \Omega(i, t)\\
\frac{F_{i}(t)}{F_\ast} =  \frac{F_{i, D}(t)}{F_\ast} + \frac{F_{i, N}(t)}{F_\ast}
\label{eqn:1}
\end{gather} where $i$ represents planet $i$, with $\theta(i, t)$ its phase angle at time $t$, $\gamma(i, t)$ its unocculted fraction at time $t$, $F_{i, D}(t)$  the flux contribution of its day side at time $t$, $F_{i, N}(t)$ the one of its night side, $F_\ast$ the average flux of the star, and $\delta(i)$ is an angle offset to take into account a possible West or East shift of the planet's flux peak relative to the substellar point. $F_{i, N, max}$ and $F_{i, D, max}$ are the maxima of the fluxes for the planet's night and day sides. For each time $t$, the occulted (by the star or another planet) fraction of the planet $\Omega(i, t)$ is computed using the eclipse model of ref. \cite{MandelAgol2002}.   \\
In a second analysis (analysis \# 2), we assumed that both planets were airless and adopted the following model from ref. \cite{Agol2007} for their phase curves:

\begin{equation}
 \frac{F_{i}(t)}{F_\ast} =  \frac{F_{i, D, max}}{F_\ast} \times \cos(\theta(i, t)/2)^{\gamma(i)}
 \label{eqn:2}
\end{equation} where $\gamma(i)$ is a free parameter. Assuming $\gamma(i) \sim 3$, this analytical function was shown to be a good approximation for the phase function of a Lambert sphere by ref. \cite{Agol2007}.
As analysis \#1 led to the firm conclusion that b shows no significant heat redistribution (see below) but to more ambiguous results for planet c, we performed a third analysis for which we assumed the `airless' model from eq. (\ref{eqn:2}) for planet b only.
In all three analyses, the global model for the light curve was then:
\begin{equation}
F(t) = B(t, X) \times \left(1+ \Sigma^{N}_{i=1}  \frac{F_{i}(t)}{F_\ast} - E(i, t)\right)
\label{eqn:5}
\end{equation} where $B(t, X)$ represents the baseline model depending on time $t$ and on an array $X$ of external parameters, $N$ is the number of planets considered, and $E(i, t)$ is the fraction of the star occulted by planet $i$, taking into account limb-darkening (as computed under the formalism of ref. \cite{MandelAgol2002}). In other words, the light curve model is computed as the sum of the flux of the star (taking into account the transits) and the fluxes of the planets (taking into account their occultations) multiplied by a baseline + normalisation model $B(t, X)$. In our analysis, only planets b, c and g were considered. We did some test analyses including the flux contribution of the other planets assuming the extreme case of airless null albedo bodies for all of them, and the results were consistent with the ones obtained from our nominal analysis.

\begin{figure}[ht!]
   \centering
    \includegraphics[width=\textwidth]{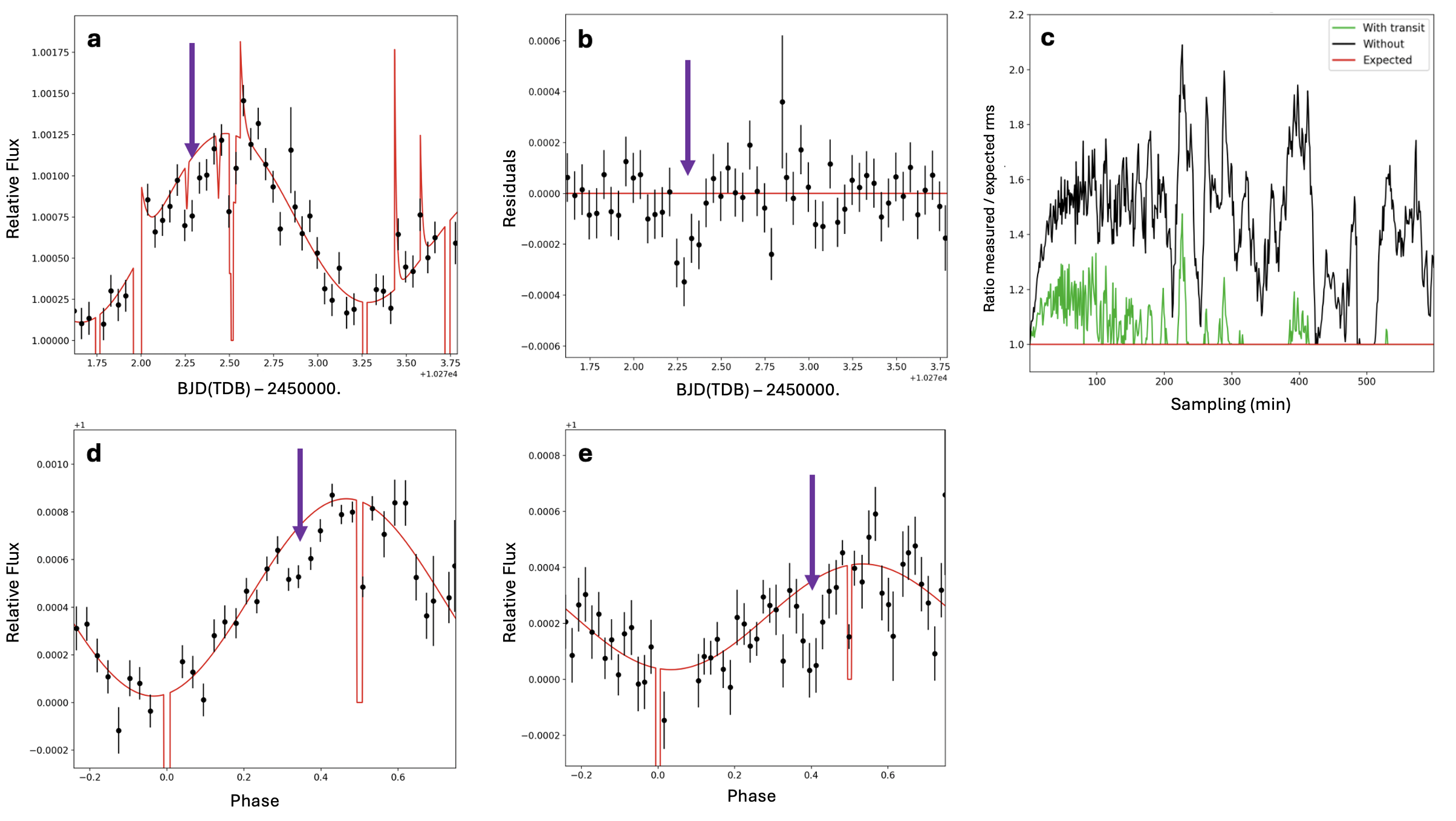}
    \caption{\textbf{Additional transit-like structure in the light curve of Program 3077.} \textbf{a}, detrended 3077 program light curve obtained by MG in a version of analysis \#1 not assuming the transit of a putative planet i, with the best-fit planet model deduced from the same analysis superimposed in red.  \textbf{b}, residuals of that version of analysis \#1. \textbf{c}, modified Allan diagram showing the ratio of the measured standard deviations of the residuals of the best-fit model and  of the one expected for a fully white noise as a function of the sampling (i.e. for different binnings), without (black) and with the inclusion of  a low-amplitude transit in the global model (green). The red line shows what the ratio should be for a fully white noise, i.e. 1. \textbf{d}, Best-fit phase curve for planet $b$ from that version of analysis \#1. The data points are phase-folded and corrected from the systematics and the contribution of the other planets. The best-fit phase curve model is superimposed in red. \textbf{e}, same for planet $c$. For panels \textbf{a}, \textbf{b}, \textbf{d} and \textbf{e},  the data points are binned per 60 minutes, each binned point is the average value of the individual points within the bin, and the error bar is the average of the individual errors divided by the square root of the number of points within the bin. The purple arrows show the location of the putative transit of a putative planet i.
   }
   \label{fig:planet_i}
\end{figure}

For analysis \#1, we  performed a preliminary Markov Chain of 20,000 steps to assess the need to rescale the photometric errors for white and red noise (e.g. \cite{Gillon2012}). The level of red noise of the residuals was anomalously high (see modified Allan's diagram in \ref{fig:planet_i}c, black line). A visual examination of the residuals binned to different samplings showed four structures consistent with low-amplitude flare-like structures with decay times of a few dozens of minutes (\ref{fig:planet_i}a). As the raw light curve showed a clear double flare (\ref{fig:miniflare}a) with a similar exponential decay time, it is reasonable to assume that it also contain lower-amplitude `mini-flares' responsible for a part of the red noise. Furthermore, using spectroscopic time-series observations taken by the Near-InfraRed Imager and Slitless Spectrograph (NIRISS) and Near-InfraRed SPECtrograph (NIRSPEC) instruments aboard JWST, \citeMethods{Howard2023} revealed a frequency of $\sim$3 low-energy ($\sim 10^{30}$ erg)  flares per day for TRAPPIST-1, and that the continuum of these flares was well described by black body emission with an effective temperature below 5300 K. Assuming a black body temperature of 5000K, we computed that the amplitudes of the four low-amplitude flare-like structures in our MIRI light curves are also consistent with energies $\sim 10^{30}$ erg. Based on these considerations, we modeled these four structures (see \ref{fig:miniflare}b) with a simple flare-model composed of an instant flux increase followed by an exponential decrease (3 parameters per flare: time and amplitude of the flux increase + exponential decay constant).

The residuals also showed a $\sim$2hr transit-like structure centered on 2460272.26 JD and with an amplitude $\sim$ 200 ppm (\ref{fig:lc_det}b).  To represent this structure, we added to our global model the transit of a putative planet i (eighth planet) with an orbital period $\sim$30d and a size $\sim$0.2 $R_\oplus$. As can be seen in \ref{fig:planet_i}, adding this transit to our global model drastically reduces the red noise level of the residuals. Given the very low amplitude of this potential transit, we used informative prior PDFs on the impact parameter and orbital period to ensure the MCMC convergence (see \ref{tab:priors}).

\begin{table}[ht!]
\begin{tabular}{@{}lll@{}}
\toprule
Parameter & Prior PDF & Source \\
\midrule
\textbf{Star} &  &  \\
\midrule
$M_\ast$ & $\mathcal{N}(0.0898, 0.0023^2) M_\odot$ & \citeMethods{Agol2021} \\
$R_\ast$ & $\mathcal{N}(0.1192, 0.0013^2) R_\odot$ & \citeMethods{Agol2021} \\
$T_{eff}$ & $\mathcal{N}(2566, 26^2) K$ & \citeMethods{Agol2021} \\
$[Fe/H]$ & $\mathcal{N}(0.04, 0.08^2)$ dex & \cite{Gillon2016} \\
$\pi_\ast$ & $\mathcal{N}(80.2123, 0.0716^2)$ mas &
\citeMethods{2020yCat.1350....0G} \\
$u_{1, \rm F1500W}$ & $\mathcal{N}(0.019, 0.035^2)$ &
\citeMethods{Bourque2022} \\
$u_{2, \rm F1500W}$ & $\mathcal{N}(0.095, 0.05^2)$ & \citeMethods{Bourque2022} \\
$F_{\ast, \rm 15\mu m}$ & $\mathcal{N}(2.496, 0.080^2)$ mJy & this work \\
\midrule
\textbf{Planet b} &  &  \\
Impact parameter $b$ & $\mathcal{N}(0.095, 0.065^2)$ $R_\ast$ &
\citeMethods{Agol2021} \\
Orbital period $P$ & $\mathcal{N}(1.51088432, 0.00000015^2)$ d & \cite{Ducrot2020}  \\
Orbital eccentricity $e$ & $\mathcal{N}(0, 0.005^2)$ & \citeMethods{Agol2021} \\
Phase curve offset $\delta$ &  $\mathcal{N}(0, 45^2)$ deg (analyses \#1) & \\
Phase curve parameter $\gamma$ & $\mathcal{N}(3.3, 0.5^2)$ (analysis \#2 and \#3) & \cite{Agol2007} \\
\midrule
\textbf{Planet c} &  &  \\
\midrule
Impact parameter $b$ & $\mathcal{N}(0.109, 0.061^2)$ $R_\ast$ & \citeMethods{Agol2021} \\
Orbital period $P$ & $\mathcal{N}(2.42179346, 0.00000023^2)$ d & \cite{Ducrot2020}  \\
Orbital eccentricity $e$ & $\mathcal{N}(0, 0.003^2)$&  \citeMethods{Agol2021} \\
Phase curve offset $\delta$
& $\mathcal{N}(0, 45^2)$ deg (analyses \#1 and \#3)  & \\
Phase curve parameter $\gamma$ & $\mathcal{N}(3.3, 0.5^2)$ (analysis \#2)  &  \cite{Agol2007} \\
\midrule
\textbf{Planet g} &  &  \\
Impact parameter & $\mathcal{N}(0.379, 0.018^2)$ $R_\ast$  &
\citeMethods{Agol2021} \\
Orbital period $P$ & $\mathcal{N}(12.353556, 0.0000034^2)$ d &
\cite{Ducrot2020} \\
Orbital eccentricity &  $N_p(0, 0.0013^2)$  & \citeMethods{Agol2021} \\
\midrule
\textbf{Candidate Planet i} &  &  \\
Impact parameter & $\mathcal{N}(0.5, 0.5^2)$ $R_\ast$  & \\
Orbital period $P$ & $\mathcal{N}(30, 5^2)$ d &  \\
\botrule
\end{tabular}
\caption{\textbf{Prior probability distribution functions (PDFs) used in the MCMC analyses of MG. } $\mathcal{N}(a, b^2)$ = normal distribution of mean $a$ and standard deviation $b$. $N_p$ = normal distribution discarding negative values. }\label{tab:priors}
\end{table}

We  relaunched a Markov Chain of 20,000 steps that resulted in a drastic decrease in the level of red noise in the data (\ref{fig:planet_i}). We  then rescaled the photometric errors for white  and red noise  as described in ref. \cite{Gillon2012}, and  performed two chains of 100,000 steps each (with the first 20\% as burn-in). The convergence of the analysis was checked using the Gelman and Rubin statistical test \cite{Gelman1992}. For the other analyses, we used the same rescaling factors than for analysis \#1, which made possible to estimate their relative probability using their best-fit BIC as a proxy for the marginal likelihood of their models. This methodology is justified by the fact that all three analyses led to the same standard deviations of the residuals, i.e. the three models fitted equally well the data.

A quadratic limb-darkening law \cite{MandelAgol2002} was assumed for the star. We assumed normal prior probability distribution functions (PDFs) for the linear and quadratic coefficient $u_1$ and $u_2$ based on the output of the \texttt{ExoCTK} limb-darkening calculator \citeMethods{Bourque2022} for a star of $T_{eff} = 3500$K and $\log_{10} (g_\ast$ [cm/s$^{-2}$]) = 5 dex. As these values correspond to a M-dwarf of earlier-type than TRAPPIST-1, we adopted for the PDFs standard deviations ten times larger than the internal errors of \texttt{ExoCTK}.

The jump parameters of the analyses, that is, the parameters perturbed at each step of the MCMC chains, included (1) for the star, the logarithm of the mass, the logarithm of the density, the effective temperature and the metallicity, the combinations $q_1 = (u_1 + u_2)^2$ and $q_2 =  0.5 u_1 (u_1 + u_2)^{-1} $ of its quadratic limb-darkening coefficients $u_1$ and $u_2$  \citeMethods{Kipping2013}, the absolute stellar flux at 15 $\mu$m $F_\ast$, and (2) for planets, the square of the planet-to-star radius ratio (for planets b, c, and g), the maximal dayside planet-to-star flux ratio  (only for planets b and c), the maximal nightside planet-to-star flux ratio (only for planets b and c, and only for the `atmosphere' model), the phase curve offset $\delta$ (only for planet b and c, and only for the atmosphere model), the cosine of the orbital inclination (for the three planets), the mid-transit timing(s) (for all planets), and the Lagrangian parameters $e\cos{\omega}$ and $e\sin{\omega}$ (with $e$ the orbital eccentricity and $\omega$ the argument of pericenter). Informative normal prior distributions were assumed for some physical parameters (see \ref{tab:priors}).

\begin{figure}
   \centering
    \includegraphics[width=\textwidth]{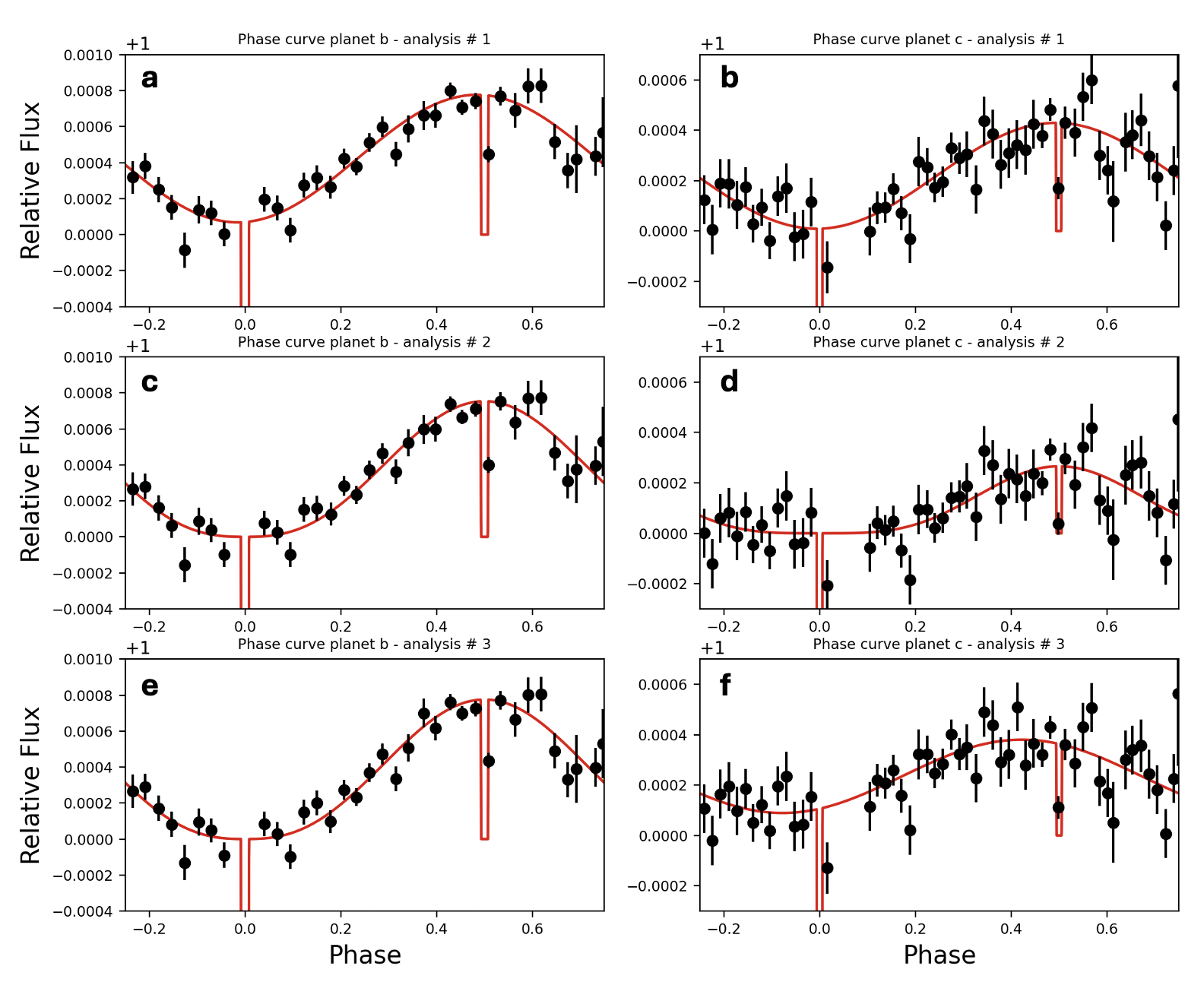}
    \caption{\textbf{Best-fit phase curve models for TRAPPIST-1\,b and c.} Phase-curve models for planets b (panels \textbf{a}, \textbf{c}, and \textbf{e}) and c (panels \textbf{b}, \textbf{d}, and \textbf{f})  from analysis \#1 -assuming planet may have some heat redistribution- (panels \textbf{a} and  \textbf{b}), \#2 -assuming both planets are airless- (panels \textbf{c} and  \textbf{d}), and \#3 -assuming b only is airless- (panels \textbf{e} and  \textbf{f}). The data points are the light curves corrected from the systematics and the contribution of the other planets, phase-folded, and binned per 60 minutes. Each binned point is the average value of the individual points within the bin, and the error bar is the average of the individual errors divided by the square root of the number of points within the bin.
   }
   \label{extfig:lc_planet_3analyses_MG}
\end{figure}

To maximize the constraints on the planets' dayside emissions at 15 $\mu$m, we decided to globally analyze the data of programs 3077, 1177 (five occultations of planet b), and 2304 (four occultations of planet c). The data of programs 1177 and 2304 are described in ref. \cite{Greene2023Natur} and ref. \cite{Zieba2023}, respectively. Our reductions of the data of these two programs  were similar in all respects to the one of the data of program 3077 described above, using the same version of \texttt{Eureka!} than used for the double phase curve data. \ref{baseline} shows the baseline function and the photometric aperture selected for each light curve. For each occultation light curve of program 1177 and 2304, we also assumed priors on the timings of the surrounding transits based on the transit timing forecasts of ref. \citeMethods{Agol2021}. We did not use any priors on the transit timings for program 3077, as the light curve include two transits of planet b and one transit of planet c. The resulting fits are shown on \ref{extfig:lc_planet_3analyses_MG}. \\

\begin{figure}
  \centering
    \includegraphics[width= 0.85\textwidth]{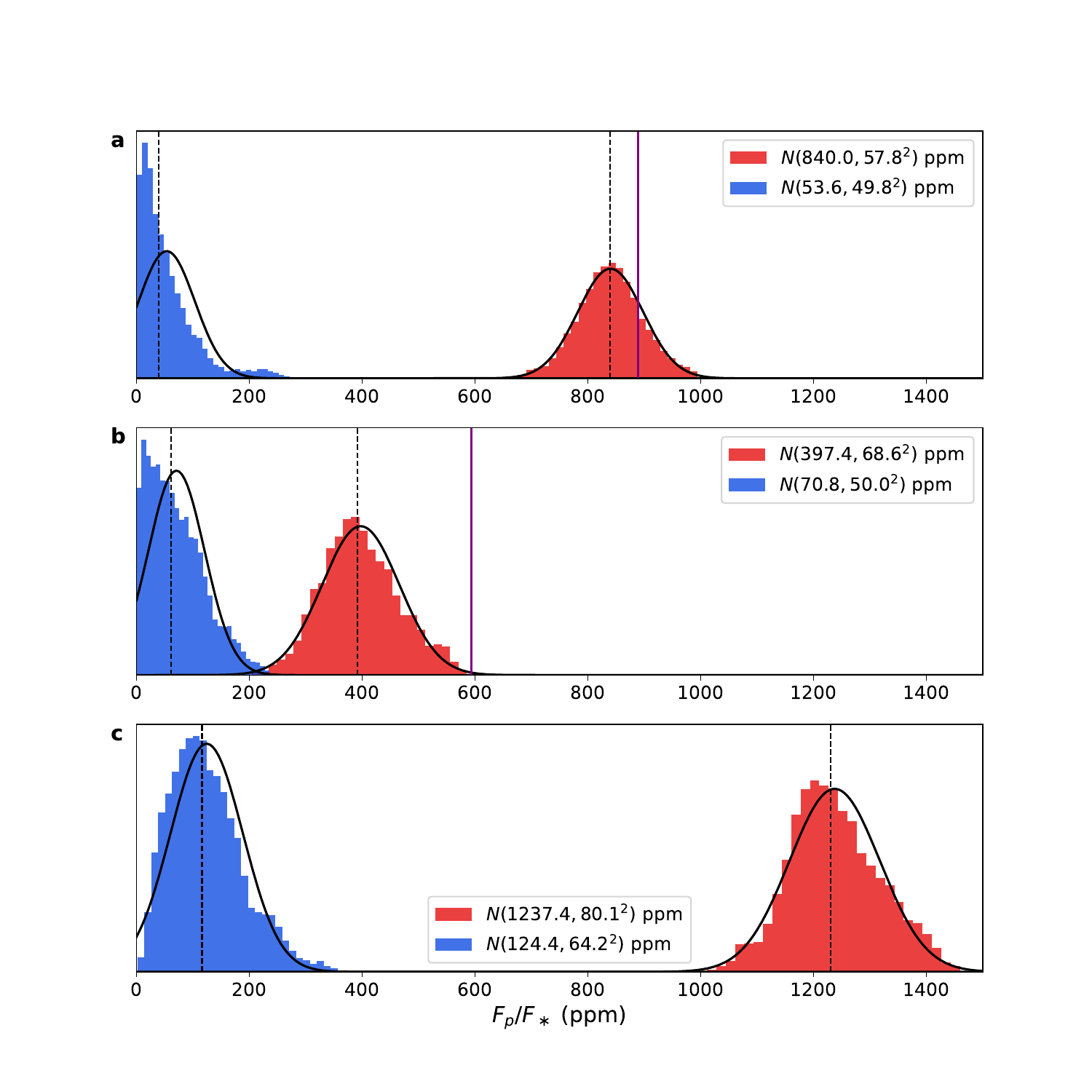}
    \caption{\textbf{Posterior probability distribution functions (PDFs) for the day- and night-side fluxes.} Analysis \#1 posterior PDFs for the relative fluxes of the day (red) and night (blue) sides of TRAPPIST-1\,b (panel \textbf{a}), c (panel \textbf{b}), and the sum of the two planets  (panel \textbf{c}). The symbol $\mathcal{N}(\mu , \sigma^2$) stands for normal centered on $\mu$ with variance $\sigma$. The median of the distributions are shown as vertical dashed lines. The best-fit normal distributions are shown as dark solid lines. For both planets (panels a and b), the relative dayside flux for an airless null-albedo planet is shown as a purple vertical line (889ppm for b and 593ppm for c).}
   \label{fig:zeta}
\end{figure}

\textbf{Results of analysis \#1.} \ref{tab:results} shows the median and 1-$\sigma$ errors of the posterior PDFs of the parameters derived in analysis \#1. \ref{fig:lc_det} shows the detrended program 3077 light curves with the best-fit model as obtained in analysis \#1. \ref{extfig:lc_planet_3analyses_MG} (top panels) shows for planet b and c the phase-folded light curve corrected from systematics and from the contribution of the other planets, binned per 60 min,  with the best-fit phase curve model. \ref{fig:corner_plot} shows the corner plot of the phase curve parameters of both planets, and the posterior PDFs for the sum of their day- and night-side fluxes. \ref{fig:zeta} compares for each planet and for their sum the posterior PDF for the day- and night-side fluxes.

This analysis leads for planet b to a night-side flux consistent with zero, and significantly smaller than the day-side flux. For none of the accepted MCMC steps was the night-side flux larger or equal to the dayside flux, leading to a full heat redistribution probability lower than 0.0006\%. The phase curve peak offset is very well constrained and consistent with zero ($-6.5 \pm 6.4$ deg). The phase curve parameters inferred for planet b are thus fully consistent with an airless planet scenario. The inferred brightness temperatures of the day- and night-sides are, respectively, $490 \pm 17$ K and $197 \pm 41$ K. On its side, the measured 15 $\mu$m transit depth for b, $7316 \pm 117$ ppm, is fully consistent with the ones measured by {\it Spitzer} at 3.6 and 4.5 $\mu$m ($7236 \pm 72$ ppm and $7206 \pm 77$ ppm respectively, \cite{Ducrot2020}).

The results for c are more ambiguous, even if they decisively discard a full heat redistribution ($<$0.0006\% probability). The night side flux of the planet is consistent with zero but also (at 3-$\sigma$) with values as large as 232 ppm.  The phase curve peak offset, is consistent with zero ($10_{-22}^{+25}$ deg), but also with values as large as +76 deg (at 3-$\sigma$). The inferred brightness temperatures of the day and night sides are, respectively, $369 \pm 23$ K and $220_{-47}^{+38}$ K.  Here again, the measured $15\mu m$ transit depth, $7098_{-149}^{+133}$ ppm,  is fully consistent with the ones measured by {\it Spitzer} at 3.6 and 4.5 $\mu m$ ($7210 \pm 140$ ppm and $7027 \pm 68$ ppm respectively, \cite{Ducrot2020}).
The posterior PDFs of the sum of their day- and night-side fluxes strongly reveal a poor heat distribution efficiency (see \ref{tab:results}, \ref{fig:zeta} and \ref{fig:corner_plot}). The  sum of the night-side fluxes is consistent with zero at the 2-$\sigma$ confidence level (\ref{fig:zeta})
The transit depth measured at 15 $\mu$m for planet g, $7563_{-124}^{+138}$  ppm, is also consistent with those measured at 3.6 and 4.5 $\mu$m by {\it Spitzer}:  $7240 \pm 240$ ppm and $7450 \pm 110$ ppm, respectively \cite{Ducrot2020}.

As can be seen in  \ref{fig:lc_det}, two planet-planet occultations (PPOs) \citeMethods{Luger2017_ppo} were adjusted by the global model before the double b+c occultation. To assess the significance of the detection of these PPOs, we performed the very same analysis except that the option to include PPOs in the global model was turned off. It led to a BIC of 7792.4 $vs$ 7787.3 with the PPOs. It corresponds to a Bayes factor of 12.8 in favor of the model with PPOs, which corresponds to a strong but far from decisive strength of evidence \citeMethods{bayesfactor}. We cannot thus conclude to a robust detection of PPOs in our program 3077 light curve.
The transit depth measured for the candidate planet i is $306_{-80}^{+70}$ ppm, i.e. its significance level is $\sim$4$\sigma$. If real, this planet would have a radius $\sim$ 0.2 $R_\oplus$, similar to Triton. From the delays between its transit and the one of g, we compute that, if this planet had the same orbital period than g, the distance between both planets would be $\sim$6 times g's Hill radius, i.e. it couldn't be a moon of g.  \\

\textbf{Results of analysis \#2.} This test analysis assuming airless models for both planets leads to a standard deviation of 862 ppm for the residuals of 3077 program's light curve, the exact same value than for analysis \# 1. The resulting day-side fluxes are lower than those derived in analysis \#1, but they are consistent with them at the $\sim$1-$\sigma$ level. The best-fit BIC of analysis \#2 is 7773.3 $vs$ 7787.4 for analysis \#1, which corresponds to a Bayes factor $> 1152$, decisively \citeMethods{bayesfactor} in favor of the airless model for both planets. Nevertheless, cautions in the conclusions is advised, as there is no guarantee that the whole set of values derived for the parameter $\gamma$ of eq. (\ref{eqn:2}) correspond to physical solutions for the surface of a rocky planet. We thus limit to the conclusion that the data are well represented by a model assuming two airless planets.
The transit depth measured for the candidate planet i is here $289_{-117}^{+326}$ ppm, i.e. its significance level is less than 3$\sigma$.

We also performed a variant of analysis \#2 assuming a non-zero night-side flux that was let free in the MCMC. The resulting values were $67_{-48}^{+74}$ ppm for b and $72_{-48}^{+60}$ ppm. In other words, the analysis did not reveal any significant night-side flux for both planets. For the sum of both planets, the inferred night-side flux was $147_{-63}^{+84}$ ppm, which is not significantly larger than zero.   \\

\textbf{Results of analysis \#3.} This analysis leads again to a standard deviation of 862 ppm for the residuals of 3077 program's light curve. The derived day-side flux of planet b is consistent with those derived in analyses \#1 and \#2. On their side, the phase curve parameters derived for planet c are consistent with those derived in analysis \#1. A full heat redistribution is again disfavored, but much less decisively than for analysis \#1  (0.4\% probability $vs$ $<0.0006$\%). The BIC of this analysis is 7780.1, which, when compared to analysis \#2, leads to a Bayes ratio of 30 in favor of the double airless scenario, strongly supporting this latter \citeMethods{bayesfactor}. But here again, caution in the conclusions is required, for the reasons mentioned above.

The transit depth measured for the candidate planet i is here $285_{-109}^{+210}$ ppm, i.e. its significance level is again less than 3$\sigma$, as in analysis \#2.
Considering the results of three analyses, the significance of this putative transit appears too low to claim a detection.

MG’s nominal analysis (\#1) and test analyses (\#2 and \#3) yield consistent results, leading to the conclusion that efficient heat redistribution is firmly ruled out for TRAPPIST-1\,b and appears unlikely for TRAPPIST-1\,c.


\clearpage
\subsection*{Data reduction and analysis (ED)}

For this second reduction, we employed the \texttt{Eureka!} pipeline \cite{Bell2022} starting from the calints.fits files. In that context, we started using \texttt{Eureka!} from Stage 3 which corresponds to the aperture photometry extraction step.
We used a subarray region for of [186, 326] in the x and y directions. We masked pixels flagged in the data quality flags array (DQ array\footnote{Data Quality flags in details here \url{https://jwst-pipeline.readthedocs.io/en/latest/jwst/dq_init/index.html}}), interpolated defective pixels, and conducted aperture photometry on the star. We choose an aperture of 11 pixels and for each integration, we recorded the center and width of the point spread function (PSF) in the x and y directions by fitting a 2D Gaussian. Background computation involved an annulus of 40 to 60 pixels (centered on the target), with the result subtracted. We identified a bright group of pixels at 20 pixels from the center of the target  which is why we started our background annulus only at 40 pixels. Once the flux was extracted we processed to stage 4, we applied a 4$\sigma$ sigma-clip to identify outliers deviating from the median flux, calculated using a 100-integrations-width boxcar filter.
In the final light curve we identified by eye five transits (partial c, b, g, b and c), the double eclipse (b+c) and two consecutive high energy flares. The planet-planet occultations and the eclipse of d are too shallow to be seen by eye in the light curve.\\

\begin{figure}[!ht]
    \centering
    \includegraphics[width=\textwidth]{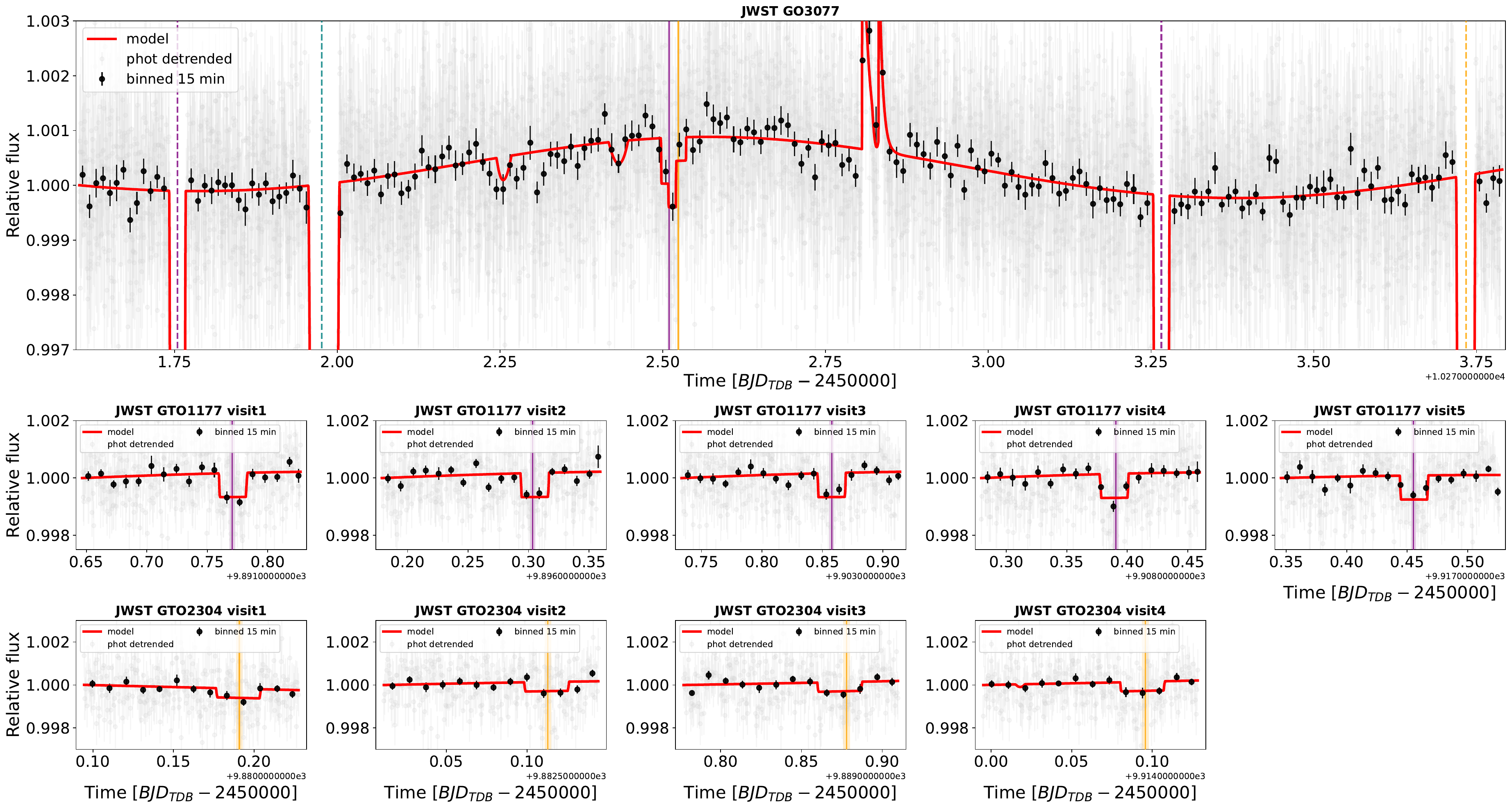}
    \caption{\textbf{Detrended light curves from ED's analysis \#1.} Detrended light curves for each JWST program that observed TRAPPIST-1 b and c at 15$\mu m$ using MIRI F1500W. The top panel is the phase curve from GO 3077, second row of panels are the 5 visits from GO 1177 (PI: Greene) and the third row are the 4 visits from GO 2304 (PI: Kreidberg). The best fit from ``Analysis 7 planets \#1" is shown with a red curve. The expected transit and eclipse timings for each planet from \protect\citeMethods{Agol2021} are shown in vertical dashed and plain lines respectively (purple for b, orange for c, and green for g). }
    \label{fig:joint_fit_elsa}
\end{figure}

\begin{figure}[!ht]
    \centering
    \includegraphics[width=0.99\textwidth]{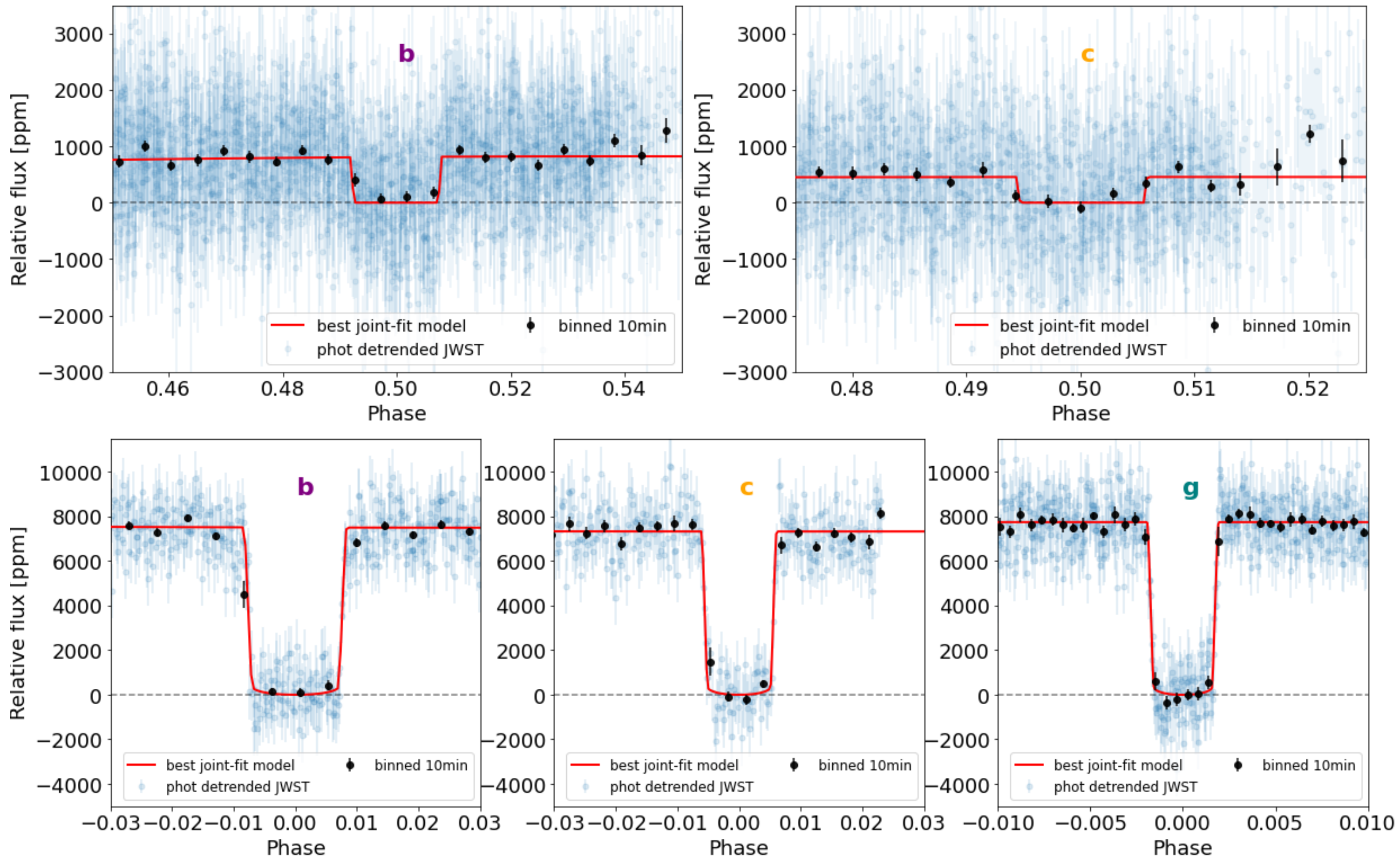}
    \caption{\textbf{Phase folded eclipse and transit light curves for planet b, c and g.} These phase folded light curve result from the best fit analysis ``Analysis 7 planets \#1" by ED. }
    \label{fig:phase_folded_elsa}
\end{figure}

For the analysis of the phase curve we also used \texttt{trafit}. In order to maximize the observational constraints on planet b and c, we proceeded to a global analysis of all MIRI observations of TRAPPIST-1 b and c at 15 $\mu m$. This included 5 eclipses of planet b (GTO 1177) and 4 eclipses of planet c (GO 2305) and the double phase curve presented in this work. We used the light curves from the ED reductions in ref. \cite{Greene2023Natur} and ref. \cite{Zieba2023}. The eclipse model from ref. \cite{MandelAgol2002} was employed to depict photometric time series, and it was multiplied by a baseline model to account for additional astrophysical and instrumental factors contributing to photometric fluctuations. The other astrophysical model included stellar photometric activity such as flares. The instrumental systematics model included PSF width ($PSF_{sx}$, $PSF_{sy}$), PSF position ($PSF_{x}$, $PSF_{y}$) and background. Initially, baselines were chosen for each light curve by running one chain with all possible polynomial combinations for $PSF_{sx}$, $PSF_{sy}$, $PSF_x$, $PSF_y$, and background, we then chose the one minimizing the BIC. In all the different analyses, described below, orders of the polynomial were all inferiors to 3. We also modeled the expected persistent effect on MIRI \citeMethods{2024A&A...683A.212D} using two ramp models (we modeled these ramp as a quadratic function of $ln(dt)$ like in ref. \citeMethods{2008ApJ...686.1341C}) at the beginning of each exposure, as the phase curve was divided in two exposures (see section \ref{sec:observations}). Once the baselines have been selected, we run one chain of 50000 steps to derive the errors correction factor of each light curves. These correction factors are a way to evaluate the need for re-scaling of the photometric errors through the consideration of a potential under- or over-estimation of the white noise of each measurement and the presence of time-correlated (red) noise in the light curve, more details in ref. \cite{Gillon2012}. A large set of jump parameters are randomly perturbed at each step, and a model is derived from their values and compared with the data. In our global analysis the jump parameters for the star were:\\
- the log of the mass, the stellar effective temperature, the log of the density, and the metallicity [Fe/H]. For each parameter we assumed a normal prior based on the value from ref. \citeMethods{Agol2021} (see \ref{tab:priors}). We also fitted for the limb darkening coefficients with priors $\mathcal{N}(0.019,0.07^2)$ and $\mathcal{N}(0.095,0.1^2)$ for $u_1$ and $u_2$ respectively. \\
For the planets we used the following jump parameters:\\
- the radius ratios $R_p/R_{\star}$, the eccentricity $e$, and the impact parameter $b$, with priors from ref. \citeMethods{Agol2021}.\\
- the TTVs (with fix $T_0$ and $P$), with a prior centered at 0.0 with an error of 3 min, 5 min, and 8 min for planets b, c and g respectively. \\
- the day and night flux ratios of planet b and c at 15$\mu m$, with priors $\mathcal{N}(863,90^2)$ ppm and $\mathcal{N}(45,500^2)$ ppm for $F_{b,~D,~15\mu m}$ and $F_{b,~N,~15\mu m}$ respectively; and priors $\mathcal{N}(421,90^2)$ ppm and $\mathcal{N}(45,500^2)$ ppm for $F_{c,~D,~15\mu m}$ and $F_{c,~N,~15\mu m}$ respectively. \\
- Finally our last jump parameter was the phase curve offset for b ($\delta_{PC,~b}$) and c ($\delta_{PC,c}$) with broad priors, $\mathcal{N}(0,30^2)^{\circ}$, for both planet.\\
We fit for the eclipse timing variations (ETVs) in addition to TTVs and fix the periods and reference transit timings of the planets to the predictions by ref. \citeMethods{Agol2021}.
The results from this global analysis are presented in \ref{tab:ED_posteriors} under the name ``Analysis 7 planets \#1", the light curves and their best-fit model are shown on \ref{fig:joint_fit_elsa}. and the resulting phase-folded eclipse and transit light curves are shown in \ref{fig:phase_folded_elsa}.

Then, we performed two additional global analyses where we changed a couple of parameters. First, we proceeded similarly to MG and assumed that planet b is an airless planet that we can model with a quasi-Lambertian phase curve model from ref. \cite{Agol2007}, with a restricted normal prior for $\gamma$, $\mathcal{N}(2.95, 0.1^2)$. We used a restrictive prior centered on 2.95 because it is the value of $\gamma$ that provides the closest approximation to the Bidirectional Reflectance Distribution Function (BRDF), and therefore the most suited way to model planet b as an airless planet. The results from this second global analysis are presented in \ref{tab:ED_posteriors} under the name ``Analysis 7 planets \#2".
Second, as we identified a structure that we could not associate to any instrumental systematic origin we tried to model it as a smaller outer 8th planet that we named planet i (similarly to MG analyses). For this planet we fitted for the following parameters : the transit depth with a wide prior $\mathcal{N}(500, 400^2)$ ppm, the impact parameter with $\mathcal{N}(0.5,0.1^2)$, the transit mid-time with $\mathcal{N}(246~0272.277, 0.02^2)$ $BJD_{TDB}$, and the period with a wide prior $\mathcal{N}(30, 10^2)$ days. In this analysis we also assumed that planet b was an airless planet that we modeled with a quasi-Lambertian phase curve. The results from this second global analysis are presented in \ref{tab:ED_posteriors} under the name ``Analysis 8 planets".  \\

\begin{table}[ht]
\footnotesize
\begin{tabular}{@{}llll@{}}
\toprule
Analyse & Analysis 7 planets \#1  &  Analysis 7 planets \#2 & Analysis 8 planets  \\
\midrule
\textbf{Planet $b$} &  &   \\
Mid-transit timing (\# 1) &  271.75438 $\pm$ 1.23E-04 & 271.75438 $\pm$ 8.69E-05   & 271.75438 $\pm$ 1.12E-04 \\
($\rm BJD_{TDB}$ - 2460000) & &  \\
Mid-transit timing (\# 2) & 273.26591 $\pm$ 1.11E-04 & 273.26591 $\pm$ 8.11E-05 & 273.26590 $\pm$ 9.83E-05  \\
($\rm BJD_{TDB}$ - 2460000) & &  \\
Impact parameter $b$ & $0.1037_{-0.07}^{+0.05}$ $R_\ast$ & $0.086_{-0.003}^{+0.004}$ $R_\ast$  & $0.095_{-0.008}^{+0.004}$ $R_\ast$  \\
Inclination $i$ & $89.7125 \pm 0.0215$ $^{o}$ & $89.7608 \pm 0.0124$ $^{o}$  & $89.7368 \pm 0.0217$ $^{o}$ \\
Semi-major axis $a$ & 0.011526 $\pm$ 7.1E-05 au & 0.011600 $\pm$ 6.4E-05 au  & 0.011465 $\pm$ 6.0E-05 au \\
Transit depth $dF$ & $7271 \pm 140$ ppm & $7334 \pm 90$ ppm  & $7315 \pm 128$ ppm \\
Eccentricity $e$ & $0.0003$ $\pm$ 2.67E-04 & $0.00015$ $\pm$ 4.44E-04 & $0.00075$ $\pm$ 2.35E-03  \\
$F_{\rm b, D, max}$ & $791$ $\pm$ 68 ppm & $788_{-41}^{+65}$ ppm  & 810 $\pm$ 40 ppm   \\
$F_{\rm b, N, max}$ & $62_{-45}^{+78}$ ppm & 0 ppm (fixed) & 0 ppm (fixed) \\
Peak offset $\delta_{PC,~b}$ & 8 $\pm$ 12 deg & 0 deg (fixed) & 0 deg (fixed)  \\
$\gamma$ & -- & $2.61 \pm 0.15$ & $2.99 \pm 0.075$  \\
$T_{\rm brightness, D}$ & $479 \pm 22$ K & $472 \pm 20$ K & $486 \pm 32$ K \\
$T_{\rm brightness, N}$ & $217 \pm 50$ K & 0 K (fixed)  & 0 K (fixed)  \\
\midrule
\textbf{Planet $c$} &  &    \\
Mid-transit timing & 273.73336 $\pm$ 1.35E-04 & 273.73335 $\pm$ 1.03E-04 & $273.73336 \pm$ 1.23E-04   \\
($\rm BJD_{TDB}$ - 2460000) & &  \\
Impact parameter $b$ & $0.104 \pm 0.075$ $R_\ast$ & $0.08 \pm 0.065$ $R_\ast$ & $0.103 \pm 0.010$ $R_\ast$   \\
Inclination $i$ & $89.7881 \pm 0.0153$ $^{o}$ & $89.8283 \pm 0.0134$ $^{o}$  & $89.7921 \pm 0.0192$ $^{o}$ \\
Semi-major axis $a$ & 0.01578 $\pm$ 9.7E-05 au & 0.01588 $\pm$ 7.7E-05 au  & 0.01570 $\pm$ 8.2E-05 au \\
Transit depth $dF$ & $7098$ $\pm$ 149 ppm & $7122$ $\pm$ 109 ppm  & 7077 $\pm$ 125 ppm  \\
Eccentricity $e$ & $0.00024$ $\pm$ 1.74E-04 & $0.00001$ $\pm$ 2.54E-04 & $0.00018$ $\pm$ 6.10E-04 \\
$F_{\rm c, D, max}$ & $405 \pm 71$ ppm & $465 \pm 74$ ppm & $473_{-45}^{+40}$ ppm  \\
$F_{\rm c, N, max}$ & $125 \pm 90$ ppm & $92_{-64}^{+93}$ ppm & $138_{-193}^{+129}$ ppm  \\
Peak offset $\delta_{PC,~c}$ & $-1_{-28}^{+32}$ deg & $14_{-14}^{+13}$ deg & $-31_{-21}^{+27}$ deg  \\
$\gamma$ & - & - & -\\
$T_{\rm brightness, D}$ & $372 \pm  21$ K & $388 \pm  25$ K & $395 \pm 36$ K  \\
$T_{\rm brightness, N}$ & $258 \pm  79$ K & $239 \pm 102$ K & $267 \pm 169$ K \\
\midrule
\textbf{Planet $g$} &  &  \\
Mid-transit timing & $271.97933 \pm 0.00015$ & $271.97934 \pm 0.00011$  &  $271.97935 \pm 0.00011$ \\
($\rm BJD_{TDB}$ - 2460000) & & \\
Impact parameter $b$ & 0.378 $\pm$ 0.021 $R_\ast$ & 0.355 $\pm$ 0.014 $R_\ast$ &  0.379 $\pm$ 0.0137 $R_\ast$ \\
Inclination $i$ & $89.7420 \pm 0.0155$ $^{o}$ & $89.7591 \pm 0.0408$ $^{o}$  & $89.7453 \pm 0.0127$ $^{o}$ \\
Semi-major axis $a$ & 0.04678 $\pm$ 2.9E-04 au & 0.04708 $\pm$ 2.6E-04 au  & 0.04653 $\pm$ 2.4E-04 au \\
Transit depth $dF$ & $7680$ $\pm$ 146 ppm & $7615_{-90}^{+82}$ ppm & $7739_{-122}^{+95}$ ppm \\
Eccentricity $e$ & $0.000017 \pm$ 1.66E-04 & $0.0000459 \pm$ 1.54E-04 & $0.0001786 \pm$ 1.54E-04 \\
\midrule
\textbf{Candidate planet $i$} &  &   \\
Mid-transit timing & - & -  & $272.289611 \pm$ 3.50E-02    \\
($\rm BJD_{TDB}$ - 2460000) &  &   & \\
Impact parameter $b$ & - &  - & $0.50 \pm 0.12$ $R_\ast$ \\
Orbital inclination $i$ & - &  -  & $89.80 \pm 0.18$  \\
Period $P$ & -  & - & $25 \pm 1.3$ d  \\
Transit depth $dF$ & - &  - & $340_{-118}^{+111}$ ppm  \\
Radius $R_p$ & - & -  & $0.239_{-0.046}^{+0.036}$ $R_\oplus$   \\
\botrule
\end{tabular}
\caption{\textbf{Resulting planetary parameters from the global MCMC analyses done by ED}}
 \label{tab:ED_posteriors}
\end{table}

\textbf{Results from the three global analyses} \\
 ``Analysis 7 planets \#1" favours a large flux contrast between the day and the night side of TRAPPIST-1b and almost no offset of the phase curve, very similar to what we expect for a bare-rock planet. For TRAPPIST-1 c we measured a larger day-side temperature than night-side, however considering the still low SNR on the later it is not possible to confirm whether the heat redistribution is fully inefficient or not on this planet. Similarly, the offset of TRAPPIST-1c's phase curve can vary between 1 and 30 degrees from one analysis to another, but the associated error bars are large such that it is always consistent with zero. Our best-fits for each analysis favor  planet-planet occultations between planet b and c, however the detection is not at the 3-$\sigma$ level, and therefore not significant yet. Finally, the transit depth that we derive for planet b, c and g are all consistent with the values obtained from Spitzer IRAC observations at 3.6 $\mu m$ and 4.5 $\mu m$ \cite{Ducrot2020}.
Our conclusions concerning the fitting the double-flare event can be found in section \ref{sec:flares}.

\subsection*{Data reduction and analysis (TJB)}

Our third independent reduction followed a similar process as that described by ref.~\cite{Greene2023Natur}. We re-reduced all MIRI F1500W eclipse observations of TRAPPIST-1 b and c and reduced the novel phase curve observation using the same reduction procedure. Specifically, we used version 0.11.dev77+g1e43198f.d20240130 of the \texttt{Eureka!} data analysis pipeline \cite{Bell2022}, with \texttt{jwst} version 1.13.4 \citeMethods{JWSTpipeline}, CRDS version 11.17.19 \citeMethods{crds}, and using the `jwst\textunderscore 1188.pmap' CRDS context. The \texttt{Eureka!} Control Files and \texttt{Eureka!} Parameter Files we used are available for download on Zenodo (\url{https://zenodo.org/uploads/15005073}). We began with the uncalibrated FITS files and ran \texttt{Eureka!}'s Stages 1--2 with the default \texttt{jwst} settings with the exception of turning off the Stage 1 EMI correction step (since initial experimentation showed that this time-consuming step offered little-to-no benefit for these data), turning on the Stage 1 `firstframe' and `lastframe' steps (which discard the first and last groups of each integration, respectively), setting the Stage 1 jump rejection threshold to 10$\sigma$, and turning off the Stage 2 `photom' step.

In \texttt{Eureka!}'s Stage 3, we first masked pixels marked as `DO\textunderscore NOT\textunderscore USE' in the data quality array, converted the data units from data number per second to electrons, masked outlier pixels in the background region using a double-iteration 5$\sigma$ clipping procedure, and interpolated bad pixel values using a bilinear interpolation. We performed centroiding on each integration using \texttt{Eureka!}'s `fgc' method on the 10 pixels surrounding the manually selected initial centroid guess for later use as covariates in Stage 5. We then performed a multi-step aperture optimization process on the phase curve observations using the median absolute deviation of the resulting lightcurves as a metric to select the lightcurve with the lowest noise before fitting. We first varied the photometric aperture radius from 3 to 11 pixels in 1-pixel steps and chose an aperture radius of 5-pixels, varied the sky annulus inner-radius from 10 to 24 pixels in 2-pixel steps and chose an inner-radius of 16 pixels, and finally varied the sky annulus width from 10 to 40 pixels in steps of 2-pixels before choosing a width of 30 pixels. Repeating the aperture and annulus selection process did not result in any changes. We then used these same aperture and annulus parameters for all nine eclipse-only observations. Finally, in \texttt{Eureka!}'s Stage 4 we performed sigma clipping on the time-series by iteratively replacing 3.5$\sigma$ outliers compared to a smoothed version of the signal computed using a 20-integration wide boxcar filter, with a maximum of 20 iterations.

We then proceeded to simultaneously fit all nine eclipse-only lightcurves and the phase curve with \texttt{Eureka!}'s Stage 5. To model the transits of TRAPPIST-1 b, c, and g and the eclipse signals of TRAPPIST-1 b and c, we used the \texttt{batman} package \citeMethods{batman2015} to compute light curves with quadratic limb-darkening using formulae from ref. \cite{MandelAgol2002}.  We neglected the planet-planet occultations expected to be present in the phase curve observation; numerous attempts at fitting all ten lightcurves with the \texttt{starry} package \citeMethods{Luger2019starry,Luger2021starry} (which models the planet-planet occultations) and \texttt{PyMC3}'s No U-Turns Sampler \citeMethods{pymc3} failed to converge, perhaps due to very high dimensionality of the fit. For the stellar limb-darkening, we freely fit the reparameterized quadratic limb-darkening coefficients ($q_1$, $q_2$) of ref.~\citeMethods{Kipping2013} assuming that the coefficients were consistent between the transits of TRAPPIST-1 b, c, and g. We used broad, minimally informative Gaussian priors on the planetary radii ($R_{\rm p}/R_*$) and eclipse depths ($F_{\rm p, day}/F_*$). For both TRAPPIST-1 b and c, we assumed the planets' orbital period ($P$), inclination ($i$), and semi-major axes  ($a/R_*$) were sufficiently constant across all epochs and shared each of these parameters between the different epochs, using Gaussian priors based on ref.~\citeMethods{Agol2021}. However, the gravitational effects of the other planets in the system are not negligible and result in transit timing variations (TTVs) as well as eclipse timing variations (ETVs); to model these variations in the eclipse timing, we used a different mid-eclipse time parameter for each epoch using the predictions of ref.~\citeMethods{Agol2021} as Gaussian priors (see \ref{tab:Agol_ETV_Predictions}). For the transits present in the phase curve observation, we also adopted Gaussian priors based on the predictions of ref.~\citeMethods{Agol2021}, with the exception of assuming that the change in orbital period was small enough over the phase curve observation that we could assume that the two TRAPPIST-1 b transits were separated by one orbital period. We estimate that the light travel time difference between transit and eclipse is only $\sim$11 seconds for TRAPPIST-1 b and $\sim$16 seconds for TRAPPIST-1 c \citeMethods{Agol2021}, so we neglect this effect in our modelling.

For our phase variation model, we considered combinations of two different models. One model was the standard first-order sinusoidal phase variation model (with free parameters `AmpCos1' and `AmpSin1' as described by ref.~\citeMethods{Bell2024a}) which is the result of a first-order Fourier series decomposition of any planetary map \citeMethods{Cowan2008}. The second model was the quasi-Lambertian model described above (ref.~\cite{Agol2007} and Equation \ref{eqn:2}; with the free parameter $\gamma$) which assumes an airless planet. We performed one fit using the quasi-Lambertian model for both TRAPPIST-1 b and c (Setup \#1), one fit using the quasi-Lambertian model for TRAPPIST-1 b and the sinusoidal model for TRAPPIST-1 c (Setup \#2), and one fit with sinusoidal phase variations for both TRAPPIST-1 b and c (Setup \#3). For cases where we used the sinusoidal phase variation model, we imposed a prior that each planet's phase variations remained positive. For all other TRAPPIST-1 planets (including TRAPPIST-1 g which transits during the observations), we assumed there was negligible thermal emission.

\begin{figure*}
    \centering
   \includegraphics[width=0.75\linewidth]{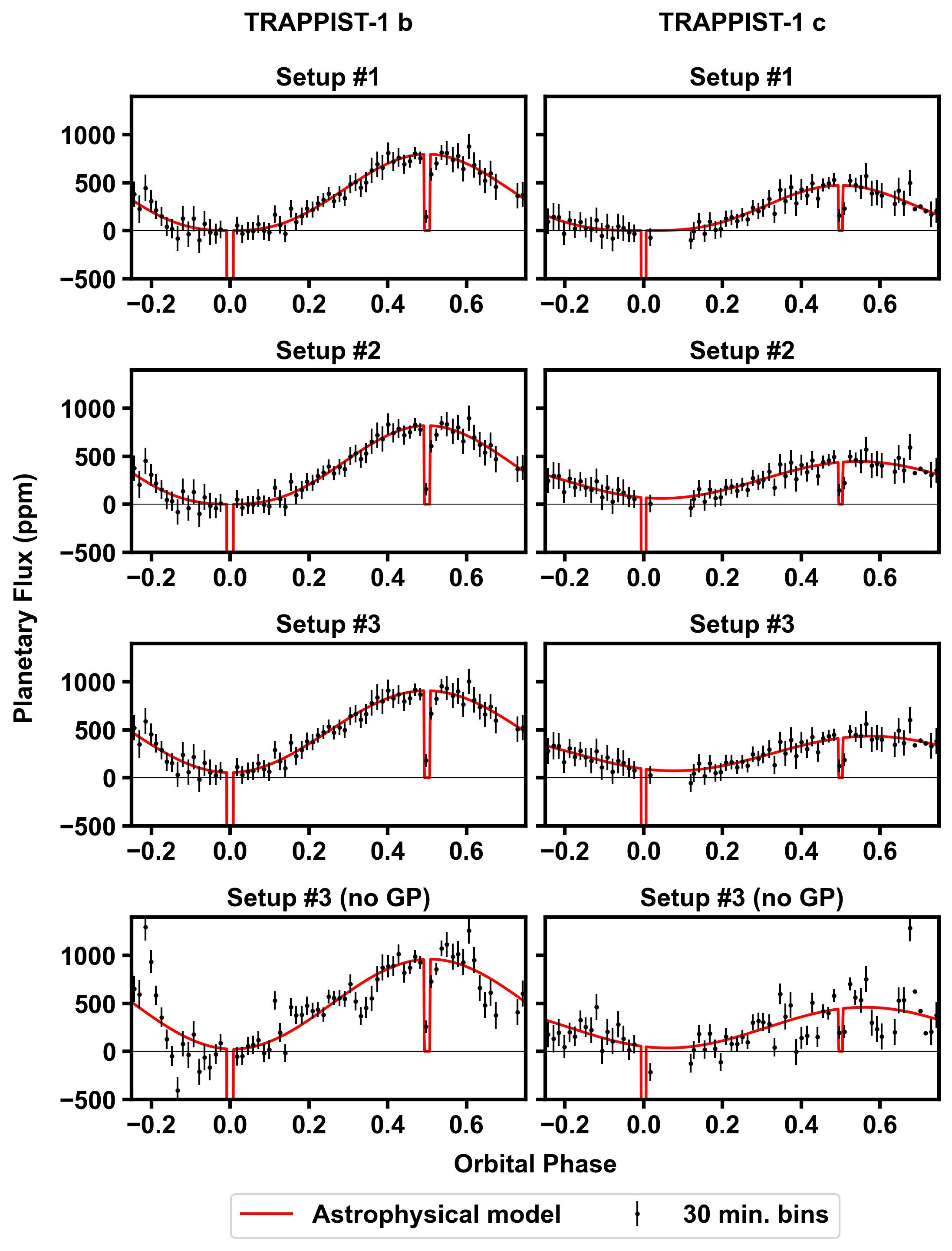}
    \caption{\textbf{Separated phase variations from TRAPPIST-1 b and c from TJB's reduction and analysis with \mbox{Setup \#1}}. The small cluster of points with small uncertainties around phase 0.7 on the TRAPPIST-1 c panels are caused by the many repeated TRAPPIST-1 b eclipse observations which happened to mostly occur at this particular orbital phase of TRAPPIST-1.  Note that in the top 3 rows, correlated variations have been subtracted from the phase curve, as estimated by a maximum-likelihood prediction from the Gaussian process.}
    \label{fig:tjb_separatePhaseCurves}
\end{figure*}

Beyond the planetary signals in the data, there were also numerous obvious stellar and instrumental noise sources present in the lightcurves. To remove the initial settling ramp seen at the beginning of most MIRI observations, we trimmed the first 100 integrations from all ten lightcurves; for the phase curve observation, we removed an additional 700 integrations (800 in total) as there was an unusually strong ramp at the start of these observations and we wanted to limit any correlations between the initial ramp and the phase variations of TRAPPIST-1 b and c. For the phase curve observation, we also removed integrations 3160--3300 as these were affected by a pair of what appeared to be stellar flares. To capture the impact of red noise in the observations on our astrophysical parameters of interest, we simultaneously fitted the astrophysical model along with a polynomial in time (with a constant term and a linear slope term), a linear decorrelation against the PSF's position and width in both the $x$ and $y$ directions, and a Gaussian Process (GP; as implemented in \texttt{celerite2} \citeMethods{celerite1,celerite2}) as a function of time using a Mat\'ern-3/2 kernel \citeMethods{RasmussenWilliams2006} with fitted coefficients of $\ln{A}$ (which controls the correlation strength and is unitless) uniformly constrained between -24 and -10 and $\ln{\tau}$ (which controls the correlation lengthscale, where $\tau$ has units of days$^{-1}$) uniformly constrained to be within -7 and 0. The systematic models for each lightcurve were treated independently. To evaluate the importance of the GP, we also performed an additional fit using Setup \#3 without using a GP during fitting. With this test, we found that the residual white noise and red noise was greatly increased without the inclusion of the GP (e.g., see \ref{fig:tjb_separatePhaseCurves}), and several key parameters of interest appeared to be overly constrained without accounting for the presence of this red noise. To demonstrate what impact the GP had on our fitted and computed astrophysical parameters, we tabulate the results of this fit along with our other fits, although we emphasize that this is only for demonstrative purposes and the results of this particular no-GP fit should not be trusted as there is significant red noise which is unaccounted for in the statistical model. Finally, we also allowed for a white-noise multiplier for each lightcurve to account for background noise (which was not included in the initial estimates of the uncertainties), an incorrect assumption for the gain in Stage 3, and any other white noise term.

We performed sampling using the \texttt{emcee} Affine Invariant Markov chain Monte Carlo (MCMC) Ensemble sampler \citeMethods{Goodman2010,For2013} to obtain estimates for the best-fit values and uncertainties for each fitted parameter. In total, we had between 122 and 124 free parameters for each of our fits, as is documented in \ref{tab:tjb_fitted_values}. We used 500 walkers, each of which repeatedly took 20,000 steps (10 million total samples) until the chain had visually converged and the results from the first 5 million and last 5 million samples gave the same best-fit values and uncertainties. This required 70 million total posterior samples for Setup \#1, 40 million total posterior samples for Setup \#2, 30 million total posterior samples for Setup \#3 with the GP, and 20 million total posterior samples for Setup \#3 without the GP. We then used the median of the final 5 million samples as our best-fit estimate and the 16th and 84th percentiles as our uncertainty estimates. We had hoped to use the \texttt{dynesty} nested sampling algorithm \citeMethods{speagle_dynesty_2020} to enable more robust statistical model comparisons using the computed Bayesian evidence, but an initial attempt at fitting Setup \#3 with the GP using \texttt{dynesty} had still not converged after 300 million posterior evaluations over 20 days of wall-clock runtime on 28 CPU threads (with $d\log{z}=578$, still far from the stopping criteria of $d\log{z}<0.1$). As a result, we have to resort to computing the Bayesian Information Criterion (BIC) for all four of our \texttt{emcee} fits following the methods of ref.~\citeMethods{Bell2019}.

\begin{figure*}
    \centering
    \includegraphics[width=\linewidth]{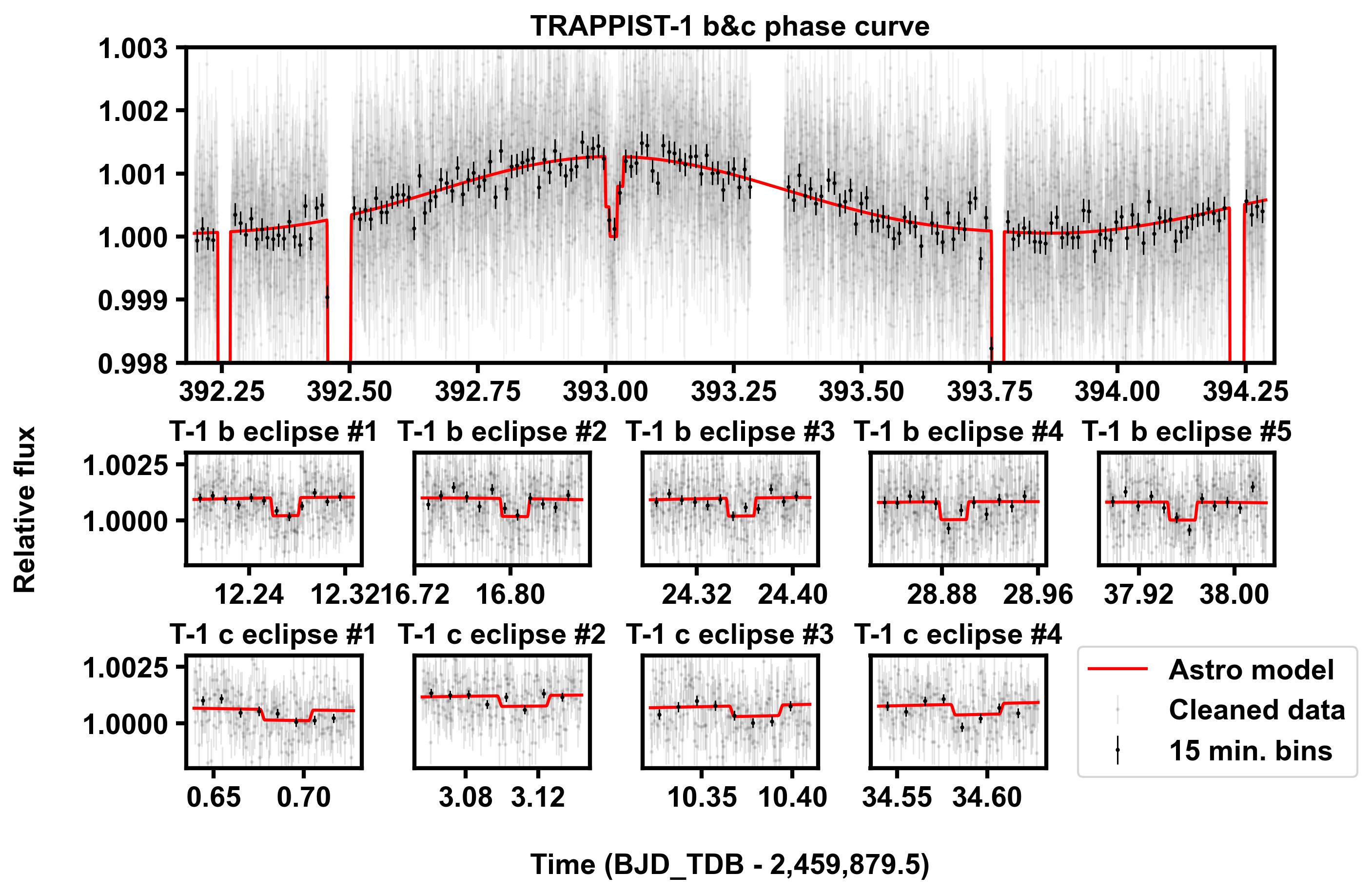}
    \caption{\textbf{Cleaned lightcurves and fitted models from TJB's reduction and analysis with \mbox{Setup \#1}.}}\label{fig:tjb_lightcurves}
\end{figure*}

The results of our fits are summarized in  \ref{tab:tjb_fitted_values}, \ref{fig:tjb_lightcurves} and \ref{fig:tjb_GP_lightcurves}. As shown in \ref{fig:tjb_AllanPlots}, our residuals (after having removed the red noise component modelled with our simultaneously fitted GP) show no unmodelled red noise. Based on the $\Delta$BIC between our different fits we find that our significantly preferred model from the TJB suite was Setup \#1 which used the quasi-Lambertian (airless) model for the phase variations of both TRAPPIST-1 b and c. Our measured eclipse depths ($F_{\rm p, day, T\text{-}1b}/F_*$=796$\pm$77 ppm, $F_{\rm p, day, T\text{-}1c}/F_*$=472$\pm$85 ppm) are consistent to within 1$\sigma$ with previously published values ($F_{\rm p, day, T\text{-}1b}/F_*$=861$\pm$99 ppm, ref. \cite{Greene2023Natur}; $F_{\rm p, day, T\text{-}1c}/F_*$=421$\pm$91 ppm, ref. \cite{Zieba2023}). While the quasi-Lambertian (airless) model assumes zero-nightside flux, those fits which used the sinusoidal phase variation model found nightside flux values that were generally consistent with 0 ppm for both TRAPPIST-1 b and c. Similarly, while the quasi-Lambertian (airless) model assumes the phase variations are symmetric about the eclipse, those fits with sinusoidal phase variations were consistent with no phase curve offset. Together, these findings favour the scenario where neither TRAPPIST-1 b nor c have an atmosphere that transports appreciable amounts of heat to the nightside.

Finally, to assess the evidence for the potential new candidate identified in MG's analyses, we repeated a fit built upon our fiducial Setup \#1 (with GP) with the addition of an additional transiting body. For this candidate, we adopted the following minimally informative priors: $R{\rm p}/R_* = \mathcal{N}(0.01414, 0.1^2)$, $P = \mathcal{N}(30, 10^2)$ days, $t_0 = \mathcal{N}(60271.76, 0.1^2)$ (in units of BJD\textunderscore TDB - 2,400,000.5), $i = \mathcal{U}(85, 90)$ degrees, $a/R_* = \mathcal{N}(100, 10000^2)$, and $e = 0$, where $\mathcal{N}(\mu, \sigma^2)$ is a Normal prior with mean $\mu$ and standard deviation $\sigma$ and $\mathcal{U}(\ell_1, \ell_2)$ is a Uniform prior with a lower bound of $\ell_1$ and an upper bound of $\ell_2$. Our posteriors closely matched our priors, and the BIC test significantly prefers the \textit{exclusion} of the potential candidate with a $\Delta$BIC of 67. As a result, with the reduction and analysis methods described in this subsection, there is no clear evidence for the inclusion of a new transiting exoplanet.


\begin{table*}
    \small
    \centering
    \begin{tabular}{ccc}
        \toprule
         & Predicted Timing \citeMethods{Agol2021} & TJB Fitted Timing \\
        Epoch & (BMJD$_{\rm TDB}$) & (BMJD$_{\rm TDB}$) \\ \midrule
        \textbf{TRAPPIST-1 b} \\
        Eclipse \#1 & 59891.2730$\pm$0.0031 & 59891.27049$^{+0.00100}_{-0.00081}$ \\
        Eclipse \#2 & 59895.8051$\pm$0.0031 & 59895.80380$^{+0.00170}_{-0.00083}$ \\
        Eclipse \#3 & 59903.3600$\pm$0.0031 & 59903.35691$^{+0.00074}_{-0.00100}$ \\
        Eclipse \#4 & 59907.8922$\pm$0.0031 & 59907.88971$^{+0.00100}_{-0.00099}$ \\
        Eclipse \#5 & 59916.9579$\pm$0.0031 & 59916.95650$^{+0.00090}_{-0.00076}$ \\
        Phase Curve Eclipse & 60272.0120$\pm$0.0031 & 60272.01087$^{+0.00190}_{-0.00069}$ \\
        Phase Curve Transits & 60271.25424$\pm$0.00089 & 60271.254440$^{+0.000076}_{-0.000077}$ \\ \midrule
        \textbf{TRAPPIST-1 c} \\
        Eclipse \#1 & 59879.6903$\pm$0.0036 & 59879.6906$^{+0.0025}_{-0.0026}$ \\
        Eclipse \#2 & 59882.1125$\pm$0.0036 & 59882.1122$^{+0.0074}_{-0.0019}$ \\
        Eclipse \#3 & 59889.3781$\pm$0.0036 & 59889.3801$^{+0.0016}_{-0.0056}$ \\
        Eclipse \#4 & 59913.5960$\pm$0.0036 & 59913.5945$^{+0.0019}_{-0.0015}$ \\
        Phase Curve Eclipse & 60272.0236$\pm$0.0037 & 60272.0210$^{+0.0033}_{-0.0022}$ \\
        Phase Curve Transit & 60273.23403$\pm$0.00085 & 60273.23311$^{+0.00012}_{-0.00012}$ \\ \midrule
        \textbf{TRAPPIST-1 g} \\
        Phase Curve Transit & 60271.4761$\pm$0.0015 & 60271.47926$^{+0.00013}_{-0.00013}$ \\ \bottomrule
            \end{tabular}
    \caption{\textbf{Predicted and fitted mid-occultation times for the transits and eclipses of TRAPPIST-1 b, c, and g.} Predicted mid-eclipse and mid-transit times from ref.~\protect\citeMethods{Agol2021} are compared to the fitted timings from TJB's joint fit using \mbox{Setup \#1}. BMJD$_{\rm TDB}$ is the date in the Barycentric Julian Date in the Barycentric Dynamical Time system minus 2,400,000.5 days. Note that both the predicted and fitted timings neglect the light travel time effect which is only $\sim$11 seconds and $\sim$16 seconds for TRAPPIST-1 b and c, respectively.}\label{tab:Agol_ETV_Predictions}
\end{table*}

\begin{table*}
    \footnotesize
    \centering
    \begin{tabular}{ccccc}
        \toprule
        & \textbf{Setup \#1} & Setup \#2 & Setup \#3 & Setup \#3 \\
        & \textbf{(with GP)} & (with GP) & (with GP) & (without GP) \\ \midrule
        \textbf{TRAPPIST-1 (star)} \\
        Limb darkening; $q_1$ & \textbf{0.036$^{+0.034}_{-0.021}$} & 0.036$^{+0.034}_{-0.021}$ & 0.035$^{+0.033}_{-0.021}$ & 0.035$^{+0.034}_{-0.021}$ \\
        Limb darkening; $q_2$ & \textbf{0.19$^{+0.29}_{-0.14}$} & 0.19$^{+0.31}_{-0.14}$ & 0.20$^{+0.34}_{-0.15}$ & 0.19$^{+0.33}_{-0.14}$ \\ \midrule
        \textbf{TRAPPIST-1 b} \\
        P.C.~Model & \textbf{Quasi-Lambertian} & Quasi-Lambertian & Sinusoidal & Sinusoidal \\
        $R_{\rm p}/R_*$ & \textbf{0.08355$^{+0.00086}_{-0.00087}$} & 0.083584$^{+0.00087}_{-0.00090}$ & 0.08357$^{+0.00082}_{-0.00086}$ & 0.08466$^{+0.00063}_{-0.00067}$ \\
        $P$ (days) & \textbf{1.511536$^{+0.000072}_{-0.000110}$} & 1.511538$^{+0.000070}_{-0.000100}$ & 1.511547$^{+0.000066}_{-0.000092}$ & 1.511563$^{+0.000055}_{-0.000084}$ \\
        $a/R_*$ & \textbf{20.81$\pm$0.12} & 20.81$\pm$0.12 & 20.82$\pm$0.11 & 20.79$\pm$0.11 \\
        $i$ (degrees) & \textbf{89.74$\pm$0.12} & 89.75$\pm$0.13 & 89.75$\pm$0.13 & 89.77$\pm$0.13 \\
        $F_{\rm p, day}/F_*$ (ppm) & \textbf{796$\pm$77} & 819$\pm$80 & 907$\pm$94 & 959$\pm$65 \\
        $F_{\rm p, night}/F_*$ (ppm) & \textbf{$\equiv$0} & $\equiv$0 & 53$\pm$59 & 23$\pm$24 \\
        Phase Offset ($^{\circ}$E) & \textbf{$\equiv$0} & $\equiv$0 & 1$\pm$10 & -2.5$\pm$4.6 \\
        Quasi-Lambertian $\gamma$ & \textbf{2.61$^{+0.53}_{-0.47}$} & 2.62$^{+0.51}_{-0.48}$ & --- & --- \\ \midrule
        \textbf{TRAPPIST-1 c} \\
        P.C.~Model & \textbf{Quasi-Lambertian} & Sinusoidal & Sinusoidal & Sinusoidal \\
        $R_{\rm p}/R_*$ & \textbf{0.0839$\pm$0.0011} & 0.0836$^{+0.0011}_{-0.0012}$ & 0.0836$^{+0.0011}_{-0.0012}$ & 0.08411$\pm$0.00082 \\
        $P$ (days) & \textbf{2.42138$^{+0.00091}_{-0.00092}$} & 2.42130$^{+0.00096}_{-0.00080}$ & 2.42136$^{+0.00089}_{-0.00081}$ & 2.42189$^{+0.00074}_{-0.00087}$ \\
        $a/R_*$ & \textbf{28.44$\pm$0.17} & 28.44$\pm$0.17 & 28.46$\pm$0.17 & 28.44$\pm$0.17 \\
        $i$ (degrees) & \textbf{89.806$\pm$0.093} & 89.813$\pm$0.096 & 89.798$\pm$0.094 & 89.812$\pm$0.094 \\
        $F_{\rm p, day}/F_*$ (ppm) & \textbf{472$\pm$85} & 437$\pm$94 & 410$\pm$110 & 443$\pm$61 \\
        $F_{\rm p, night}/F_*$ (ppm) & \textbf{$\equiv$0} & 66$\pm$57 & 89$\pm$91 & 50$\pm$45 \\
        Phase Offset ($^{\circ}$E) & \textbf{$\equiv$0} & -16$\pm$24 & -27$\pm$31 & -22$\pm$19 \\
        Quasi-Lambertian $\gamma$ & \textbf{3.13$^{+0.46}_{-0.44}$} & --- & --- & --- \\ \midrule
        \textbf{TRAPPIST-1 g} \\
        $R_{\rm p}/R_*$ & \textbf{0.0872$\pm$0.0011} & 0.0873$\pm$0.0011 & 0.0874$\pm$0.0011 & 0.08691$^{+0.00065}_{-0.00063}$ \\
        $P$ (days) & \textbf{12.3557$\pm$0.0020} & 12.3556$^{+0.0021}_{-0.0022}$ & 12.3557$^{+0.0020}_{-0.0021}$ & 12.3556$^{+0.0021}_{-0.0020}$ \\
        $a/R_*$ & \textbf{84.21$\pm$0.48} & 84.22$\pm$0.46 & 84.23$\pm$0.45 & 84.26$\pm$0.48 \\
        $i$ (degrees) & \textbf{89.7465$\pm$0.0096} & 89.7482$\pm$0.0096 & 89.7472$\pm$0.0097 & 89.7468$\pm$0.0099 \\ \midrule
        \textbf{Gaussian Process} \\
        $\ln{A}$ (Amplitude) & \textbf{-16.72$\pm$0.24} & -16.64$^{+0.28}_{-0.25}$ & -16.63$^{+0.42}_{-0.28}$ & --- \\
        $\ln{\tau}$ (Lengthscale) & \textbf{-3.87$^{+0.47}_{-0.36}$} & -3.78$^{+0.59}_{-0.40}$ & -3.79$^{+1.00}_{-0.44}$ & --- \\ \midrule
        \# Free Parameters & \textbf{122} & 123 & 124 & 122 \\
        Residual Std.~Dev.~(ppm) & \textbf{883} & 883 & 884 & 907 \\
        $\Delta$BIC (N$_{\rm data}$=6571) & \textbf{0} & 12.3 & 17.4 & 136.8 \\ \bottomrule
    \end{tabular}
        \caption{\textbf{Fitted and inferred parameters from TJB's fits using \mbox{Setup \#1}.}} \label{tab:tjb_fitted_values}
\end{table*}

\subsection*{Data reduction and analysis (ZH)}

\begin{figure}[!ht]
   \centering
    \includegraphics[width= 1.0\textwidth]{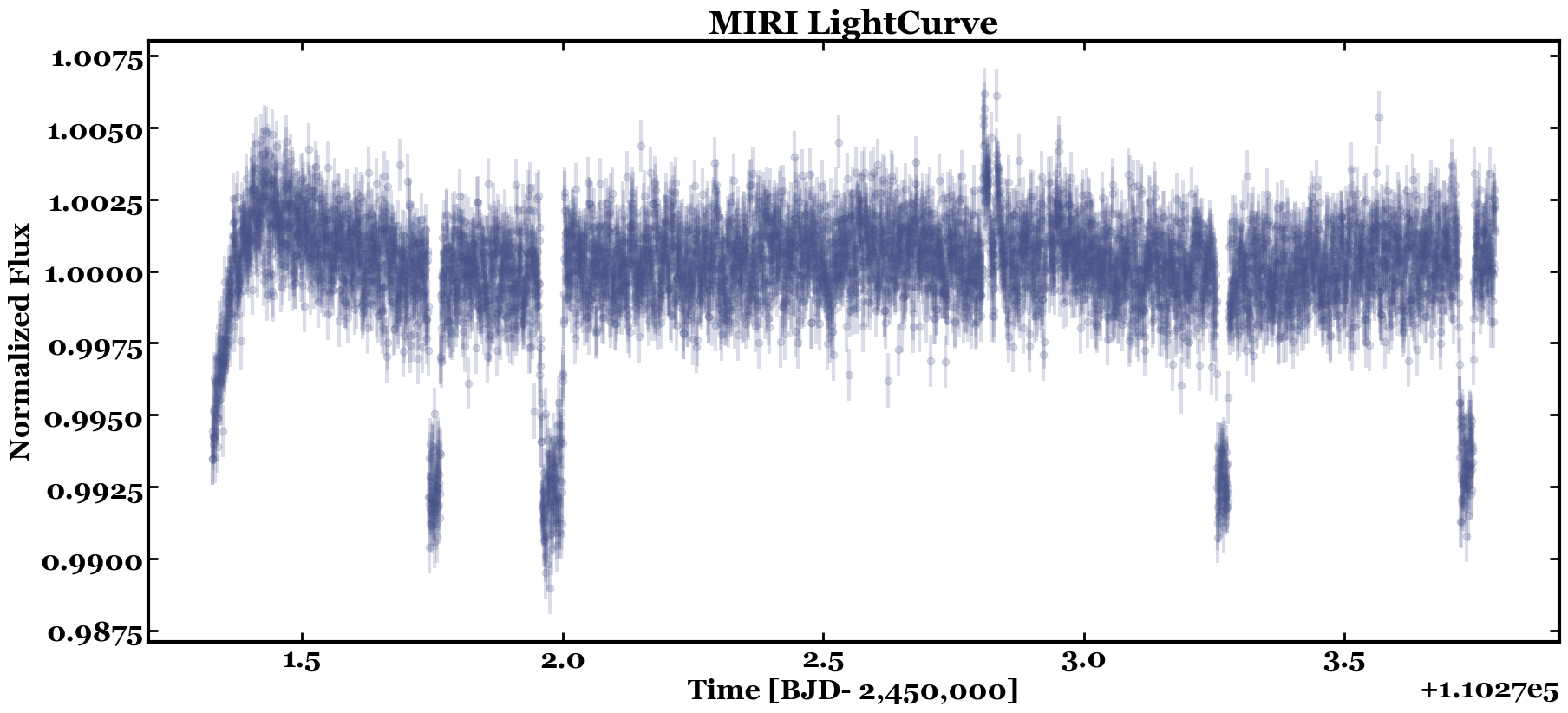}
   \caption{\textbf{Raw light curve from ZH analysis.} Raw data from Stage 3, source radii = 9.0 pixels, background annulus from 16 to 36 pixels, from ZH Analysis}
   \label{fig:ZH-DataReduction}
\end{figure}

\textbf{Data Reduction (Stage 1-3)} We carried out the data reduction and analysis starting from \textit{\_uncal.fits} files.
We used the \texttt{JWST pipeline} (version 1.13.0) and \texttt{Eureka!} (version v0.10-696df7c) pipeline  to include correction for noise appearing around 390 Hz and 10 Hz which was reported to potentially affect the MIRI detectors. The configuration of Stage 1 and Stage 2 are following the default settings except that we turned off the jump correction.  The Stage 2 of \texttt{JWST} also includes converting lightcurves into $MJy.sr^{-1}$. Starting from Stage 3, the pipeline is used for aperture photometry and light curve fitting including phase curve, eclipse model  and transit model. We first selected a window range  of [200, 300] for x and [200, 300] for y that contains the source. We then used a 2D Gaussian model to estimate the position of the centroid and experimented with various source radii. Observations of the combined phase curve of TRAPPIST-1 b and TRAPPIST-1 c lasted about 59 hrs and were taken in two exposures. Between the two observations, we notice a drift of the centroid which could possibly alter the data analysis with different radii used in this stage. As a result, we selected a range of the radii sizes from 5 to 12 pixels, incrementing in a step of 1 pixel.  The ideal size for minimizing the phase curve fitting is a 9-pixel aperture, together with a background annulus ranging from 16 to 36 pixels away from the centroid. It's worth mentioning that the selection of the background annulus had minimal influence on the resulting light curve, indicating its robustness to variations in this parameter. The resulting lightcurve is shown in \ref{fig:ZH-DataReduction}

\textbf{Data Analysis (Stage 4-5)} The stage 4 and stage 5 fitting of the light curves are also done by \texttt{Eureka!} with a slight modification to enable the modeling of double phase curves. In stage 4, the obtained light curves are normalized, and an iterative 4-$\sigma$ outlier clipping process is applied to identify significant outliers using a 10-minute moving median algorithm. Note that the normalized light curves show a large jump at the beginning and two flares around  BJD 2460272.805 and  BJD 2460272.905. We removed those data points before the phase curve fitting process. In the initial double phase curve fitting in stage 5, the forward model of the planet flux is described as follows:
\begin{equation}
F_{pi}=E_i+C_{1i}(\cos^2 ( (\omega_i t - \delta_i)/2)-1)
\end{equation}
where $E_i$ is the occultation depth of the planet and nightside flux can be calculated by $F_{nighside} = E_i - C_{1i}$.  Furthermore, we also introduced an angle $\delta_{i}$ for each planet to explore possible peak offset due to atmosphere circulation. The phase curve, occultations, transits and planet-planet occultations are captured by using the \texttt{starry} package \citeMethods{Luger2019starry,Luger2021starry}.  The total number of free parameters in the analysis is 22, including phase curve parameters and orbital parameters ($r_{pi}$,$P_i$, $\omega_i$, $ecc_i$ and $i_i$), several observational parameters such as limb-darkening coefficients $u_1$, $u_2$ (fitted by the quadratic law) and systematic variables $c_0$ and $c_1$. The priors of orbital parameters for TRAPPIST-1 b and TRAPPIST-1 c are taken from ref. \citeMethods{Agol2021} (details can be found in \ref{tab:priors}). We also took secondary occultation depth of $ 861 \pm 99 $ ppm for TRAPPIST-1 b and $421 \pm 94$ ppm for TRAPPIST-1 c as priors for our analysis taken from  \cite{Greene2023Natur} and \cite{Zieba2023}, respectively. The phase curve parameter $A_b$ for both planets follows the prior of
$A_b \sim \mathcal{N}(0.5,0.1^2)$. In addition, we also added physical constraints for $A_b$ to ensure positive emission for both dayside and nightside. For both planet, the offset angle $\delta_i$ is sampled using $\mathcal{N}(0,45^\circ$) The \texttt{PyMC3} No-U-Turn Sampler(NUTS) is used to sample parameter space and explore the optimal fit.
We ran 6 chains of
6000 tuning steps and 6000 production draws for the retrieval. We used the chi-square to compute the likelihood and used the 16th, 50th, and 84th percentiles from the samples to determine the best-fit values and their associated uncertainties.

\begin{figure}
   \centering
    \includegraphics[width= 0.9\textwidth]{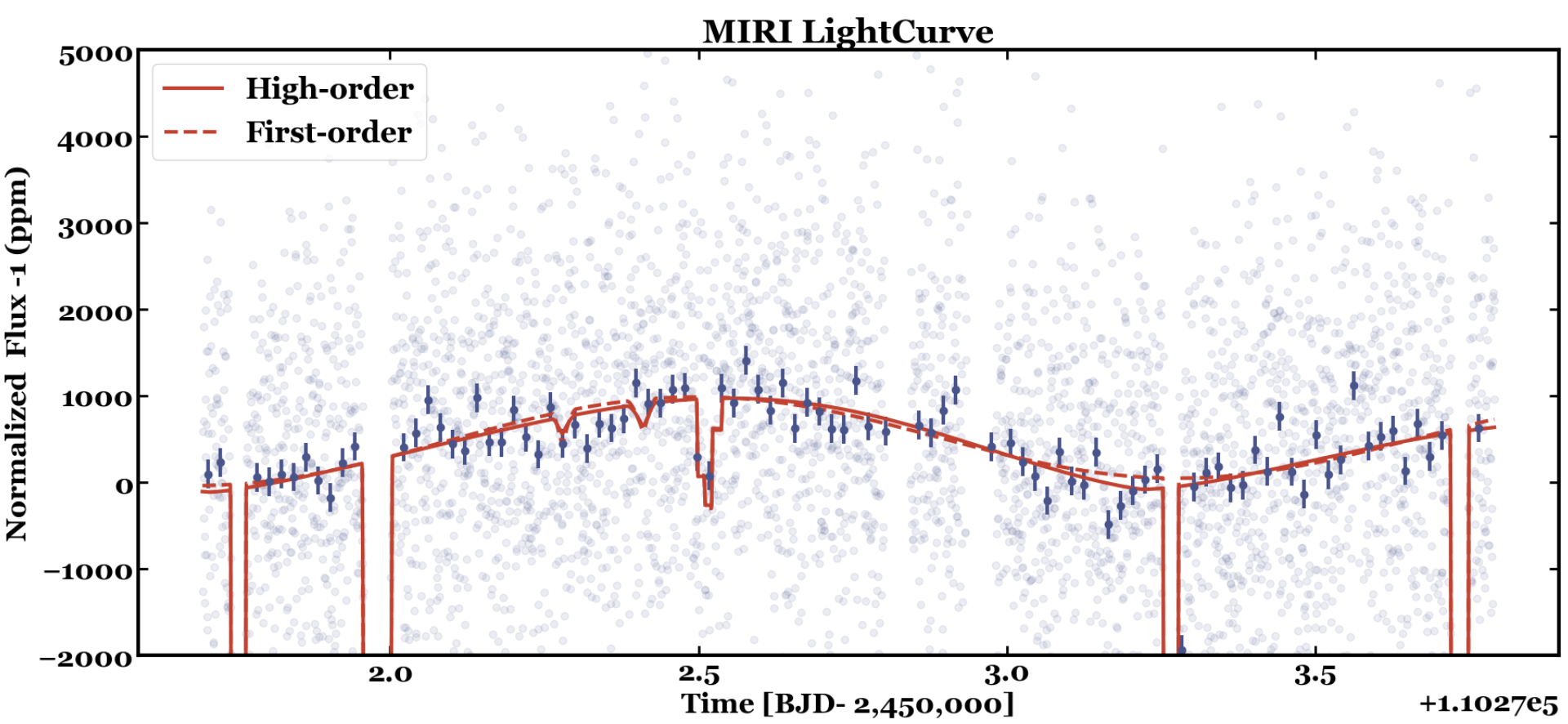}
    \caption{\textbf{ Combined phase curve of TRAPPIST-1 b and TRAPPIST-1 c fitting (High-order : quasi-lambertian, First-order : sinusoidal from ZH data reduction in \ref{fig:ZH-DataReduction}}}
   \label{fig:ZH-PhaseCurve-Fitting}
\end{figure}

The best fit parameters are listed in \ref{Tab:ZH-Summary} and the fitted combined phase curve is shown in \ref{fig:ZH-PhaseCurve-Fitting}. We obtained a dayside flux of $835_{-71}^{+75}$ ppm and nightside flux of $100_{-57}^{+66}$ ppm for TRAPPIST-1 b with a shifted angle of  $-2.02_{-5.70}^{+6.14}$ degree, which indicates an extremely poor heat-redistribution. For TRAPPIST-1 c, the dayside and nightside fluxes are $334_{-78}^{+79}$ ppm and $170_{ -92}^{+31}$ ppm respectively with a shifted angle of  $8.98_{-37.07}^{+22.86}$ degree. The best fit flux and shifted angle point to the possible existence of an atmosphere that re-distribute heat; however, the values are poorly constrained since the contribution of TRAPPIST-1 c's light curve to the total light curve is relative small. This leaves the existence of an atmosphere and the atmospheric composition of TRAPPIST-1 c still unclear based on the double phase curve retrieval but combined with previous eclipse depth of TRAPPIST-1 c, a thin atmosphere is favored (see atmospheric modeling below).

\textbf{Quasi-Lambertian Phase Curve Fitting} Alternatively, the phase variation is found to be consistent with high order cosine functions with a form similar to equation \ref{eqn:2}]:
\begin{equation}
\Phi(\alpha)=\cos ^{\gamma}((\omega_i t - \delta_i)/ 2).
\end{equation}
Here the $\gamma$ is the Quasi-Lambertian parameter from ref. \cite{Agol2007}. In the retrevial, we adopted a prior of $\gamma_i \sim N\left(2.0,2.0^2\right).$ The combined phase curve can be derived similarly as:
\begin{equation}
F_{pi}=E_i+\Sigma_i C_{1i} (\Phi_i - 1)
\end{equation}

The best fit results are listed in \ref{Tab:ZH-Summary} with TRAPPIST-1 b has the order of $\gamma_b = 2.82_{-0.56}^{+0.64}$ and $\gamma_c = 3.92_{-1.88}^{+1.89}$. The newly retrieved dayside flux of TRAPPIST-1 b is $909_{-73}^{+72}$ ppm and the nightside flux is $65_{-34}^{+56}$ ppm with a shifted angle of  $6.71_{-7.14}^{+10.09}$ degree. For TRAPPIST-1 c, the dayside and nightside fluxes are $356_{-82}^{+82}$ ppm and $135_{ -72}^{+57}$ ppm respectively with a shifted angle of  $-8.23_{-20.09}^{+33.27}$ degree.  To calculate the brightness temperature, we use the stellar flux of 2.496 $\pm$ 0.080 mJy at 14.87 $\mu$m (see below) and retrieved parameters from MCMC results. Applying Planck's law, the brightness temperature of TRAPPIST-1 b and TRAPPIST-1 c are shown in \ref{Tab:ZH-Summary}.

\renewcommand{\arraystretch}{1.2}
\begin{table}
\centering
\begin{tabular}{lll}
\hline
\textbf{Free Parameter} & \textbf{PC model: Sinusoidal}                  & \textbf{PC model: Quasi-Lambert}               \\ \hline
\textbf{Trappist-1 b}   &                                                &                                                \\
$A_b$                   & \( 0.880_{-0.098}^{+0.072} \)         & $0.929_{-0.074}^{+0.040}$                 \\
$\delta_b$ (degree)     & \( -2.0_{-5.7}^{+6.1} \)        & $6.7_{-7.1}^{+10.1}$                \\
$\gamma_b$              & -                                              &
$2.82_{-0.55}^{+0.64}$
\\
$F_{b}$ (ppm)           & \( 835_{-71}^{+75} \)     & $909_{-73}^{+72}$             \\
$P_b$ (days)            & \( 1.510884_{-1.51e-07}^{+1.50e-07} \)     & \( 1.510884_{-1.51e-07}^{+1.50-07} \)     \\
$T_{b-day}$ (K)         & 489  $\pm$ 18         & 508 $\pm$ 18        \\
$T_{b-night}$ (K)       & 243 $\pm$ 40        & 220 $\pm$ 35        \\ \hline
\textbf{Trappist-1 c}   &                                                &                                                \\
$A_c$                   & \( 0.49_{-0.30}^{+0.32} \)          & $0.62_{-0.32}^{+0.24}$                 \\
$\delta_c$ (degree)     & \(9.0_{-37}^{+23} \)       & $-8.0_{-20}^{+33}$              \\
$\gamma_c$              & -                                              & $3.9 \pm 1.9$                 \\
$F_{c}$ (ppm)           & \( 334_{-78}^{+79}\)     & $356 \pm 82$             \\
$P_c$ (days)            & \( 2.421793 \pm 2.3e-07 \) & \( 2.421793 \pm 2.3e-07 \) \\
$T_{c-day}$ (K)         & 349 $\pm$ 28     & 356 $\pm$ 28         \\
$T_{c-night}$ (K)       & 283 $\pm$ 43     & 266 $\pm$ 36         \\ \hline
$u_1$                   & \( 0.107 \pm 0.062 \)         & $0.142 \pm 0.062$                 \\
$u_2$                   & \( 0.236 \pm 0.083 \)        & $0.210 \pm 0.082$                 \\
\hline

\end{tabular}
   \caption{\textbf{Summary of MCMC derived physical parameters including brightness temperature from ZH analysis.}} \label{Tab:ZH-Summary}
\end{table}

\newpage
\subsection*{Comparison of the different reductions }

\begin{figure}[ht!]
    \centering
    \includegraphics[width=0.9\textwidth]{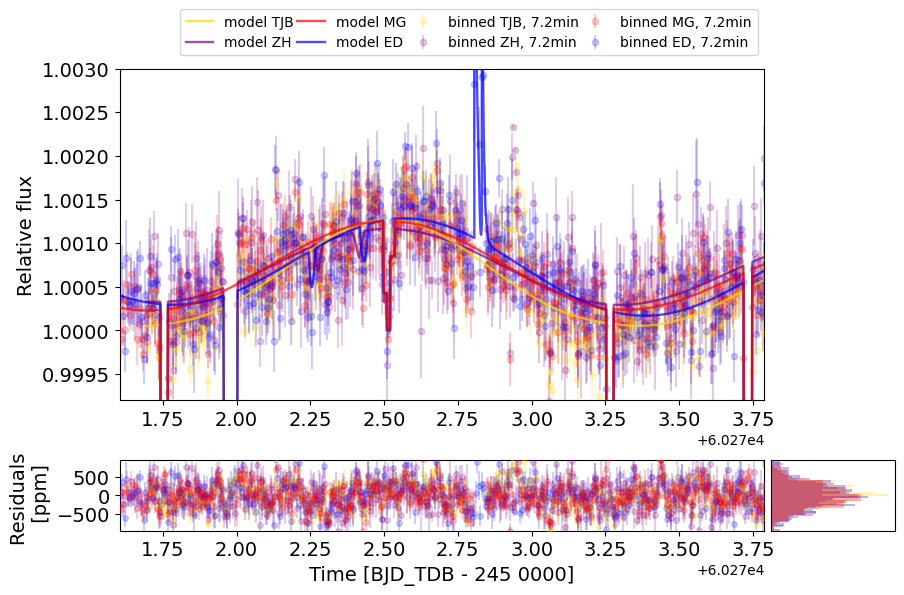}
    \caption{\textbf{Comparison of the light curves obtained from the four different data reductions.} The top panel shows the raw light curves (MG analysis \#1 in orange, ED analysis 7-planets \#1 in green, TJB setup \#1 in red, ZH quasi-Lambert in blue). The data are binned  per 14.4 minutes. Each binned point is the average value of the individual points within the bin, with error bars equal to the standard deviation of the points within the bin. The solid line represent the best fit models from each analysis. The bottom panel shows the residuals for each light curve (raw minus best-fit). }
    \label{fig:comparison}
\end{figure}

\begin{figure}[ht!]
    \centering
    \includegraphics[width=0.95\textwidth]{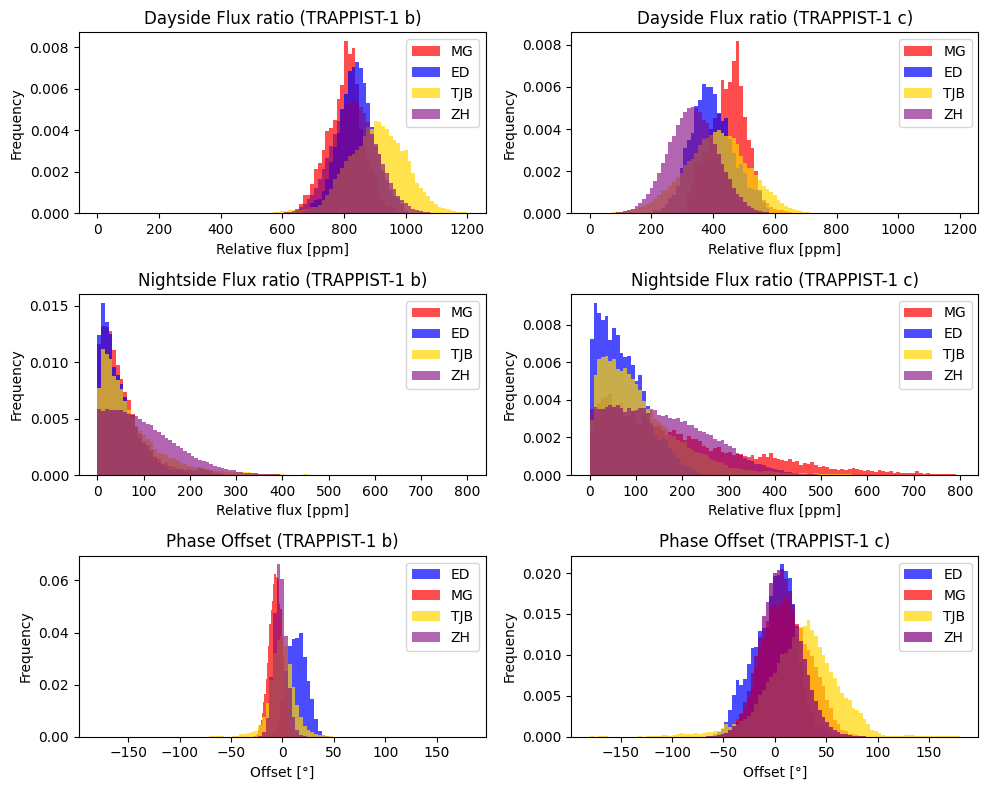}
    \caption{\textbf{Comparison of the posterior distribution functions for the dayside flux ratio $F_{p,day}/F_{\star}$, the nightside flux ratio $F_{p,night}/F_{\star}$ and the phase curve offset $\delta$. }}
    \label{fig:comparisonResiduals}
\end{figure}

We compared the different reductions and found good agreement between the four approaches. \ref{fig:comparison} shows the detrended light curves obtained from four distinct analyses with their best-fit phase curve model. The analyses shown are (1) Analysis \#1 for MG, (2) Analysis 7-planet \#1 for ED, (3) Setup \# 3 for TJB, and (4) the Sinusoidal analysis for ZH. \ref{fig:comparisonResiduals} shows the posterior distribution functions for the dayside flux ratio $F_{p,day}/F_{\star}$, the nightside flux ratio $F_{p,night}/F_{\star}$ and the phase curve offset $\delta$ from the same four analyses. 

We note that all reductions and analyses yield results that are consistent with each other within 1-$\sigma$.

\subsection*{Calibrated stellar flux (MG + TJB)}

As emphasized in ref. \cite{Fauchez2025}, translating our measured planet-to-star flux ratios into absolute planet fluxes and brightness temperatures required measurements of the absolute stellar flux $F_{\ast,15\mu m}$ and the effective wavelength of the observations.
We performed the absolute flux measurement in the calibrated images of programs 1177, 2304, and 3077, using an aperture of 25 pixels, large enough to encompass 99.9\% of the star's PSF \citeMethods{miripsf}. An annulus extending from 30 to 45 pixels from the measured PSF's center was used to measure the background. We performed this flux measurement independently for the ten data sets, discarding images taken during transit or occultation, and the images of program 3077 affected by the initial large systematic effect and by a flare. Combining the ten measurements led to a mean stellar flux $F_{\ast,15\mu m}$ of $2.448 \pm 0.040$ mJy, the error bar being the standard deviation of the ten measurements. We added
quadratically a systematic error of 3\%, which corresponds to the estimated absolute photometric precision of MIRI (P.-O. Lagage, private communication). It resulted in $F_{\ast,15\mu m}$ of $2.448 \pm 0.083$ mJy. We note that measurements' standard deviation is only 1.6\%, nearly twice smaller than the instrument systematic error, which demonstrates the photometric stability of the star at 15 $\mu$m.


We then performed a similar independent measurement of the absolute stellar flux using TJB Stage 1 outputs. After photometric calibration and aperture extraction using the same aperture and annuli as in the first approach, we removed the astrophysical variations in the data caused by the planets (transits, eclipses, and phase variations) using the best-fit model from TJB's analysis described below. The median and standard error were then calculated for each visit, and then a final value was computed using the error-weighted mean of each of the different visits giving a value of $2.54314 \pm 0.00011$ mJy (where the uncertainty is the error-weighted standard error). After adding in quadrature the systematic error of 3\%, this gives an absolute stellar flux of $2.543 \pm 0.076$ mJy. Both measurements are thus consistent with each other. We ultimately adopted the average of their values and errors - $2.496 \pm 0.080$ mJy - in our analyses. For the effective wavelength of the observations, we followed ref. \cite{Zieba2023} and adopted 14.87 $\mu$m.

\subsection*{Day-Night Atmospheric Phase Curve Modeling (VM/AL)}\label{sec:atmos_1D}

To bridge the radiative and compositional versatility advantages of 1D climate-photochemical models, and the spatial and heat redistribution capabilities of 3D models, we use the Virtual Planetary Laboratory (VPL) 1D climate-photochemical model framework \citeMethods{Lincowski2018,Robinson2018} upgraded with a new two-column (day--night) climate mode \citeMethods{Lincowski2023}, to model the thermal phase curves of TRAPPIST-1 b and c.  The new two-column climate mode explicitly calculates day and night hemispheres with layer-by-layer, day-night advective heat transport driven by simplified versions of the 3D primitive equations for atmospheric transport. The geopotential gradient and convection/advection parameters were taken from ref. \citeMethods{Lincowski2023}.  Comparison with ExoCAM 3D GCM results show that this model can adequately reproduce the day-night contrast of 3D GCM phase curves \citeMethods{Lincowski2023}, from radiatively thin (N$_2$/O$_2$) to radiatively thick (CO$_2$) atmospheres, which are comparable to the types of atmospheres modeled here. VPL Climate uses SMART \citeMethods{Meadows1996} with DISORT \citeMethods{Stamnes1988,Stamnes2000} for spectrum-resolving radiative transfer for both climate modeling and planetary spectra generation.  We adjust the stellar flux in our model for spectral simulations to match the observed stellar flux in the MIRI 15$\mu$m bandpass \citeMethods{Lincowski2023}. To generate the thermal phase curves, phase-dependent hemispherically-averaged, day--night weighted thermal flux spectra are calculated for each atmosphere, assuming either a  basaltic surface \citeMethods{Kokaly2017} or fixed albedo, and then convolved with the MIRI 1500W filter bandpass. We then generate thermal phase curves for post- or ongoing-water-loss atmospheres from 0.01 to 10~bar to compare with the observations (\ref{tab:VPL-experiments}).

To identify the most plausible planetary environments that fit the observed phase curves, we compare to the day and night data points assuming both b and c can have atmospheres, and hence a non-zero night-side flux for both planet (using secondary eclipse analyses from ref. \citeMethods{ducrot_2024,Zieba2023} and the nominal MG \#1, this work). We use the following process to quantify the data/model goodness of fits.  We first calculate the $\chi^2$ between data and model. To determine the goodness of fit, we then calculate the confidence interval $Z$ (in units of $\sigma$) as:
\begin{equation}
    Z = \Phi^{-1}(1-p/2),
\end{equation}
where $p$ is the upper-tail $p$-value for the cumulative distribution function for $\chi^2$ given the number of degrees of freedom for comparing our forward models (with no fit parameters). We list these confidence intervals in \ref{tab:VPL-experiments}, along with the best-fit phase offset for each atmosphere model.

For TRAPPIST-1 b, we model two different surfaces. The primary surface was guided by the bare rock models in this work. We halve the albedo of our basalt surface to approximate the bond albedo of Mercury, and similar to the Fe-oxidized surface, the darkest mineral surface from the bare rock models. The modeled N$_2$ atmospheres are for comparison with the GCM results, and use a fixed grey albedo of 0.1. For TRAPPIST-1 b, a tenuous (0.1~bar) pure O$_2$ atmosphere or with 0.1\% H$_2$O, both with photochemically generated ozone, provide the best joint fits (within 1.9$\sigma$) to the data of the models we considered (see \ref{fig:VPL-phasecurves}). Our 0.01~bar N$_2$, 100~ppm CO$_2$ GCM comparison case also fits at 1.8$\sigma$.
A pure 10 bar or greater oxygen atmosphere is ruled out at $>5\sigma$. A 0.1 bar O$_2$ atmosphere with 1\% H$_2$O is disfavored at 2.6$\sigma$.  Adding O$_2$ or 100 ppm or more of CO$_2$ (in 0.1 bar or greater) redistributes too much energy (\ref{tab:VPL-experiments}). Our modeled 1 bar N$_2$ with 1~ppm atmosphere for comparison with the GCM is ruled out at 3.2$\sigma$.

We note in our phase curve modeling that our thin O$_2$ and N$_2$ models with the hottest day sides fail to reach the observed peak flux on the day side. This may have less to do with the atmospheric environments and is more likely to be a limitation of the two-column model, which may fail to resolve the sharp peak substellar temperature with thin atmospheres. This effect can be seen in the comparison of the 0.01 bar N$_2$ atmosphere modeled phase curves modeled by both the day--night model and the GCM (\ref{fig:VPL-phasecurves}).

For TRAPPIST-1 c, the data quality make it impossible at this stage to discriminate between atmosphere fits. None of our atmospheres, which all fit the data point from \citeMethods{Zieba2023}, are ruled out by the phasecurve. The flattest modeled phase curve, the 0.1~bar steam atmosphere, is disfavored at only 2.3$\sigma$.

\begin{figure}
    \centering
    \includegraphics[width=0.95\textwidth]{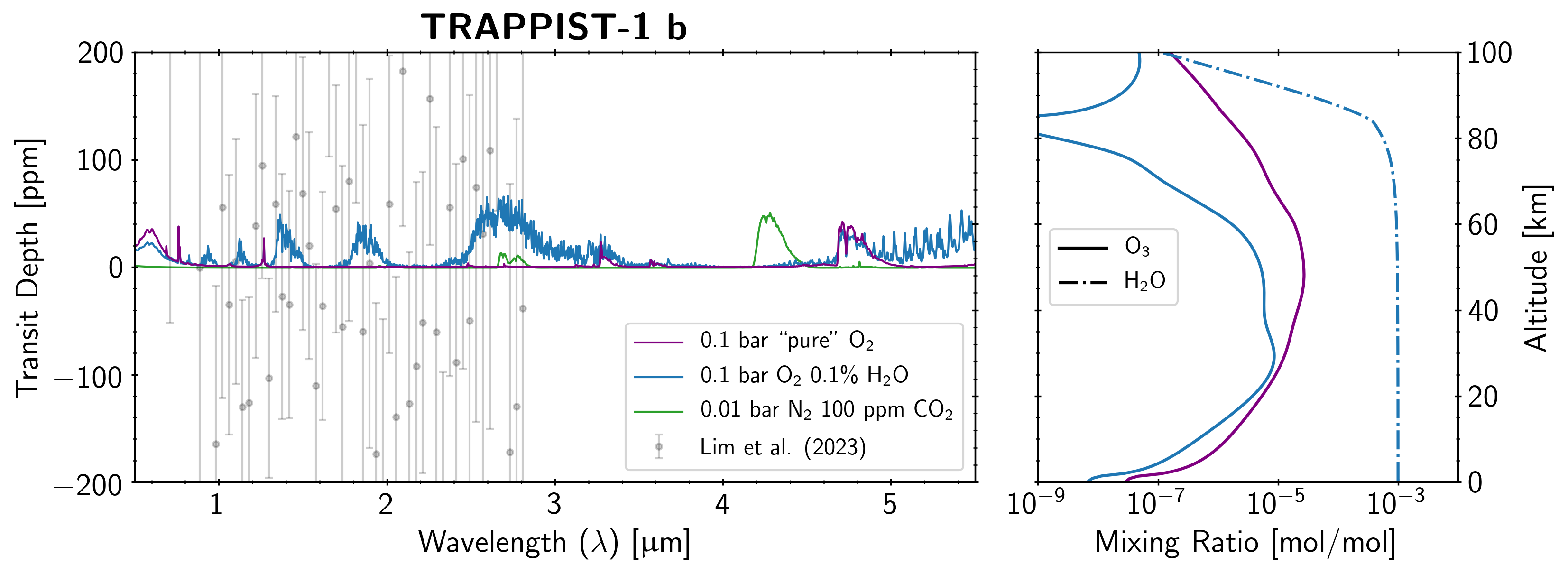}
    \caption{\textbf{Transit spectra and key gas profiles of 1.5D atmospheric models that fit the TRAPPIST-1~b phase curve measurements within 3$\sigma$}. All features shown here are significantly smaller than the 1$\sigma$ ($\sim$100~ppm) error bars from currently observed spectra, due to the stellar source noise issues. Shown are model transit spectra and gas profiles for TRAPPIST-1 b. The zero point for the transit depth is the planetary surface. The spectra demonstrate the potential for water vapor absorption at relative transit depths of 50--60~ppm. TRAPPIST-1~b may also exhibit ozone or CO$_2$  NIR features up to $\sim$50~ppm. These potential features should be taken into account if using either TRAPPIST-1 b in analyses for stellar contamination. \label{vpl_transit}}
\end{figure}

Although several tenuous atmospheres plausibly fit for TRAPPIST-1 b, and denser atmospheres are ruled out, even tenuous atmospheres may still produce relatively large transmission features (50--60~ppm), as shown in \ref{vpl_transit}. Additionally, the compositional constraints on the atmospheres for both planets are not sufficiently conclusive to categorically rule out even larger features in transit transmission observations, which may complicate attempts to use TRAPPIST-1 b to remove stellar contamination from the spectra of other TRAPPIST-1 planets \cite{DeWit:2023}, though work is underway to do so \citeMethods{Rathcke2024}.

\begin{table*}[thb!]
    {\tiny\selectfont
\centering
\begin{tabular}{lccclc}
\toprule
\textbf{Environment} & \textbf{Surface} & \textbf{Surface} & \textbf{Day / Night} & \textbf{Key Gases} & \textbf{Std. dev.} \\
~ & ~ & \textbf{Pressure} & \textbf{ Surf. Temp. [K]} & ~ & \textbf{[$\sigma$]}  \\
\midrule
\multicolumn{2}{l}{\textbf{TRAPPIST-1 b}} & & & & \\
O$_2$ pure & dark basalt & 0.1 bar & 456 / 165 & O$_2$, O$_3$ & 1.4 \\
O$_2$ pure & dark basalt & 1 bar & 461 / 252 & O$_2$, O$_3$ & 1.9 \\
O$_2$ pure & dark basalt & 10 bar & 468 / 340 & O$_2$, O$_3$ & 5.7 \\
O$_2$ outgassing & dark basalt & 0.1 bar & 493 / 288 & O$_2$, 2\% H$_2$O, 100 ppm CO$_2$, O$_3$, CO & 5.8 \\
O$_2$-H$_2$O (0.1\%) & dark basalt & 0.1 bar & 474 / 234 & O$_2$, 0.1\% H$_2$O, O$_3$ & 1.6 \\
O$_2$-H$_2$O (1\%) & dark basalt & 0.1 bar & 483 / 267 & O$_2$, 1\% H$_2$O, O$_3$ & 2.6  \\
N$_2$ & $A=0.1$ & 0.01 bar & 463 / 132 & N$_2$, 100 ppm CO$_2$ & 1.8   \\
N$_2$ & $A=0.1$ & 1 bar & 455 / 245 & N$_2$, 1 ppm CO$_2$ & 3.2  \\
\midrule
\multicolumn{2}{l}{\textbf{TRAPPIST-1 c}} & & & & \\
O$_2$ pure & basalt & 0.1 bar & 395 / 156 & O$_2$, O$_3$ & <1 \\
O$_2$ pure & basalt & 1 bar & 388 / 237 & O$_2$, O$_3$ & <1  \\
O$_2$ pure & basalt & 10 bar & 383 / 300 & O$_2$, O$_3$ & 1.7 \\
O$_2$ outgassing & basalt & 0.1 bar & 410 / 250 & O$_2$, 2\% H$_2$O, 10 ppm CO$_2$, CO, O$_3$ & <1 \\
O$_2$ outgassing & basalt & 0.1 bar & 419 / 235 & O$_2$, 2\% H$_2$O, 100 ppm CO$_2$, CO, O$_3$ & <1 \\
O$_2$-H$_2$O (0.1\%) & basalt & 0.1 bar & 400 / 203 & O$_2$, 0.1\% H$_2$O, O$_3$ & <1 \\
O$_2$-H$_2$O (1\%) & basalt & 0.1 bar & 405 / 236 & O$_2$, 1\% H$_2$O, CO, O$_3$ & <1\\
O$_2$-H$_2$O (10\%) & basalt & 0.1 bar & 433 / 267 & O$_2$, 10\% H$_2$O, 100 ppm CO$_2$, CO, O$_3$ & 1.5\\
Steam & basalt & 0.01 bar & 437 / 277 & H$_2$O, 0.1\% CO$_2$, CO, O$_3$ & <1  \\
Steam & basalt & 0.1 bar & 547 / 433 & H$_2$O, 0.1\% CO$_2$, CO, O$_3$ & 2.3  \\
\bottomrule
\end{tabular}
\caption{\textbf{VPL Climate Day-Night Model Planetary States and Results} Table showing day-night model results for a broad sweep of atmosphere types, which include different surface pressures and key gases. Calculated day/night surface temperatures and standard deviation to combined day/night goodness of fit to \protect\citeMethods{ducrot_2024} and day/night points of nominal analysis MG \#1 are shown.}\label{tab:VPL-experiments}
    \flushleft
    }
\end{table*}

\subsection*{3-D Global Climate Modeling of TRAPPIST-1b and c and associated synthetic observables}\label{sec:atmos_3D}

We performed 3-D Global Climate Model (GCM) simulations of TRAPPIST-1b and c using the state-of-the-art \texttt{Generic PCM} (Generic Planetary Climate Model), extensively described in refs. \citeMethods{Turbet:2018,Turbet:2023} in the context of the TRAPPIST-1 planets. GCM simulations enable resolving the 3-D structure of the atmosphere necessary to capture the day-night temperature contrast and to compute physically-consistent thermal phase curves.

The model includes self-consistent treatment of radiation, convection and clouds. Parameterizations of these processes are detailed in ref. \citeMethods{Turbet:2023} (and references therein). In short, radiation is computed using the correlated-$k$ approach, using opacity tables based on HITRAN and HITEMP; convection is represented through dry and moist adjustment schemes; H$_2$O and CO$_2$ cloud formation is treated using a prognostic scheme, and assumes a fixed amount of cloud condensation nuclei (CCN). TRAPPIST-1b and c are in fact hot enough that water condensation is very limited and therefore that only a tiny amount of clouds form in the simulations.
\\
We performed GCM simulations of TRAPPIST-1b and c  to assess whether phase curves could be used to identify the atmospheric scenarios that are still plausible based on constraints from occultations alone \citeMethods{Greene2023Natur,Zieba2023,ducrot_2024}. For TRAPPIST-1b, this includes thin, residual atmospheres \citeMethods{Ih2023} as well as CO$_2$-rich atmospheres with a hot stratosphere \citeMethods{ducrot_2024}, induced by absorption of high-altitude hazes. For TRAPPIST-1c, we investigated the specific case of a H$_2$O-dominated atmosphere using the simulations of ref. \citeMethods{Turbet:2023}, which was shown to be roughly compatible with occultation measurements in two independent analyses \citeMethods{Lincowski2023,Turbet:2023}. All simulations were performed assuming synchronous rotation. All planetary parameters (masses, radii, insolations) were taken from ref \citeMethods{Agol2021}. For the TRAPPIST-1 stellar spectrum, we used a PHOENIX BT-Settl spectrum as in ref. \citeMethods{Turbet:2018}. The surface albedo was set to 0.1 in all simulations. This choice is motivated by the fact that in most scenarios (except cases with hazes where the emission comes from very high in the atmosphere) a low surface albedo is necessary to match the measured occultations. A much wider range of 3-D GCM simulations will be presented in a follow-up paper (Maurel et al., submitted).
\\
In the hazy TRAPPIST-1b simulations (\textit{Haze high} and \textit{Haze low}), we followed an approach very similar to the one used in ref. \citeMethods{ducrot_2024}, which consisted in adding an extra source term of haze-induced opacity to the (correlated-k) CO$_2$ opacity tables in the GCM: $\sigma=f_\mathrm{haze}~\kappa~(\lambda_0/\lambda)^2~(M_\mathrm{CO_2}/N_a)$, with $\kappa=0.5~\mathrm{cm}^2/g$ and $\lambda_0=1~\mu \mathrm{m}$.
In the \textit{Haze high} case, we used exactly the same parameters as in ref. \citeMethods{ducrot_2024} ($f_\mathrm{haze}=7.0 \times 10^{-4}$), but used a single-scattering albedo of 0.5 by artificially decreasing the incoming stellar flux accordingly. This artificial increase in albedo is necessary -- compared to \citeMethods{ducrot_2024}, which used 1-D atmospheric simulations with higher heat redistribution than predicted by the GCM -- to prevent the dayside emission to be higher than the constraints given from the occultation observations. In the \textit{Haze low} case, we used $f_\mathrm{haze}=3.0 \times 10^{-5}$ and a single-scattering albedo of 0.2. In this case, the amount of hazes is so large that absorption of incoming stellar radiation takes place in the uppermost layers of the atmosphere.

\begin{figure}
    \centering
    \includegraphics[width=0.75\textwidth]{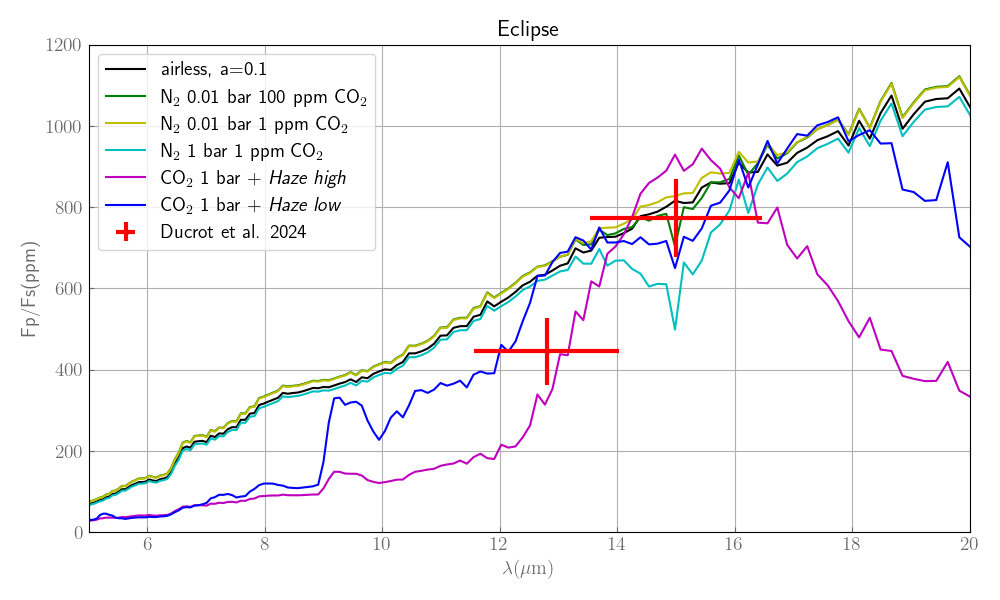}
    \includegraphics[width=0.75\textwidth]{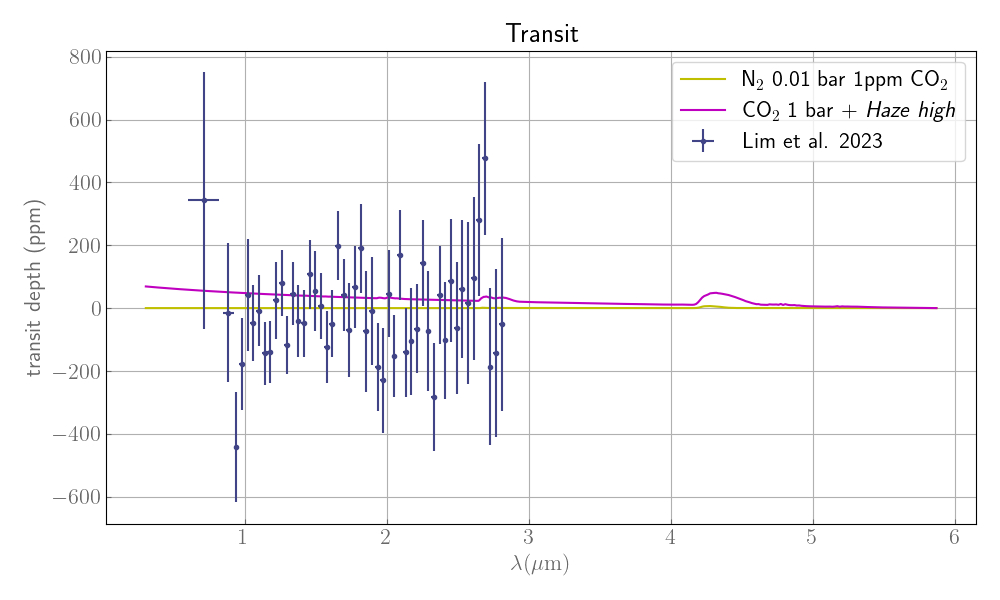}
    \caption{\textbf{ Upper panel : Occultation depth measurements compared to models} of TRAPPIST-1\,b occultation depth spectra computed for five distinct GCM simulations along with a low albedo, airless planet case. Data points (in red) were taken from ref. \cite{Greene2023Natur,ducrot_2024}. The six spectra have been selected here for their good agreement with the observed occultations at 12.8 and 15 $\mu m$. Lower panel : Transit spectra computed for the TRAPPIST-1b atmospheric scenarios fitting all available infrared MIRI data points (0.01~bar N$_2$ with 1~ppm of CO$_2$ and \textit{Haze high} cases). The transit spectrum has been offset by 103~ppm for the \textit{Haze high} case so that the minimum value is zero. For comparison, we added the transmission spectrum from ref. \cite{Lim2023}, not corrected from stellar contamination, with an offset of 7150 ppm.}
    \label{fig:eclipse_spectra_GCM}
\end{figure}

Our strategy was the following: (1) we performed a grid of GCM simulations for both thin atmospheres and CO$_2$-rich atmospheres with high-altitude hazes (for TRAPPIST-1b) and of an H$_2$O-rich atmosphere for TRAPPIST-1c; (2) we computed occultations in the MIRI F1280W and F1500W filters, using the \texttt{Pythmosph3R} 3-D radiative transfer package \citeMethods{Falco:2022}; (3) we selected the GCM simulations that best fit the MIRI observations \citeMethods{Greene2023Natur,Zieba2023} (see \ref{fig:eclipse_spectra_GCM}); (4) we computed thermal phase curves in the MIRI F1500W filter, using again \texttt{Pythmosph3R} (see \ref{fig:VPL-phasecurves}). The sigma values can be found in the legend of \ref{fig:VPL-phasecurves}. However, due to the complexity of the GCM simulations, the sigma values can not be used to compare GCM simulations to 1-D simulations, since they correspond to a very different modeling approach. We refer to Maurel et al., submitted, for more details.

For the post-processing, we used a stellar spectrum from the SPHINX model grid \citeMethods{2023ApJ...944...41I} interpolated to TRAPPIST-1's parameters. This whole spectrum was normalized by the measured flux at 15 $\mu m$. We checked the consistence with the measured flux at 12.8 $\mu m$ from ref. \citeMethods{ducrot_2024}, and found a difference of only 1.3\% between our flux integrated at 12.8 $\mu m$ and the one measured.
Note that sensitivity tests were performed by comparing the outputs of \texttt{Pythmosph3R} with the \texttt{Planetary Spectrum Generator (PSG)} \citeMethods{Villanueva:2018,Villanueva2022} for airless and CO$_2$ dominated scenarios. The two codes show a good agreement, but we note that the amplitude of the synthetized phase curve is very sensitive to the opacity tables used. A more detailed comparison is out of the scope of this paper and will be carried on as part of the MALBEC model intercomparison project \citeMethods{Villanueva:2024}.

\ref{fig:eclipse_spectra_GCM} shows the results of TRAPPIST-1b occultations computed for five different GCM simulations that fit relatively well the measurements of ref. \citeMethods{Greene2023Natur,ducrot_2024}, compared with the flux emitted by a low albedo (0.1), airless planet. At least two types of atmospheric scenarios are compatible with the data. This is the case for (1) thin and/or transparent atmospheres, in agreement with \citeMethods{Ih2023}; (2) CO$_2$-rich atmospheres with a fine-tuned upper-atmosphere haze layer, in agreement with \citeMethods{ducrot_2024}. While in the former cases most of the thermal emission comes from the surface, in the latter cases it comes from the upper-atmosphere layers (see \ref{fig:eclipse_spectra_GCM}). In this case, CO$_2$ is seen in emission due to the strong near-infrared absorption by the haze layer. For this mechanism to work though, the optical properties of the hazes need to be fine-tuned (see above).

\begin{figure}
    \centering
    \includegraphics[width=\textwidth]{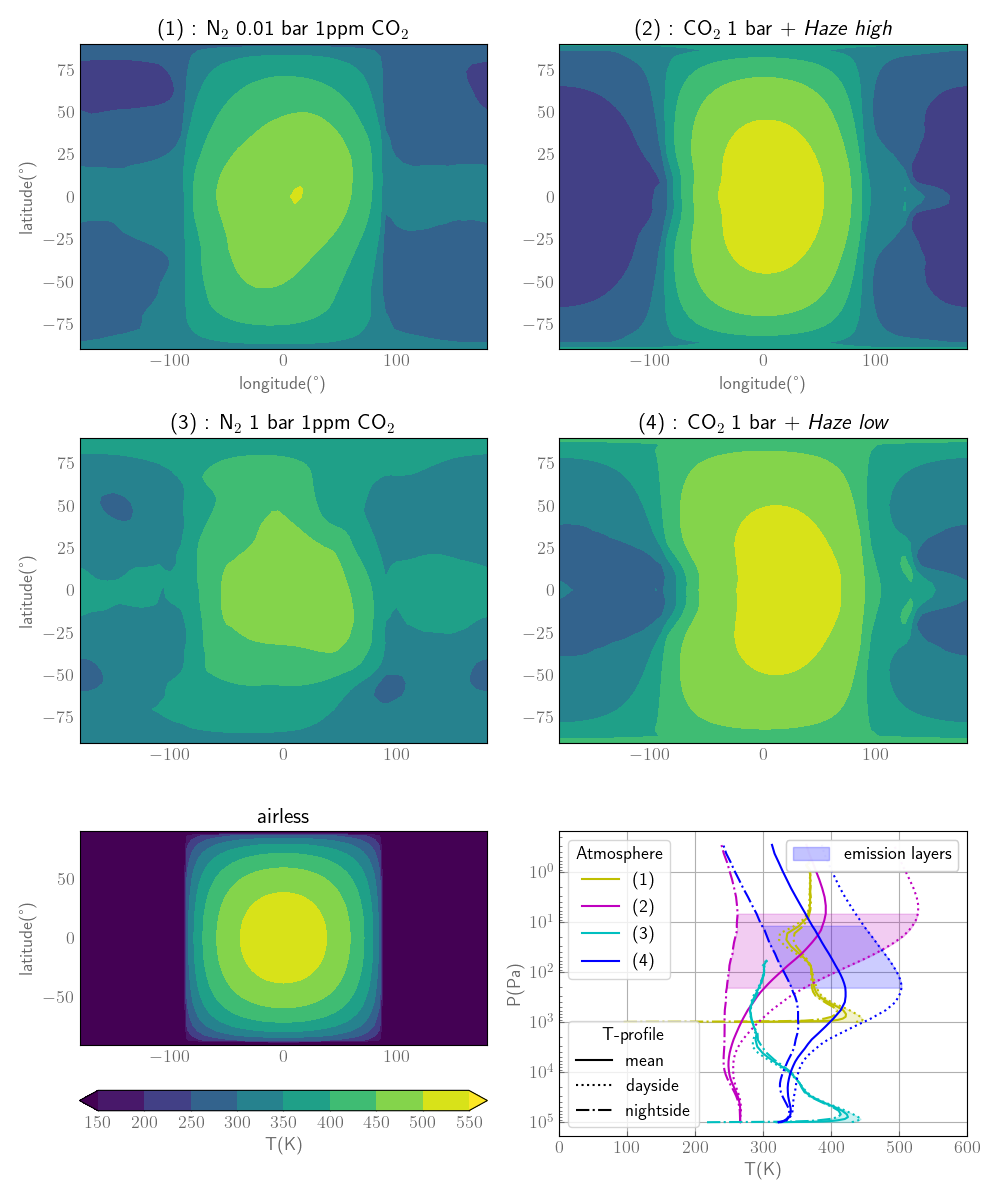}
    \caption{\textbf{Temperature maps computed for four distinct GCM simulations along with a low albedo, airless planet case.} The maps were computed by weighting vertically the 3-D field of atmospheric temperatures by their contribution to the thermal emission at 15 $\mu m$. The figure also shows the vertical temperature profiles for each case, as well as the vertical region of contribution to thermal emission. All five cases have a dayside temperature very close to the brightness temperature measured by the JWST MIRI occultation observations.}
    \label{fig:temperature_maps_GCM}
\end{figure}

\ref{fig:eclipse_spectra_GCM} also shows that all five cases have a dayside temperature (weighted by the 15 $\mu m$ emission contribution function) which is very close to the brightness temperature measured in the JWST MIRI occultation observations. However, they also show that as a result of non-negligible atmospheric heat transport, the nightside temperatures can vary significantly compared to the low-albedo, airless case (see temperature maps in \ref{fig:temperature_maps_GCM}), leaving distinct features on the thermal phase curves (see \ref{fig:VPL-phasecurves}). In the case of thin N$_2$+CO$_2$ atmospheres, a too high N$_2$ partial pressure and/or a too high a CO$_2$ mixing ratio increase heat redistribution, which can be discarded by the phase curve observations (see e.g. the N$_2$ 1bar + 1ppm CO$_2$ case). GCM simulations also show that, in some working cases, CO$_2$ can collapse on the nightside (e.g. the N$_2$ 0.01bar + 100ppm CO$_2$ case, which, after complete condensation of the CO$_2$, ends up on the N$_2$ 0.01bar + 1ppm CO$_2$ case). In the case of thick CO$_2$ atmospheres with hazes, GCM simulations show that the amount of hazes and their optical properties need to be fine-tuned to fit both the occultations and the phase curve. In particular, as soon as the hazes absorb too low in the atmosphere, strong super-rotating winds in the atmosphere are enough to flatten the phase curve and generate an offset (see e.g., the \textit{Haze low} case). For this scenario to work (see e.g., the \textit{Haze high} case), the atmosphere must emit and therefore the hazes absorb high enough that the radiative timescale is much less than the advective timescale (by winds), thus limiting heat redistribution. The difference in altitude at which thermal emission occurs (\textit{Haze high} vs \textit{Haze low} cases) is clearly visible on the pressure-temperature profiles of \ref{fig:temperature_maps_GCM}. Consequences on the difference in nightside emission temperatures are also clearly visible on the associated temperature maps.

We did a similar exercise for TRAPPIST-1c, but focusing now on the H$_2$O-dominated atmospheric scenarios using the GCM simulations of ref. \citeMethods{Turbet:2023}. The case of a H$_2$O-dominated atmosphere is particularly interesting, as it had been proposed as being compatible within less than 2$\sigma$ to occultations \citeMethods{Lincowski2023,Turbet:2023}. GCM simulations show that H$_2$O-dominated atmospheres are in fact very effective at flattening the phase curve and generating a strong offset (see \ref{fig:VPL-phasecurves}, bottom panel). Phase-curve observations therefore discard the cases of H$_2$O-dominated atmospheres. Given the past and present insolation on TRAPPIST-1c, which places it outside the water condensation zone \citeMethods{Turbet:2023}, this result suggests that TRAPPIST-1c, like TRAPPIST-1b, started out with a relatively limited amount of water \citeMethods{Bolmont2017}.

Finally, we computed transit spectra of TRAPPIST-1b with \texttt{Pythmosph3R} employing the same GCM simulations. The aim is to assess (1) the amplitude of transit spectra features produced by the different types of atmospheres which are still compatible with the different MIRI datasets (0.01~bar N$_2$ with 100~ppm of CO$_2$ and \textit{Haze high} cases), (2) the impact of using planets b to remove stellar contamination from the transit spectra of other, outer planets \citeMethods{DeWit:2023, Rathcke2024}. This is an essential prerequisite for the success of the JWST GO 6456 program \citeMethods{Allen:2024}, whose aim is to use consecutive transits of TRAPPIST-1b and e to correct for the stellar contamination of planet e using planet b. This method assumes, however, that planet b has no atmosphere, otherwise it risks introducing a new source of contamination. We find that the amplitude of transit spectra (see \ref{fig:eclipse_spectra_GCM}) is low but non-negligible, in particular for the CO$_2$-rich hazy atmospheric scenario. This is because the presence of hazes produces a significant slope in the spectrum (in the visible and near-infrared wavelengths covered by NIRSpec and NIRISS instruments), and the heating produced by this haze increases the scale height of the atmosphere, which makes CO$_2$ feature prominent -- although not as strong as the clear-sky case, due to blanketing by hazes \citeMethods{Lustig-Yaeger:2019} -- in particular at 4.3~$\mu m$. Thin atmosphere scenarios produce absorption features with low amplitudes as expected \citeMethods{Morley:2017}. For the TRAPPIST-1b transit spectra we have computed, in all cases the amplitude of the molecular features is well below the precision of the JWST NIRISS observations of ref. \citeMethods{Lim2023} (see \ref{fig:eclipse_spectra_GCM}).

\subsection*{Tidal heating and constraints from observations}
\label{subsec_tidal_heating}

Here we present potential observational constraints on the rotation and the obliquity states of the planet~b considering the effect of tidal heating.
Deviations from a circular orbit, a synchronized rotation and a null obliquity result in tidal dissipation.
According to posterior distributions, the maximum nightside fluxes at $2\sigma$ and $3\sigma$ are $196$~ppm and $270$~ppm, respectively, which correspond to brightness temperatures of $291$K and $322$K, respectively.
These upper limits on the nightside temperature allows to put an upper limit on the actual tidal heating and therefore to constrain the domain of possible rotation rate - obliquity - the eccentricity being constrained independently by transit and eclipse timing/durations.
The tidal heat flux also depends, however, on the internal structure of the planet (Bolmont et al. in prep). 

Usually, the computation of tidal heating uses a simple prescription of tides, based on an equilibrium model, such as the constant time lag model (i.e. CTL, \citeMethods{Goldreich1966, Mignard_1979}).
But such a model does not take into account the complex response of a rocky planet to the tidal forcing \citeMethods{MakarovEfroimsky_2013}.
Here, we use a frequency-dependent tidal response following the method of refs. \citeMethods{Tobie_2005, Tobie_2019}, computed with an Andrade rheology \citeMethods{Andrade_1910}.
The tidal heating is computed using the method of ref. \citeMethods{Levrard_2008} (Eq~4.)
\begin{equation}
    \dot{E}_\mathrm{tidal} = -\frac{\mathrm{d}E}{\mathrm{d}t} = -C\Omega \times \frac{\mathrm{d}\Omega}{\mathrm{d}t} - \frac{GM_pM_\star}{2a^2}\times\frac{\mathrm{d}a}{\mathrm{d}t},
\end{equation}
where $E$ is the total energy of the system, $M_p$ and $M_\star$ the masses of the planet and star respectively, $\Omega$ the rotation rate of the planet, $C$ its moment of inertia and $a$ its semi-major axis.
We use the equations for the derivative of $a$ and of $\Omega$ of ref. \citeMethods{Boue_Efroimsky_2019} (Eq.~116 and 123).

We calculate the dissipation assuming a multi-layered internal structure, based on masses and radii from ref. \citeMethods{Agol2021}, with a silicate mantle and a liquid metallic core.
The internal structures are computed using the \texttt{BurnMan} code \citeMethods{Cottaar_2014, Myhill_2021}, which enables us to explore the effects of degeneracy in possible planetary structures by testing three different scenarios (Bolmont et al., in prep).
We consider three different core sizes by varying their composition, in percentage of iron, silicium and sulfur.
The smallest core we consider has an Earth-like composition, comprising approximately 57\% of the planet's total radius.
The biggest core we consider has a much lighter composition, comprising approximately 84\% of the planet's total radius.
The different relative core sizes considered are listed in \ref{tab:relative_core_size}.
\begin{table}[h]
    \centering
    \caption{Relative size of the core to the total radius of each planet (\%)}
    \label{tab:relative_core_size}
    \begin{tabular}{|c|c|c|c|}
        \hline
        Planet & Big core & Intermediate core & Small core \\
        \hline
        T1-b & 84.3 & 70.3 & 57.0 \\
        T1-c & 85.7 & 71.6 & 57.8 \\
        \hline
    \end{tabular}
\end{table}
The mantle is assumed to be silicate with pyrolitic composition.
The bulk dissipation is neglected, assuming only the dissipation associated with shear deformation \citeMethods{Tobie_2005,Kervazo_2021}.
As shown below, a bigger core corresponds to a higher dissipation than a smaller core.
It is important to keep in mind that the retroaction of the tidal heating on the internal structure is not taken into account here.

\begin{figure}[ht!]
   \centering
   \includegraphics[width=\textwidth]{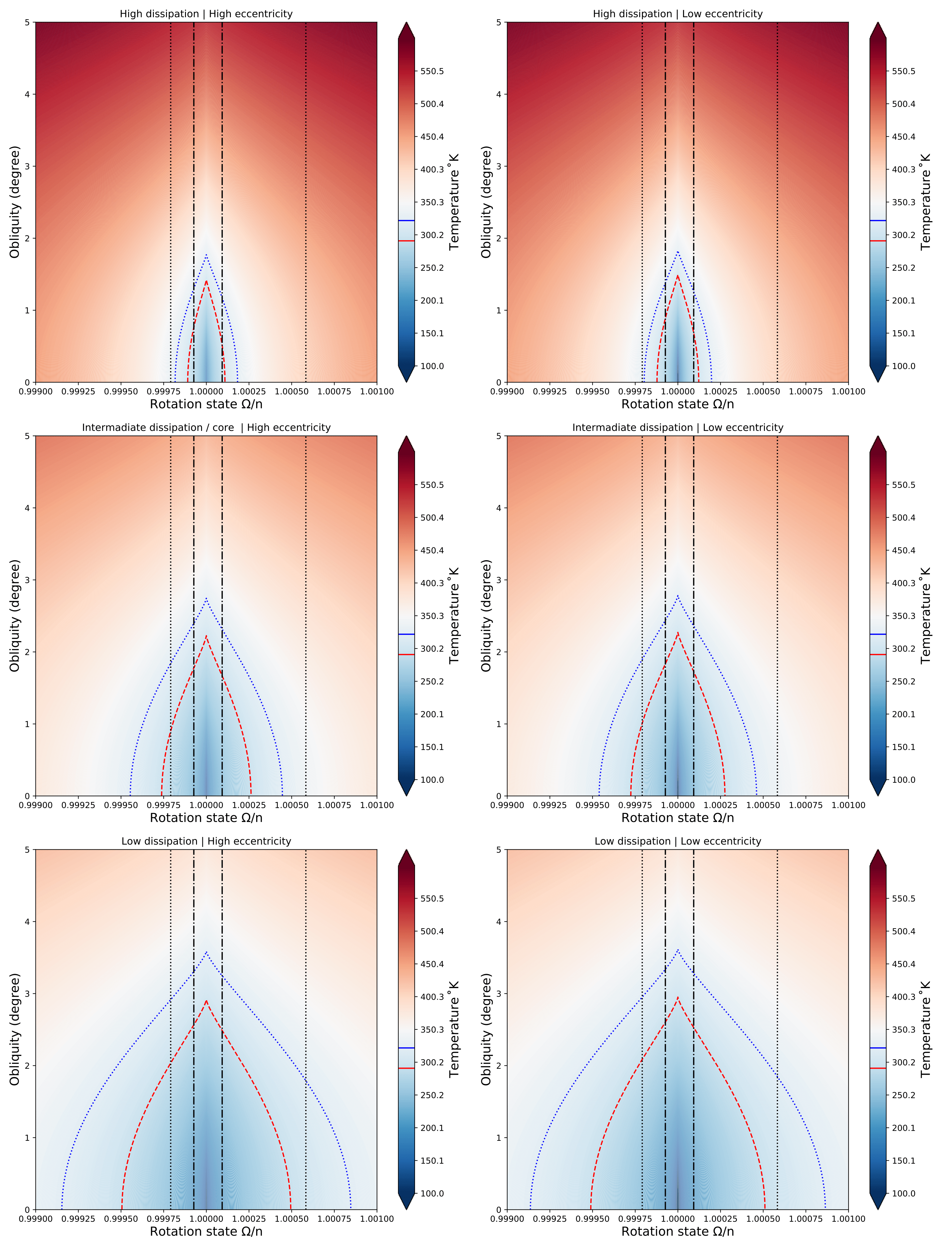}
   \caption{\textbf{Tidal temperature (in K) for the planet~b. }
   The red and blue curves represent the 196 and 270~ppm maximum emission from the night side of the planet, i.e. $291$K and $322$K. The black dash-dotted lines correspond to the maximum and minimum variations of the spin rate from the synchronization state $\Omega/n=1.0$ from ref. \protect\citeMethods{Revol_2024}. The dash-dotted lines represent the range within which 70\% of the spin variation occurs.}
   \label{fig:tidal_temperature}
\end{figure}

\ref{fig:tidal_temperature} shows color maps of the tidal temperature (given by $(\Phi_{\rm tides}/\sigma)^{1/4}$, where $\Phi_{\rm tides}$ is the tidal heat flux and $\sigma$ the Boltzmann constant) as a function of the rotational and obliquity state, for the three internal structures, that correspond to high to low dissipation cases, and for the upper and lower limit in eccentricities given in \ref{tab:results}.
The maximum night side temperatures measured at $291$K and $322$K constrain the obliquity to be up to $2^{\circ}$ for the high dissipation case and to $3.5^{\circ}$ for the low dissipation case.
The rotation state can be constrained within about 0.02\% and 0.06\% of the orbital period, i.e. $0.99979<\Omega/n<1.00058$.

\begin{figure}[ht!]
   \centering
   \includegraphics[width=0.5\textwidth]{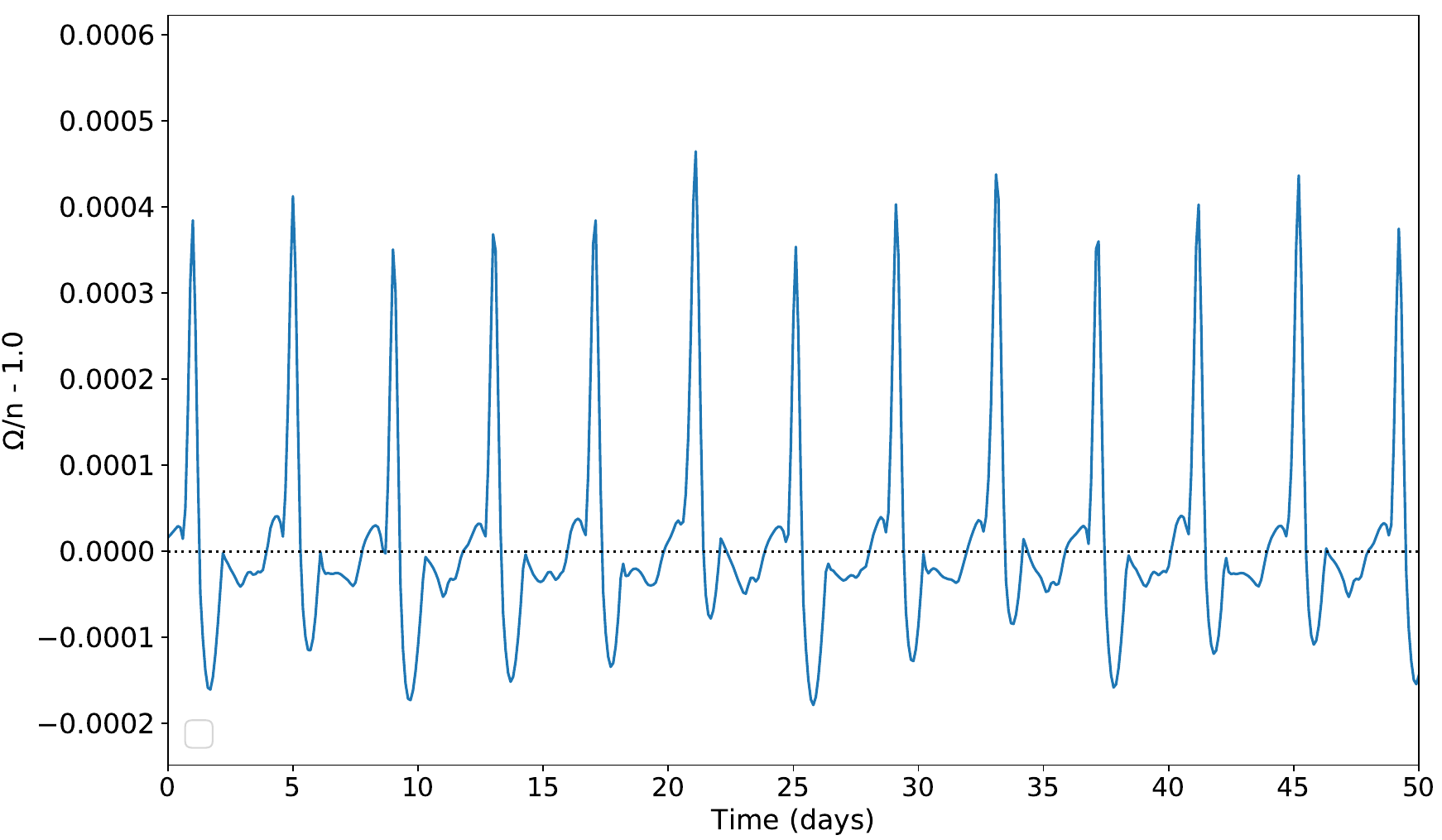}
   \caption{\textbf{Deviation of the spin state of planet~b from the synchronization} in $\Omega/n -1$, with $\Omega$ the spin rate, $n$ the mean motion. The horizontal line at $\Omega/n -1 = 0$ represents the perfect synchronization state. The simulation was made with the N-body code Posidonius \protect\citeMethods{Blanco_Cuaresma_2017,Bolmont2020b,Revol_2024}.}
   \label{fig:spin_variation}
\end{figure}

Planet-planet interactions induce short-time scale variations in the mean motion of each planet.
For planet~b, the mean motion variations are strong enough to take the spin out of the synchronization, i.e., the spin rate does not stay strictly equal to the mean motion \citeMethods{Revol_2024}.
Ref. \citeMethods{Revol_2024} used the N-body code \texttt{osidonius} to compute the spin of the TRAPPIST-1 planet using a tidal model which allows to correctly descrive the response of rocky planets.
They found that the spin oscillates around the synchronization state $\Omega/n=1.0$, see \ref{fig:spin_variation}.
The maximum and minimum deviation of that spin is represented in \ref{fig:tidal_temperature} (vertical black dotted lines).
The vertical black dashed-dotted lines correspond to the range within which the spin spends 70\% of the simulation time.
It is interesting to note that the limits of the spin oscillations are compatible with the observational constraints we get.

We compare tidal temperatures with nightside brightness temperatures, acknowledging that tidal temperatures may differ from nightside values in cases of non-purely synchronous rotation.
For such rotations, planetary inertia must be considered to accurately determine nightside temperatures.
However, in the narrow range of $\Omega/n$ (0.999 to 1.001) studied here, the departure from synchronization is negligible, allowing tidal temperatures to reliably approximate nightside brightness temperatures.

\clearpage

\subsection*{Bare rock simulations for TRAPPIST-1\,b}

Based on the phase curve of TRAPPIST-1\,b, we performed simulations assuming the planet is an airless bare rock. We use the model developed in ref. \cite{Lyu2024}, which simulates the 3D temperature profile (latitude, longitude, and depth) of an airless planet. The model takes into account instellation, reflection, thermal emission, and heat conduction in the subsurface. For the stellar spectrum we assume a Phoenix stellar model. For the planet's surface we assume it consists of regolith, like the Moon and Mercury. We then model the regolith's bi-directional reflectance and emission following \citeMethods{hapke2012theory}; this approach is motivated by laboratory measurements and solar system observations. We neglect horizontal heat conduction since heat redistribution via conduction is too small to influence a planet's phase curve \citeMethods{seager2009method}. In addition we assume both planets are in 1:1 synchronous rotation and ignore potential tidal heating (see \ref{subsec_tidal_heating}).

First, we simulate different geologically fresh surfaces. The surfaces are based on reflectance spectroscopy measurements of different materials and were compiled in ref. \citeMethods{hu2012}. For each material, we first find the planet's temperature distribution in local radiative equilibrium. We then compute phase curves in the MIRI-F1500W bandpass plus secondary eclipse spectra. Note that, since initial submission of this manuscript, other databases of surface materials have been published \citeMethods{paragas2025,first2025}. For ease of comparing our results to previous models for TRAPPIST-1\,b, here we only consider the materials from ref.~\citeMethods{hu2012}.

\begin{figure}
    \begin{minipage}[c]{0.5\textwidth}
    \includegraphics[width=\textwidth]{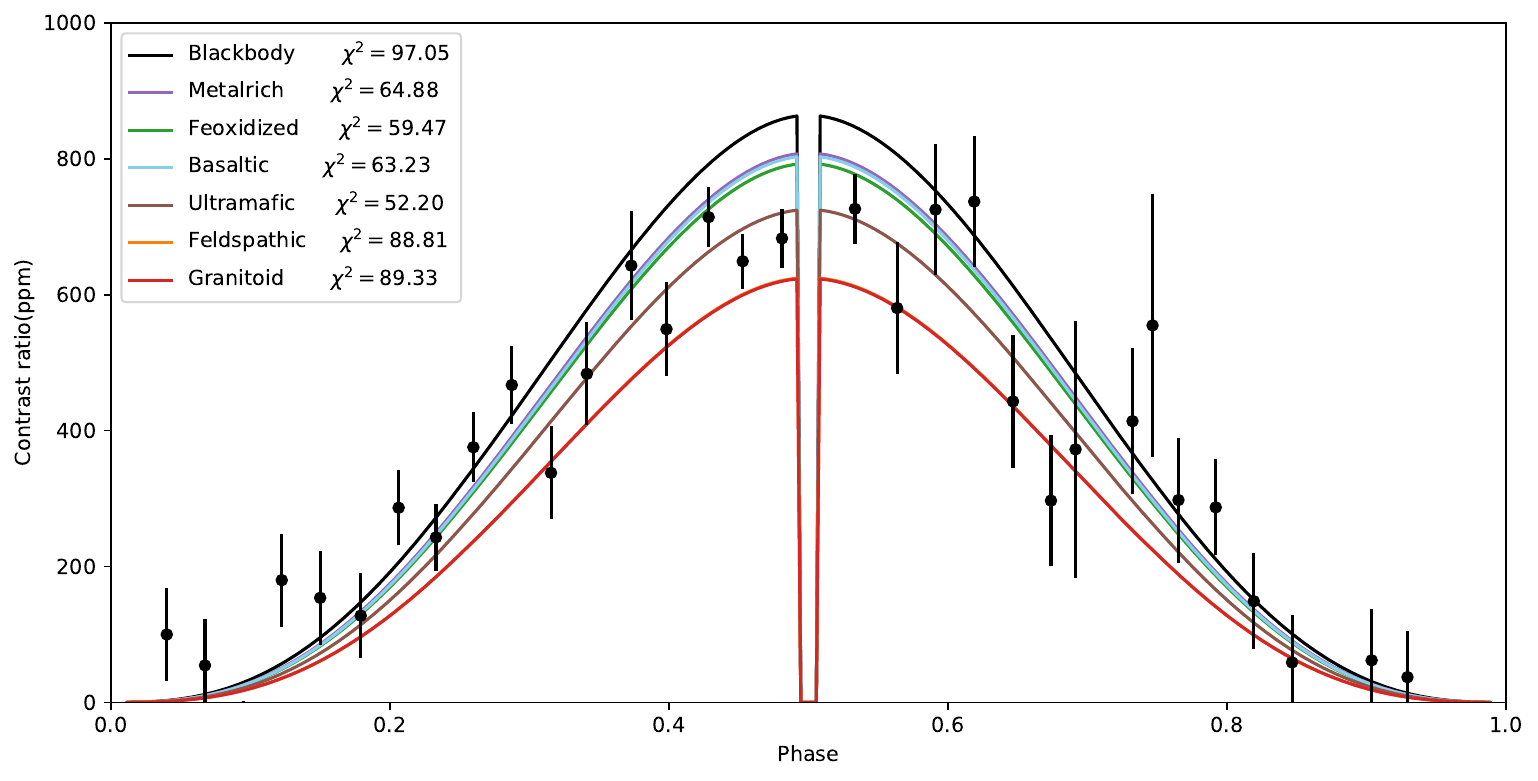}
    \end{minipage}
    \begin{minipage}[c]{0.5\textwidth}
    \includegraphics[width=\textwidth]{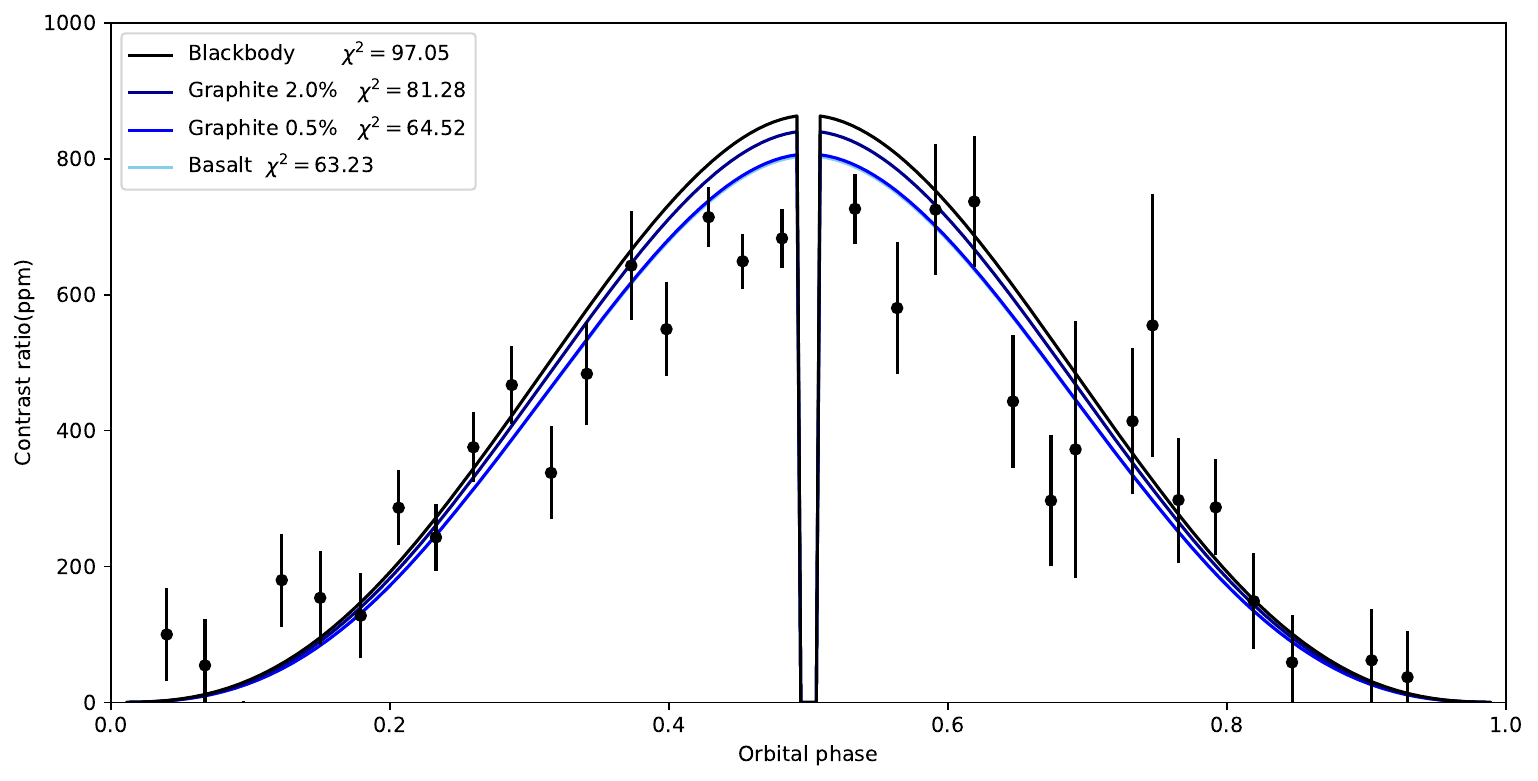}
    \end{minipage}
    \caption{\textbf{TRAPPIST-1\,b's phase curve for different geologically fresh surface materials (left) and a space weathered basalt surface (right) compared to data from MG Analysis \#2.} Left: Models for the planet's dayside emission spectrum based on geologically fresh materials (curves), compared to the 15$\mu$m phase curve from this work (black point). An ultramafic is preferred over all other fresh materials shown here. Right: Space weathering reduces the planet's albedo at short wavelengths and increases the phase curve amplitude at 15 $\mu m$.}
    \label{X.L bare rock phase}
\end{figure}

The phase curves are shown in the left panel of \ref{X.L bare rock phase}. The data are best fit by an ultramafic; all other surfaces considered here produce a higher $\chi^2$ than an ultramafic. Even the second best-fitting fresh model, a Fe-oxidized surface made of hematite (Fe$2$O$_3$), produces a significantly worse fit than an ultramafic ($\chi^2$=52.20 for an ultramafic versus $\chi^2$=59.47 for a Fe-oxidized surface).
Assigning confidence intervals based on $\Delta \chi^2$ relative to the best-fit model \citeMethods{avni1976}, suggests that any model with $\Delta \chi^2 > 6.63$ relative to the ultramafic can be ruled out at 3$\sigma$ confidence. Applying this criterion to TRAPPIST-1\,b's phase curve, we find that an ultramafic is preferred over all geologically fresh surfaces from ref.~\citeMethods{hu2012} at more than 3$\sigma$.

\begin{figure}[ht!]
    \begin{minipage}[c]{0.5\textwidth}
    \centering
    \includegraphics[width=\textwidth]{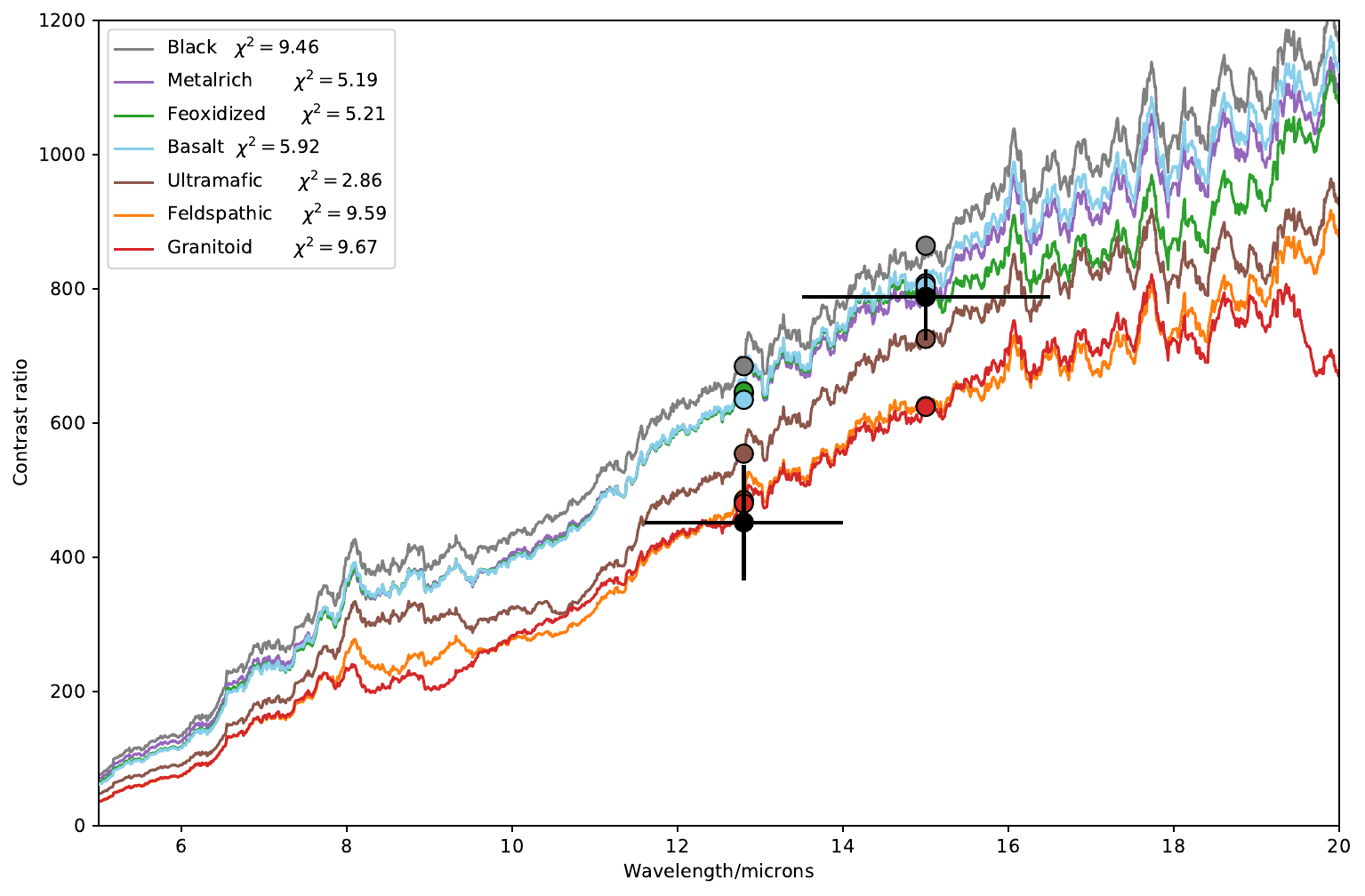}
    \end{minipage}
    \begin{minipage}[c]{0.5\textwidth}
    \centering
    \includegraphics[width=\textwidth]{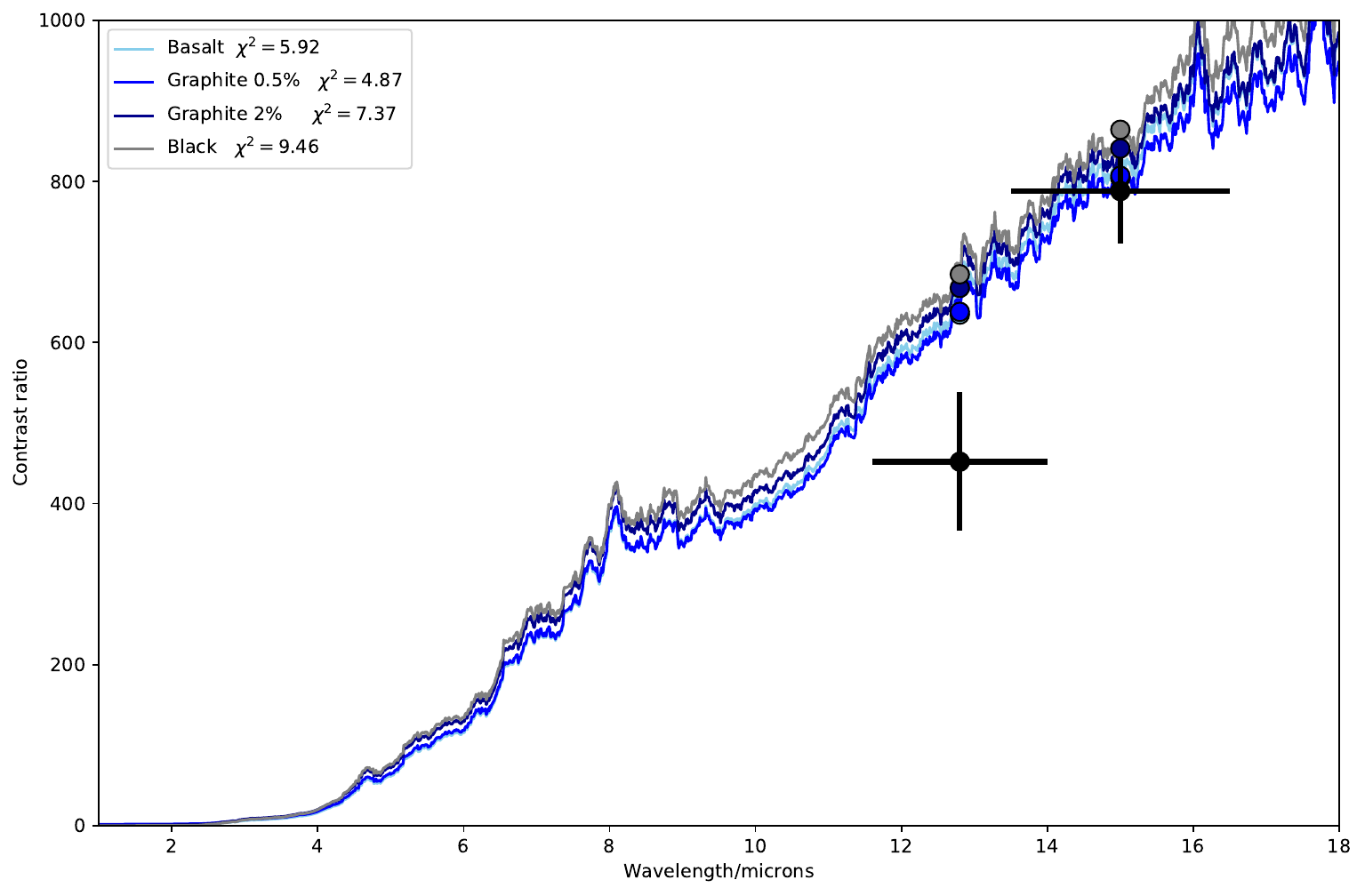}
    \end{minipage}
    \caption{\textbf{TRAPPIST-1\,b's dayside spectrum for different geologically fresh surface materials (left) and a space weathered basalt surface (right).} Left: Models for the planet's dayside emission spectrum based on geologically fresh materials (curves), compared to the contrast ratio at 15$\mu$m from this work (black point) and the reported contrast ratio at 12.8$\mu$m (from ref. \protect\citeMethods{ducrot_2024}). Right: Space weathering darkens the surface and could improve the match to the observed flux at 15$\mu$m. Progressively darker curves correspond to increasing space weathering via the addition of graphite.}
    \label{X.L spectrum}
\end{figure}

In contrast to the MIRI phase curve, which strongly indicates the planet's surface should be dark, we find that the secondary eclipse spectrum provides a much weaker constraint. The left panel of \ref{X.L spectrum} compares spectra of geologically fresh surfaces against the observed eclipse depth at 15$\mu$m (from Analysis \#2; see main text) and the planet's published eclipse depth at 12.8$\mu$m from ref.~\citeMethods{ducrot_2024}. We find that the observed secondary eclipse spectrum is best fit by an ultramafic surface ($\chi^2$=2.86), similarly to \citeMethods{ducrot_2024}. This result corresponds to the phase curve fitting. However, most surfaces and even a blackbody match the observed eclipse spectrum within 3$\sigma$ ($\Delta\chi^2<6.63$ relative to best-fit). Based solely on the eclipse measurements, the only fresh surfaces that are ruled out at more than 3$\sigma$ relative to the best-fitting surface are those with very high albedo, namely feldspathic and granitoid surfaces. This highlights the valuable additional constraints provided by the phase curve.

Taking the phase curve at face value, the question is: what physical process could explain a low albedo for TRAPPIST-1\,b? To address this question we simulate the potential impact of space weathering. In the Solar System, albedos and spectra of atmosphere-less bodies are modified and typically darkened by a range of processes called space weathering \citeMethods{pieters2016space}. Here we investigate the impact of space weathering using the same approach that was previously applied to LHS-3844 b \cite{Lyu2024}. For simplicity we assume the planet's bulk surface consists of basalt. This assumption is reasonable, as basalt is a wide-spread material in the solar system. We then simulate the effects of space weathering by adding increasing amounts of graphite, which has been previously proposed to explain the low albedo of Mercury \citeMethods{syal2015}. Note that we are not trying to constrain the exact space weathering mechanism, and space weathering via the formation of iron nanoparticles could similarly produce a low surface albedo \citeMethods{Lyu2024}.

The results are shown in the right panels of \ref{X.L bare rock phase} and \ref{X.L spectrum}. We find that moderate space weathering weakens the fit to the observed phase curve ($\chi^2$=81.28 for strongly weathered basalt versus $\chi^2$=63.23 for fresh basalt), plus increasing the mismatch to the eclipse spectrum ($\chi^2$=7.37 versus $\chi^2$=5.92). With 2\% graphite by volume added, TRAPPIST-1\,b's albedo significantly decreases, and the surface becomes increasingly indistinguishable from a blackbody. For comparison, the 2\% graphite considered here is less than the 3-6\% invoked to explain Mercury's low surface albedo \citeMethods{syal2015}.


Assuming the planet is airless, consistent with previous atmospheric analyzes, we conclude that, based on the JWST 15~$\mu$m phase curve and the materials considered here, TRAPPIST-1~b’s surface is most likely to be fresh ultramafic. However, current data are still insufficient to place strong constraints on TRAPPIST-1\,b's bulk composition if taking account of 12.8~$\mu$m observation and potential space weathering effect. Although we only simulated space weathering for basalt, space weathering would similarly darken other materials. Taking into account the potential degeneracy due to space weathering, feldspathic and granitoid might thus produce a similarly good match to the observed phase curve amplitude.



\clearpage

\subsection*{Flare characterization }\label{sec:flares}

\begin{figure}[ht!]
    \centering
    \includegraphics[width=0.8\textwidth]{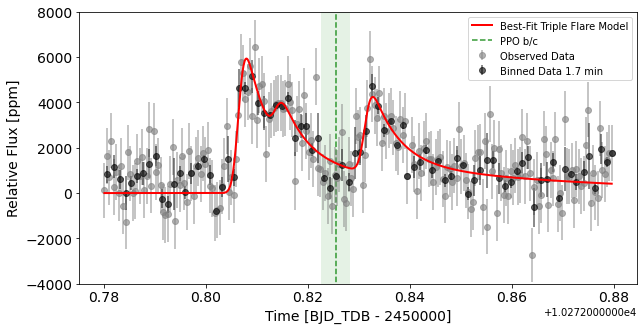}
    \caption{\textbf{Best fit triple-flare event caught during the observations.} The three flares are modeled using eq. \ref{eq:flaremodel}. We note that a planet-planet occultation (PPO) of planet b by planet c is also supposed to happen in between the two flares according to the parameters and ephemeris from ref.~\protect\citeMethods{Agol2021}.}
    \label{fig:flare_fit}
\end{figure}

During the $\simeq60$hours of observations of TRAPPIST-1 we observed a clear flares. In this section we discuss how these flares can be modeled and estimate the associated energy to compare it to existing studies on TRAPPIST-1 \citeMethods{Ducrot2020, Howard2023, Vida_2017}. The flaring event and the best-fit model from this analysis is shown in \ref{fig:flare_fit}. The flares were fitted with a three-component model using the formalism from \citeMethods{Mendoza2022} described by equation (\ref{eq:flaremodel}) and fitted with \texttt{emcee} \citeMethods{For2013}:

\begin{equation}
      F_{\mathrm{flare}, t} = \frac{\sqrt{\pi}}{2} A C * \Big( F_1 h(t, A, B, C, D_1) + F_2 h(t, A, B, C, D_2) \Big)
      \label{eq:flaremodel}
\end{equation}
with
\begin{equation}
    h(t, B, C, D) = e^{-D t + (\frac{B}{C} + \frac{D C}{2})^2} \text{erfc}\left(\frac{B-t}{C} + \frac{C D}{2}\right)
\end{equation}
where, $t$ is the relative time; $A$ is the amplitude; $B$ is the position of the peak of the flare; $C$ is the Gaussian heating timescale; $D_1$ is the rapid cooling phase timescale; $D_2$ is the slow cooling phase timescale; and $F_2 = 1 – F_1$ describes the relative importance of the exponential cooling terms.
The parameters resulting from this fit are presented in \ref{flares_stats}.

\begin{table*}[h!]
\setlength{\tabcolsep}{1.1pt} 
\small
\centering                          
\begin{tabular}{|p{1.4cm}|c c|c c|c c|c c|}        
\hline
{Flare \#}
& \multicolumn{2}{|c|}{\textbf{Timing $\pm$ 1-$\sigma$ } } & \multicolumn{2}{|c|}{\textbf{Amplitude $\pm$ 1-$\sigma$ } }
& \multicolumn{2}{|c|}{\textbf{Duration $\pm$ 1-$\sigma$ }} & \multicolumn{2}{|c|}{\textbf{Flare energy  }} \\ 
 & \multicolumn{2}{|c|}{\textbf{[JD - 2460000]}} & \multicolumn{2}{|c|}{\textbf{[ppt]}}
& \multicolumn{2}{|c|}
{\textbf{\textbf{(min)}}} & \multicolumn{2}{|c|}{\textbf{\textbf{(erg)}}} \\
 & \multicolumn{2}{|c|}{\textbf{B in eq. (\ref{eq:flaremodel})}} & \multicolumn{2}{|c|}{\textbf{A in eq. (\ref{eq:flaremodel})}}
& \multicolumn{2}{|c|}
{\textbf{\textbf{D1 in eq. (\ref{eq:flaremodel})}}} & \multicolumn{2}{|c|}{} \\
\hline
\hline
1&272.808004 & ~~4.44E-04 & 6.24 & 0.17 & 8.74 & 2.11 & 1.389047e+33 & \\ 
2&272.815411 & ~~9.6E-04 & 1.89 & 0.91 & 9.92 & 4.47 & 1.526377e+33 & \\ 
3&272.832834 & ~~3.93E-04 & 3.55 & 0.37 & 8.97 & 2.47& 2.974541e+33  & \\ 
\hline                                   
\end{tabular}
   \caption{\textbf{Output flare parameters from fit using Mendoza et al. 2022.} Timing, amplitudes, and duration of the flares are measured through a MCMC analysis of the light curve centered on the flaring events.}
\label{flares_stats}
\end{table*}
Assuming that the star is a blackbody radiator, the bolometric flare luminosity can be defined as equation:
\begin{equation}
    L_{\mathrm{flare,bol}} = \sigma_{SB} T_{\mathrm{flare}}^{4} A_{\mathrm{flare}},
    \label{boloc_lumi}
\end{equation}
where $\sigma_{SB}$ is the Stefan-Boltzmann constant, $T_{\mathrm{flare}}$ is the black-body temperature of the flare, and $A_{\mathrm{flare}}$ is the area of the flare.
Then, to estimate $A_{\mathrm{flare}}$, we use the observed luminosity of the star ($L_{\star}$) and the luminosity of the flare ($L_{\mathrm{flare}}$) defined by equation (\ref{lum_flare}), where the integration is made for the F1500W MIRI filter band.

\begin{equation}
    L_{\mathrm{flare,F1500W}}(t) = A_{\mathrm{flare}}(t) \bigintsss_{F1500W} R_{\lambda} B_{\lambda}(T_{\mathrm{flare}}) d\lambda
    \label{lum_flare}
\end{equation}
and the flare amplitude in relative flux $\frac{\Delta F(t)}{F(t)}$ is defined as equation (\ref{ampli_flare}) :
\begin{equation}
    \frac{\Delta F(t)}{F(t)} = \frac{L_{\mathrm{flare,F1500W}}(t)}{L_{\star}}.
    \label{ampli_flare}
\end{equation}

In equation (\ref{lum_flare}), $R_{\lambda}$ stands for the spectral response function of the F1500W filter and $B_{\lambda}$ is the Planck function. From equations (\ref{lum_flare}) and (\ref{ampli_flare}), we can derive $A_{\mathrm{flare}}$, see equation (\ref{A_flare}) :

\begin{equation}
    A_{\mathrm{flare}} =  \frac{\Delta F(t)}{F(t)} \pi R_{\star}^{2}  \frac{\int R_{\lambda} B_{\lambda}(T_{\mathrm{eff}}) d\lambda}{\int R_{\lambda} B_{\lambda}(T_{\mathrm{flare}}) d\lambda}
    \label{A_flare}
\end{equation}

Finally, the total bolometric energy of the flare ($E_{\mathrm{flare,bol}}$) is defined as the integral of $L_{\mathrm{flare,bol}}$ over the flare duration, equation (\ref{E_flare}) :

\begin{equation}
    E_{\mathrm{flare,bol}} = \int L_{\mathrm{flare,bol}}(t) dt
    \label{E_flare}
\end{equation}

\begin{figure}
    \centering
    \includegraphics[width=0.45\textwidth]{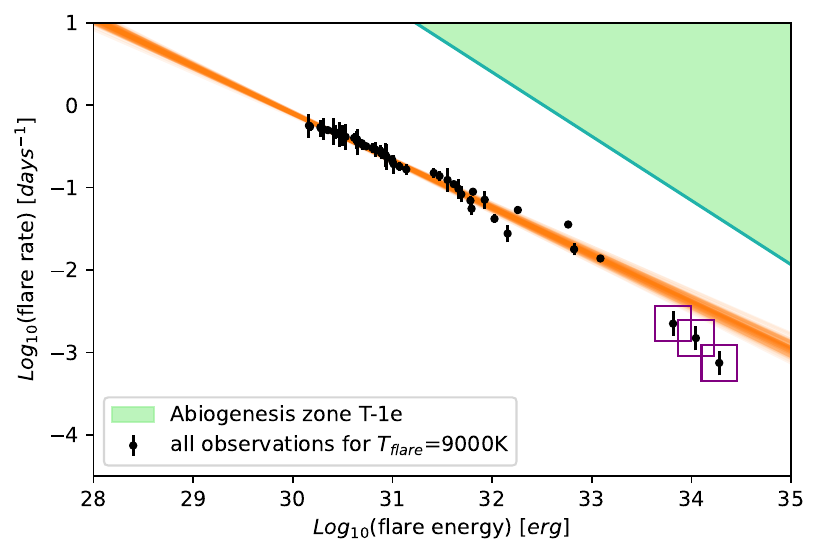}
    \includegraphics[width=0.45\textwidth]{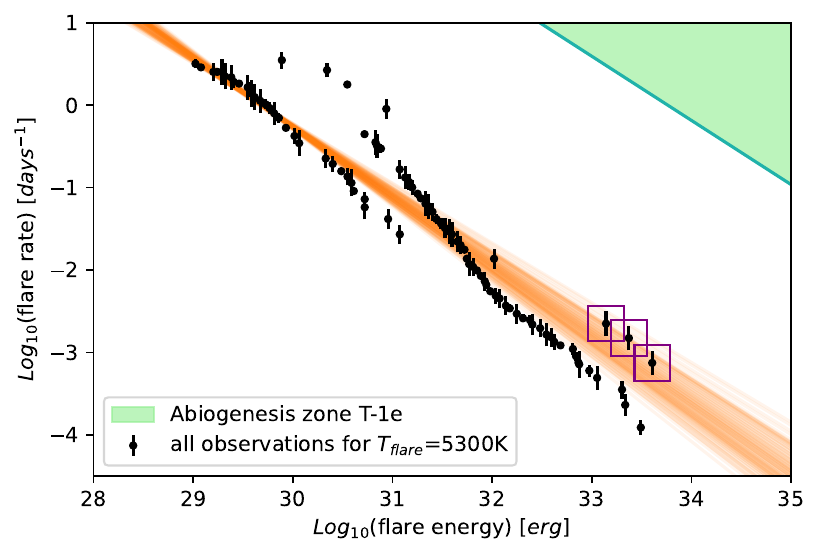}
    \caption{\textbf{Flare frequency distribution plot for TRAPPIST-1}. {\it Left:} FFD using data from ref. \protect\citeMethods{Ducrot2020} and refs. \protect\citeMethods{Vida_2017, Paudel2019} and the two newly observed flares from MIRI F1500W observations, assuming a 9000 K blackbody for the flares. Orange lines show the results from 200 best fit from the MCMC. {\it Right: }Same but using data from \protect\citeMethods{Howard2023} and  assuming a 5300 K blackbody for the flares. }
    \label{fig:FFD}
\end{figure}

From eq. (\ref{E_flare}) we derive the energy of the three flares, see \ref{flares_stats}. We note that this energy is dependent on the temperature that we assume for the flare. It is common practice to assume a temperature of 9000 K in the literature \citeMethods{Gunther2019, Howard2018}. However, a recent study by refs. \citeMethods{Howard2023, Maas2022} show that the spectra of the flares observed with MuSCAT1 and JWST/NIRISS instruments are well fitted with black body temperatures lower than the commonly used 9000K. Most recently, \citeMethods{Howard2023} shows that the flares are well-described by a blackbody emission with an effective temperature 5300 K. To address both of these scenarios we compute the flare energies and frequencies assuming first that the temperature of the flare is 9000 K and then assuming it is 5300 K. In \ref{fig:FFD} we show the resulting flare frequency distribution (FFD) for these two scenarios.
In both cases the three flares that we detect with MIRI at 15$\mu m$ are among the most energetic ever observed. These observations are remarkable but also highlight the possible presence of several "mini flares" in the data set, adding more red noise. Furthermore, having gained valuable insights from this experience, we highly recommend performing simultaneous observations of TRAPPIST-1 in a bluer band whenever possible, preferably from the ground, for all future JWST observations (GO 6456). This approach would enable the detection of lower energy flares that might not be significantly visible in the NIR light curve but which could introduce correlated noise into the observations. In addition these observations are underlining the important of programs such as GO 7068 that aim to provide a library of M-dwarfs flares\footnote{\url{https://www.stsci.edu/jwst-program-info/download/jwst/pdf/7068/}}.

Finally we run a MCMC analysis to derive that the occurrence rate as a function of flare energy can be described as:
\begin{equation}
    \log(\nu) = \beta \log(E) + \alpha
\end{equation}
Where $\beta = -0.8467 \pm 0.05 $ for $T_{flare}=9000$ K and  $\beta = -0.84 \pm 0.04 $ for $T_{flare}=5300$ K. In comparison ref. \citeMethods{Ducrot2020} found $\beta = -0.63 \pm 0.14$, fitted from our measurements, ref. \citeMethods{Paudel2019} found $\beta = -0.61 \pm 0.02$, ref. \citeMethods{Maas2022} found $\beta = -0.45 \pm 0.20$, and ref. \citeMethods{Vida_2017} found $\beta = -0.59$.

\section*{Data availability} The program 3077 JWST data used in this work are available on the MAST online database (\url{https://mast.stsci.edu/portal/Mashup/Clients/Mast/Portal.html}).

\section*{Code availability}The code \texttt{TRAFIT} used to analyse the light curves is a \texttt{Fortran 2003} code that can be obtained from the first author on reasonable request.  The \texttt{Generic PCM} (and documentation on how to use the model) can be downloaded from the SVN repository \url{https://svn.lmd.jussieu.fr/Planeto/trunk/LMDZ.GENERIC/}. More information and documentation on the model are available on \url{http://www-planets.lmd.jussieu.fr}.
The N-body code \texttt{Posidonius} and its documentation are available from \url{https://github.com/revolal/posidonius/tree/kaula\_v2}.

\section*{Authors' contributions}
MG co-led the project with ED. Both of them wrote the JWST proposal of program 3077, interacted with STScI to set up observations, performed an analysis of the data, and wrote a significant part of the paper.
T.J.B. contributed one of the independent analyses using the \texttt{Eureka!}\ pipeline and contributed significantly to the writing and editing of the paper.
X.L. and D.D.B.K. designed the strategy and experiments for the bare rock models and space weathering simulations of TRAPPIST-1 b and c, and wrote the corresponding parts of the paper.
V.S.M. and A.P.L. designed the strategy and experiments for day-night climate-photochemical-spectral modeling of the TRAPPIST-1 b and c planets. A.P.L. conducted the day-night climate-photochemical simulations, and prepared the associated secondary eclipse spectra and phase curves of TRAPPIST-1 b and c. V.S.M. and A.P.L. interpreted the results and wrote the corresponding parts of the paper. E.A., A.P.L., V.S.M., and J.L-Y. developed the calculation of the confidence of phase curve model fits to the data.
A.M. and M.T. performed the 3D GCM simulations of TRAPPIST-1b and c for the JWST proposal of program 3077 and the present paper. A.M., M.T., T.F. and J.L. computed the associated eclipses, phase curves and transit spectra for the proposal and the present paper. A.M., M.T., F.S., J.L. and T.F. designed the strategy for 3D climate modeling of the TRAPPIST-1b and c planets, and wrote the corresponding parts of the paper.
AH and DB performed a data reduction and analysis of the photometric time-series.
BOD conducted an independent reduction of the data and analysed the photometric time-series.
TG contributed to the design of the observations and contributed the JWST program 1177 observations used in the joint analyses.
Z.H. and C.D. performed an independent data reduction and analysis of the JWST/MIRI data and wrote the corresponding part of the manuscript.
AR performed N-body simulations of TRAPPIST-1 including tides with \texttt{Posidonius} to estimate the rotation of the planets. AR and EB performed estimations of the tidal heating that planets b and c are experiencing, within the framework of a realistic dissipation response. AR, EB and FS wrote the corresponding part of the paper.\\


\noindent \textbf{Acknowledgments}
{\small We thank Olivia Lim for sharing their transmission spectra of TRAPPIST-1 b obtained from the reduction and analysis of JWST/NIRISS data and published in their paper ref. \cite{Lim2023}.
This work is based on observations made with the NASA/ESA/CSA JWST. The data were obtained from the Mikulski Archive for Space Telescopes at the Space Telescope Science Institute, which is operated by the Association of Universities for Research in Astronomy, Inc., under NASA contract NAS 5-03127 for JWST. These observations are associated with programs 1177, 1279, 2304 and 3077. MG is F.R.S-FNRS Research Director. His contribution to this work was done in the framework of the PORTAL project funded by the Federal Public Planning Service Science Policy  (BELSPO) within its  BRAIN-be: Belgian Research Action through Interdisciplinary Networks program.
The authors thank the Belgian Federal Science Policy Office (BELSPO) for the provision of financial support in the framework of the PRODEX Programme of the European Space Agency (ESA) under contract number 4000142531.
T.J.B.\ and T.P.G.\ acknowledge funding support from the NASA Next Generation Space Telescope Flight Investigations program (now JWST) via WBS 411672.07.05.05.03.02. T.J.B. also acknowledges funding support from STScI provided with GO 3077 through JWST-GO-03077.015-A.
C.D. and Z.H. acknowledge funding support from NASA Exoplanets Research Program (grant no. 80NSSC23K1115), the Space Telescope Science Institute (GO-3077; JWST-GO-03077.019-A) and the Alfred P. Sloan Research Fellowship.
X.L. and D.D.B.K. acknowledge support from the National Natural Science Foundation of China (NSFC) under grants 12473064 and 42250410318.
M.T. acknowledges support from the Tremplin 2022 program of the Faculty of Science and Engineering of Sorbonne University. M.T., A.M. and J.L. thank the Generic PCM team for the teamwork development and improvement of the model. M.T. acknowledges support from the High-Performance Computing (HPC) resources of Centre Informatique National de l'Enseignement Supérieur (CINES) under the allocations No. A0120110391 and A0140110391 made by Grand Équipement National de Calcul Intensif (GENCI).
V.S.M., A.P.L.\ and E.A.\ are part of the Virtual Planetary Laboratory Team, which is a member of the NASA Nexus for Exoplanet System Science, and financed through NASA Astrobiology Program grant 80NSSC18K0829. V.S.M. and A.P.L. made use of the advanced computational, storage, and networking infrastructure provided by the Hyak supercomputer system at the University of Washington.
B.-O. D. acknowledges support from the Swiss State Secretariat for Education, Research and Innovation (SERI) under contract number MB22.00046.
A.R. and E.B. acknowledge the financial support of the SNSF (grant number: 200021\_197176 and 200020\_215760). This work has been carried out within the framework of the NCCR PlanetS supported by the Swiss National Science Foundation under grants 51NF40\_182901 and 51NF40\_205606. The N-body computations including tides with \texttt{Posidonius} were performed at University of Geneva on the Baobab and Yggdrasil clusters.
The GO 3077 team offers a special thanks to Sarah Kendrew and the JWST Science Operations' Planning and Scheduling Team at the Space Telescope Science Institute for their dedicated efforts to optimize and schedule the long MIRI observation that made the entirety of this work possible.}



\newpage
\section*{Supplementary Figures}
\renewcommand{\figurename}{\hspace{-4pt}}
\renewcommand{\thefigure}{Supplementary Fig.~\arabic{figure}}
\renewcommand{\theHfigure}{Supplementary Fig.~\arabic{figure}}
\renewcommand{\tablename}{\hspace{-4pt}}
\renewcommand{\thetable}{Supplementary Table \arabic{table}}
\renewcommand{\theHtable}{Supplementary Table \arabic{table}}
\setcounter{figure}{0}
\setcounter{table}{0}

\begin{figure}[!ht]
   \centering
   \includegraphics[width=\textwidth]{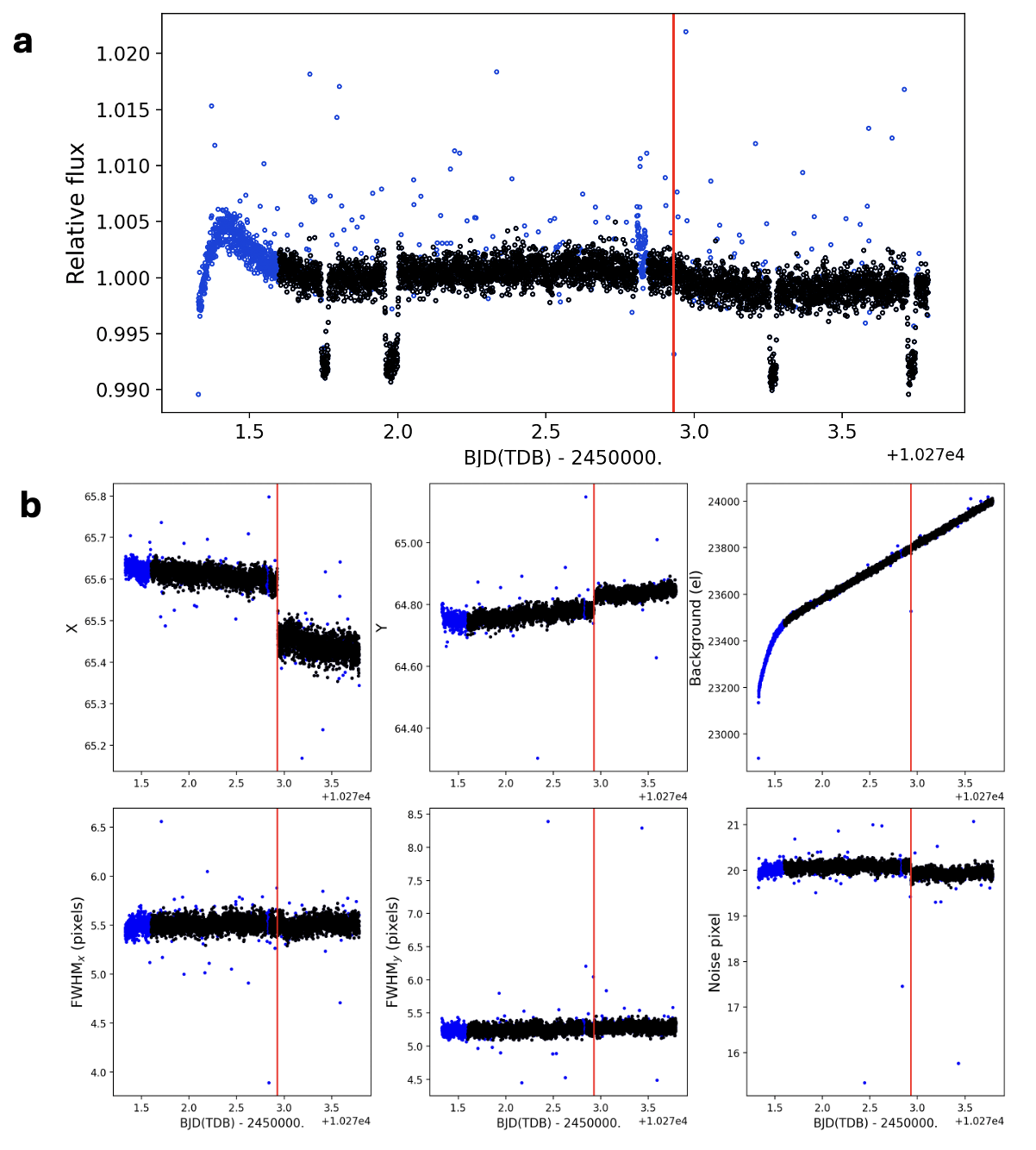}
   \caption{\textbf{Program 3077 raw light curve and external parameters evolution.} \textbf{a}, raw light curve obtained by MG with a photometric aperture of 4 pixels radius (see text for details). \textbf{b}, evolution of the external parameters: $x$ and $y$ positions of the target's point-spread function (PSF) center, background, PSF's full-width at half-maximum  in the $x$ and $y$ directions, and  noise-pixel (as defined in \protect\citeMethods{Deming2015}). Blue points are those discarded before the MCMC analysis. The vertical red line shows the gap between visit 1 and 2.}
   \label{fig:lc_raw}
\end{figure}

\begin{figure}
   \centering
    \includegraphics[width=\textwidth]{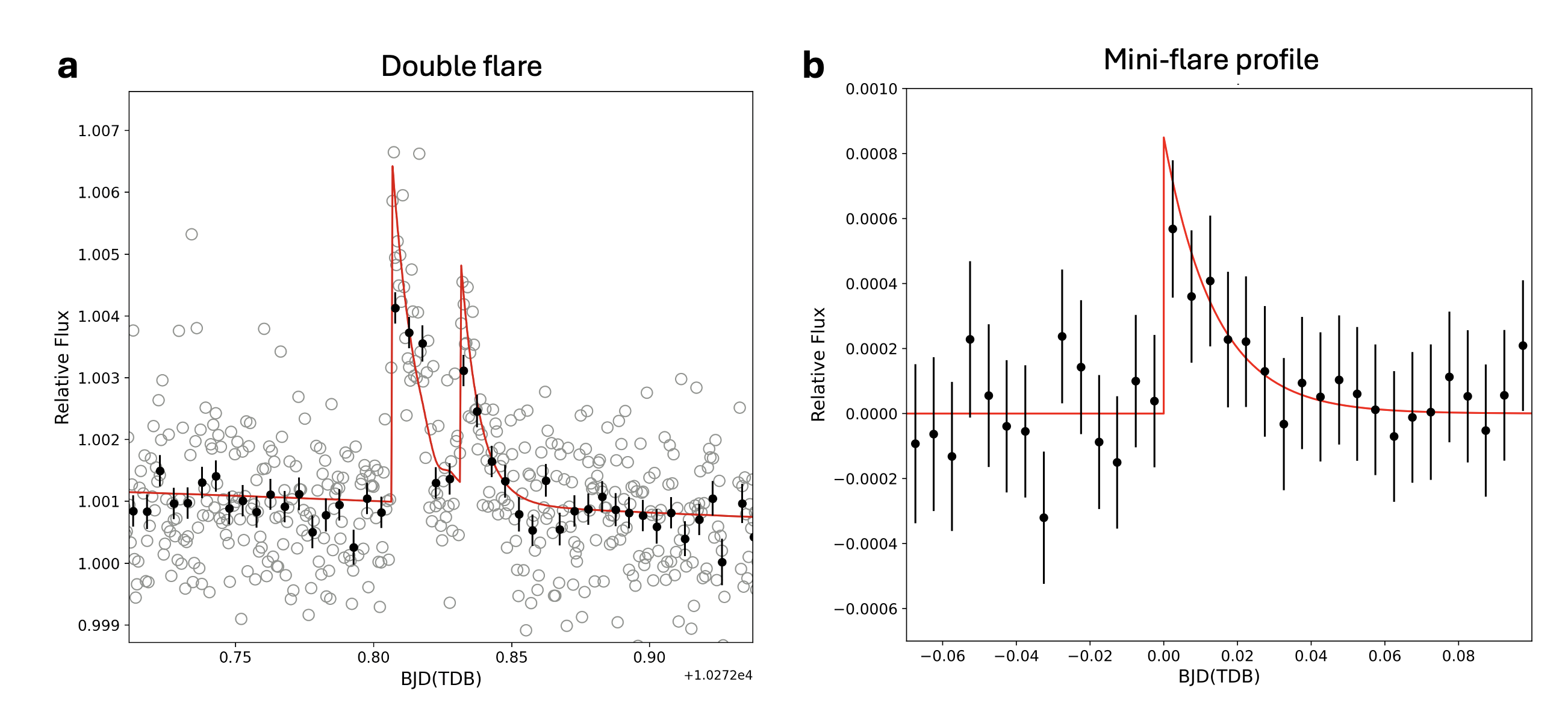}
    \caption{\textbf{Flares in the light curve of program 3077.} \textbf{a}, double flare-like structure in the light curve of program 3077. The black points are flux measurements binned per 7.2 minutes. \textbf{b}, Stack of the four low-amplitude flare-like structures in the residuals of MG's analysis \#1, binned per 7.2 minutes. For both panels, the error bars are the averages of the individual errors divided by the square root of the number of points within the bin, and the best-fit flare models are overimposed in red.
   }
   \label{fig:miniflare}
\end{figure}

\begin{table}
\begin{tabular}{@{}llll@{}}
\toprule
Light curve & Aperture & Baseline function & $\sigma$ \\
& (pixels) & & (ppm) \\
\midrule
3077 (23 Nov 2023) & 4  &  2 ramps + $p(t) + p(x^2) + p(y^2)$ & 862 \\
1177.1 (8 Nov 2022) & 3.5 & ramp + $p(t) + p(y)$  & 907 \\
1177.2 (12 Nov 2022) & 4 & ramp + $p(t^2)$ & 865 \\
1177.3 (20 Nov 2022) & 3.5 &  ramp  & 876 \\
1177.4 (24 Nov 2022) & 4 &  ramp + $p(t)$ & 1040 \\
1177.5 (3 Dec 2022) & 4 & ramp + $p(y)$ & 908 \\
2304.1 (27 Oct 2022) & 3.5 & ramp + $p(t)$ & 954  \\
2304.2 (30 Oct 2022) &  3.5  & ramp + $p(t) + p(x) + p(y)$  & 913 \\
2304.3 (6 Nov 2022) & 3.5  & ramp & 985  \\
2304.4 (30 Nov 2022) & 3.5  &  ramp + $p(t) + p(x^2)$ & 921  \\
\bottomrule
\end{tabular}
    \caption{\textbf{Light curves used in this work. } First column: light curve (program + date). Second column: photometric extraction aperture. Third column: baseline functions used for each light curve of program 3077, 1177, and 2304 in the global MCMC analysis of MG. The last column gives the standard deviation of the residuals of the best-fit model. $p(a^b)$ represents a polynomial of degree $b$ of the parameter $a$. The ramp model used for all light curves of programs 1177 and 2304 is a quadratic function of the logarithm of the time difference between the considered data point and the first data point. }
 \label{baseline}
\end{table}

\begin{figure}
   \centering
    \includegraphics[width=\textwidth]{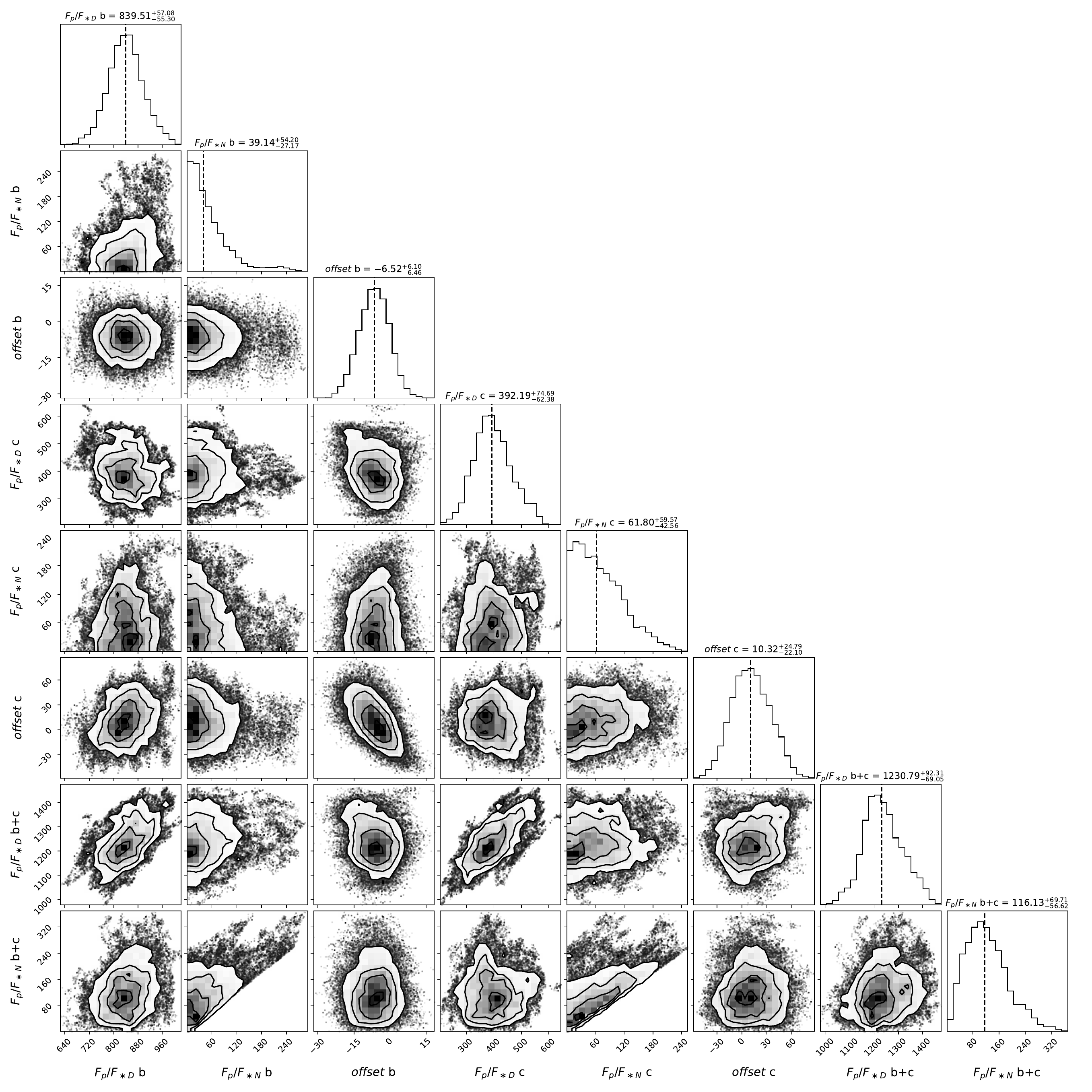}
    \caption{\textbf{Corner plot \protect\citeMethods{corner} for the phase curve parameters of planet b and c inferred from MG's analysis \# 1.} The posterior PDFs for the sum of the day- and night-side fluxes of planets b and c are also shown.}
   \label{fig:corner_plot}
\end{figure}


\begin{figure*}
    \centering
    \includegraphics[width=\linewidth]{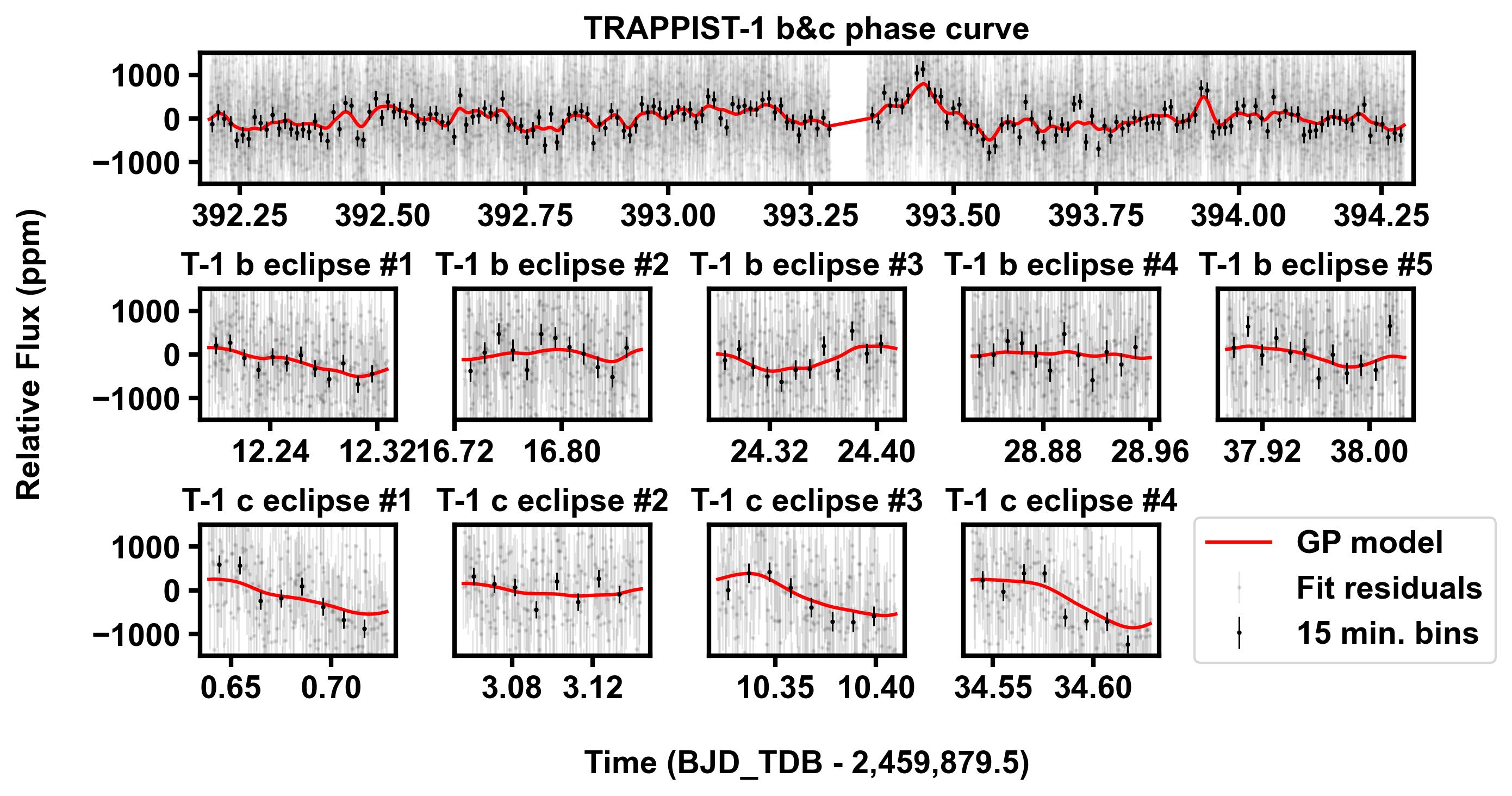}
    \caption{\textbf{Fitted GP time-series models from TJB's reduction and analysis with \mbox{Setup \#1}.}}\label{fig:tjb_GP_lightcurves}
\end{figure*}

\begin{figure*}
    \centering
    \includegraphics[width=\linewidth]{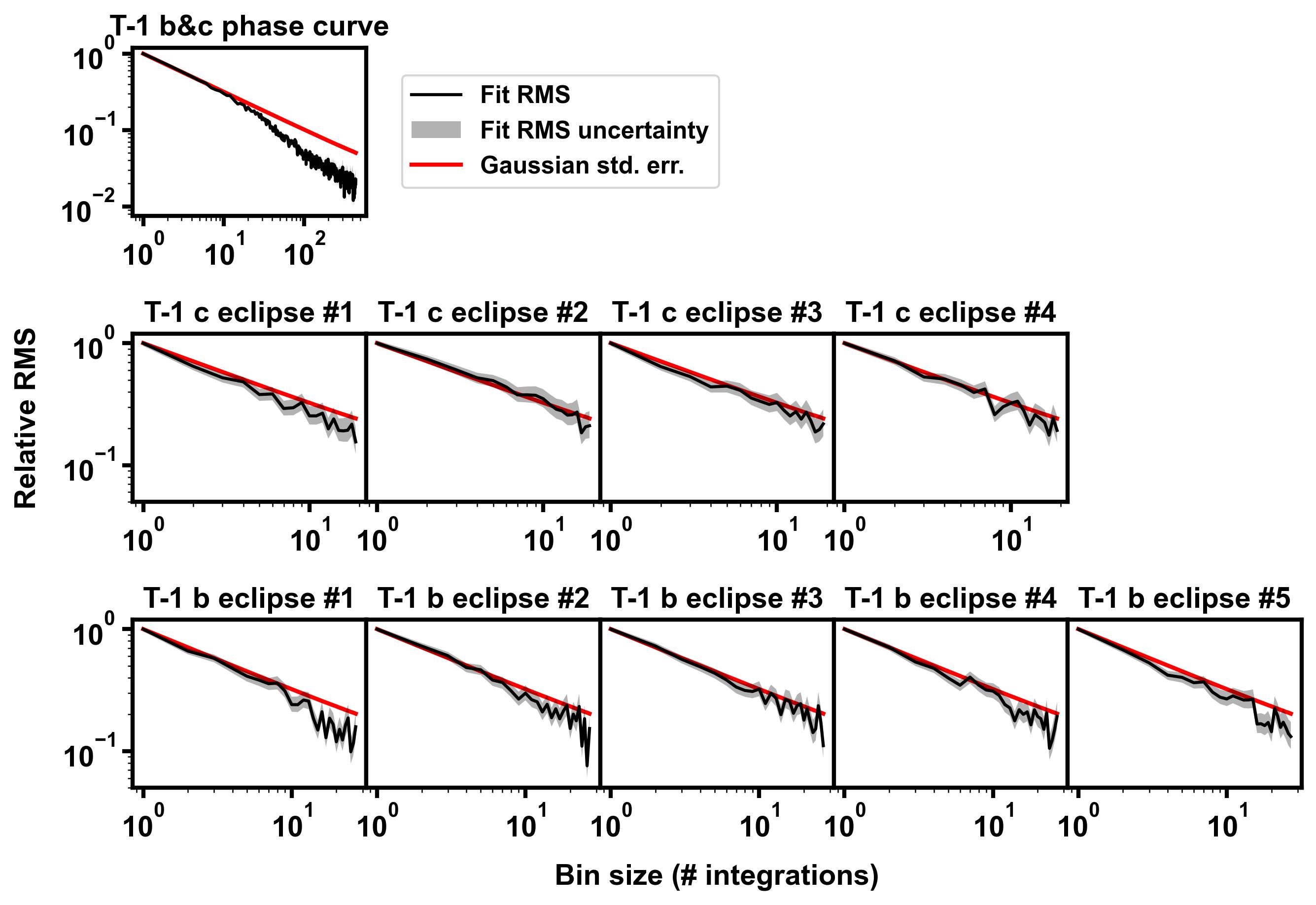}
    \caption{\textbf{Allan variance plots from TJB's reduction and analysis with \mbox{Setup \#1}}.  Note that the reduction in RMS at long timescales (Bin size $>$20 integrations) in the phase curve panel is due to the subtraction of correlated variations as estimated from the GP. }\label{fig:tjb_AllanPlots}
\end{figure*}


\begin{figure}
   \centering
    \includegraphics[width= 0.4\textwidth]{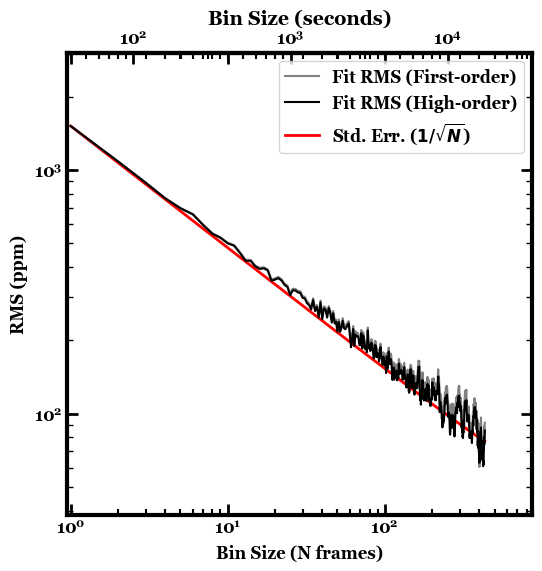}
    \caption{\textbf{Allan plot from ZH data analysis}}
   \label{fig:ZH_Fitting_AllanPlot}
\end{figure}

\end{document}